\documentclass[11pt]{JHEP3}
\usepackage{amsmath,amssymb,amsfonts,bbm,latexsym,axodraw,cite}

\def\clap#1{\hbox to 0pt{\hss#1\hss}}

\newcommand{\be}{\begin{equation}}
\newcommand{\ee}{\end{equation}}
\newcommand{\bea}{\begin{eqnarray}}
\newcommand{\eea}{\end{eqnarray}}

\newcommand{\hsp}{\hspace{-2mm}/}

\newcommand{\ra}{\rightarrow}

\newcommand{\cK}{{\cal K}}
\newcommand{\cL}{{\cal L}}
\newcommand{\cJ}{{\cal J}}

\newcommand{\cN}{{\cal N}}
\newcommand{\cO}{{\cal O}}

\newcommand{\cI}{{\cal I}}
\newcommand{\bZ}{{\bf Z}}

\newcommand{\Li}{\mbox{Li}}

\preprint{DESY 06-028; DAMTP-2005-85}

\title{Higher derivatives  and brane-localised kinetic 
 terms in gauge theories on orbifolds}
\author{
Dumitru M. Ghilencea$^{\diamond,\,}$\footnote{New address after 1 March 2006.}\,\,\,,
Hyun Min Lee$^\dagger$ and Kai Schmidt-Hoberg$^\dagger$ \\

	$^\diamond$\upshape D.A.M.T.P., Centre for Mathematical Sciences,
	University of Cambridge, \\
	Wilberforce Road, Cambridge CB3 OWA, United Kingdom.

\vspace{0.2cm}
	$^\dagger$\upshape Deutsches Elektronen-Synchrotron DESY,\\ 
        Notkestra\ss e 85, 22607 Hamburg, Germany.\\ 
	\upshape e-mail: {\ttfamily hyun.min.lee@desy.de, 
        kai.schmidt.hoberg@desy.de }

\vspace{0.2cm}
	$^*$\upshape Rudolf Peierls Centre for Theoretical Physics,
         University of Oxford, \\
	1 Keble Road, Oxford, United Kingdom.\\
	\upshape e-mail: {\ttfamily d.ghilencea1@physics.ox.ac.uk}}

\abstract{
We perform a detailed analysis of one-loop corrections to the self-energy
of the (off-shell) gauge bosons 
in six-dimensional ${\cal N}\!=\!1$ supersymmetric gauge 
theories on orbifolds. 
After discussing the Abelian case in the standard Feynman diagram approach, 
we extend the analysis to the non-Abelian case by employing the  method
of an orbifold-compatible one-loop effective action for a classical 
background gauge field. We find that bulk higher derivative and brane-localised
 gauge kinetic terms are required to cancel one-loop divergences
of the gauge boson self energy. 
After their renormalisation we study the momentum dependence 
of both the higher derivative coupling $h(k^2)$ 
and the {\it effective} gauge coupling $g_{\rm eff}(k^2)$.
For momenta  smaller than the compactification scales, we obtain 
the 4D logarithmic running of $g_{\rm eff}(k^2)$,
with suppressed power-like corrections, while the higher derivative
coupling is constant. We present in detail the threshold
corrections to the low energy gauge coupling, due to the massive bulk modes.
At momentum scales above the compactification scales, the
higher derivative operator becomes important and leads to 
a power-like running of $g_{\rm eff}(k^2)$ with respect to the momentum scale.
The coefficient of this running is at all scales
equal to the renormalised coupling of the higher derivative operator 
which ensures the quantum consistency of the model. 
We discuss the relation to the similar one-loop correction 
in the heterotic string, to show that the higher derivative operators
are relevant in that case too, since the field theory limit 
of the one-loop string correction does not commute with the infrared regularisation
of the (on-shell) string result.
}


\begin{document}
\section{Introduction}

In recent years, the study  of additional compact space dimensions in an
 effective field theory framework \cite{add} has been popular in the
 particle physics community, since this provides new possibilities
for searching for physics beyond the Standard Model. Although string theory 
may present a better set-up for such studies,   effective field
 theories also allow a fully consistent investigation of 
quantum effects associated with (large) extra dimensions, and may 
even capture effects not seen by the {\it on-shell} string. 
Since no additional space dimensions are observed at low energies, 
these have to be  compactified at sufficiently high
 scales\footnote{Non-compact, infinite extra dimensions are also 
possible \cite{randall}.}. 
In field theory approaches only simple 
covering spaces  are usually considered, such as $S^1$, $T^2...$,  
sufficient however to capture the main effects investigated. 
To obtain 4D chiral fermions from  bulk fields
  discrete symmetries must act (non-freely) upon
 the extra dimensions, resulting in orbifolds such as 
$S^1/\mathbbm{Z}_2$ or $T^2/\mathbbm{Z}_N$ ($N=2,3,4,6$).
 These orbifolds have fixed points, invariant  under subgroups of the
 discrete group action. 
Since the bulk fields satisfy boundary 
conditions at the orbifold fixed points, momentum conservation 
does not hold in the extra dimensions. 
 Ultimately, brane-localised (either 4D or 
 higher derivative) interactions  are required as counterterms 
\cite{ggh,schmaltz,Ghilencea:2004sq,Ghilencea:2005hm,Ghilencea:2005nt}, 
to ensure the quantum consistency  of the models. 
New bulk interactions, in addition to the original ones, 
are also generated dynamically 
\cite{Ghilencea:2005nt,santamaria,nibbelink,nibbmark,
Ghilencea:2003xj,Alvarez:2006we}
as higher  dimensional (derivative) terms.

In this paper we consider the one-loop correction 
to the self-energy of gauge bosons in 6D ${\cal N}=1$ supersymmetric
Abelian and non-Abelian gauge theories coupled to hypermultiplets 
on the $T^2/\mathbbm{Z}_2$ orbifold, within the component field formulation. 
We find that one-loop divergences are generated which require the
addition of new counterterms. These involve 
new, brane-localised 4D interactions as well as higher derivative, bulk
gauge interactions, not present in the original action. We provide a careful
study of the role of these operators in the running of the gauge
coupling at high and low momentum scales. We also discuss  
the link between these one-loop corrections and those in string theory. 
These are the main purposes of this paper. 
Recent work on this topic can be found in \cite{nibbelink,nibbmark}
in the superfield formalism (for related studies 
see also \cite{Taylor:1988vt}).

In the Abelian case, we use the Feynman diagram approach to consider
bulk scalar and fermion contributions to the self-energy 
of the gauge bosons. We find that the fermions give rise to a bulk
divergence only, requiring a  bulk higher derivative
counterterm. At the technical level, the origin of
 this divergence is the presence of infinite double sums over the modes
and  a re-summation of their individual divergent contributions
\cite{Ghilencea:2004sq,Ghilencea:2005hm,Ghilencea:2005nt,nibbelink,nibbmark,
Ghilencea:2003xj,Ghilencea:2002ak}. In contrast, bulk complex scalars bring in
both  bulk and brane corrections. Their divergent part must be cancelled
by bulk higher derivative and  brane-localised gauge kinetic counterterms, 
respectively. Both fermionic and bosonic contributions 
also contain finite Lorentz violating mass terms in the bulk.
For a hypermultiplet 
there are neither brane contributions nor bulk Lorentz
violating mass terms. Thus, even after compactification, 
the Lorentz invariance in these mass corrections is protected 
by the initial supersymmetry. Nonetheless, one still needs a bulk 
higher derivative counterterm, which reflects the non-renormalisable nature 
of the initial, higher dimensional field theory.

The above analysis is extended  to the  non-Abelian case by employing a
background field method which is made consistent with the orbifold boundary
conditions. This formalism can be generalised to other orbifold actions, 
such as Wilson lines. 
The results show that  a hypermultiplet generates only a  bulk 
loop correction, just like in the Abelian case, while a vector multiplet
generates both bulk and brane-localised contributions. These
contributions contain divergent terms which are cancelled 
by bulk higher derivative  and brane-localised gauge kinetic counterterms.
After the renormalisation of these operators, the  running of the 
one-loop {\it effective} coupling $g_{\rm eff}(k^2)$, which is the coupling
of the zero mode gauge bosons, is controlled by 
finite terms coming from 
both bulk and branes. This will be discussed in detail.

In the limit of external momenta $k^2$ smaller than the
compactification scale(s), the higher derivative gauge kinetic term 
is suppressed. In this case, after considering both bulk and brane
one-loop effects, 
we show that the {\it effective} gauge coupling  has 
a 4D {\it logarithmic} running with respect to the momentum $k^2$, 
with the 4D ${\cal N}\!=\!1$ beta function. This is an interesting
result and a consistency check of our calculation.
The logarithmic running in momentum originates from both bulk and brane
contributions. We also establish a relation between
the high scale physics ($g_{\rm tree}$) and $g_{\rm eff}(k^2\!\ll\! 1/R^2_{5,6})$,
which involves re-summing  threshold corrections due to infinitely many
massive Kaluza-Klein modes. We provide detailed  expressions of these 
corrections including finite terms. This relation is little dependent
on the role of the higher derivative operator, 
strongly suppressed at such low momentum scales.
The  running of the effective coupling with respect to $k^2$
can  be extended to larger values of $k^2$, closer to 
compactification scales ($k^2\sim 1/R^2_{5,6}$), 
to reach the regime of dimensional cross-over \cite{O'Connor:zj}.  
In this case the higher derivative
operator brings in an important contribution to the effective gauge coupling.  
After its renormalisation,  there are non-negligible power-like corrections 
in momentum scale to $g_{\rm eff}(k^2)$.
The coefficient of the power-like running 
is the renormalised coupling $h(k^2)$ of the higher
derivative operator, which below the compactification scales is
constant while far above them  it runs logarithmically with respect to
the momentum scale. 
At even higher momentum scales $k^2\gg  1/R_{5,6}^2$
we show that $g_{\rm eff}(k^2)$  has a power-like running with respect to
the high momentum scale, with a coefficient equal to the renormalised 
coupling of the higher derivative operator. 

The link  of these corrections to similar results from string theory 
is addressed. We discuss the relation of our result to  string corrections in 
the type I strings \cite{ADB} and  heterotic toroidal orbifolds
\cite{Dixon:1990pc2,Mayr:1993mq} with $\cN=2$ sub-sectors.
 Although  the {\it on-shell} (heterotic) string calculation to the gauge 
boson self-energy
 misses contributions associated with higher derivative operators, 
we show that there  are remnant effects  of their presence,  even 
in the (on-shell) string result. 
These effects are related to the fact that the  infrared
 regularisation  of the (heterotic) string loop corrections 
and their   $\alpha'\rightarrow 0$  limit  do not commute, leaving a
troublesome  UV-IR mixing in the effective  field theory regime of
 the (heterotic) string ($\alpha'\rightarrow 0$). This stresses the
 importance of investigating the role  of such operators  in string theory,
 too. 

The results for the self-energy of the gauge bosons in our 
component field formulation are fully consistent with those 
obtained in the superfield formulation. 
Nevertheless, the gauge fixing term and 
the associated ghost Lagrangian considered are not invariant 
under the original supersymmetry transformation. This is related to 
the well-known fact that the Wess-Zumino gauge 
is not consistent with a super-covariant gauge fixing \cite{wz}. 
This problem is very common in similar works,  and
becomes manifest in the fact that the anomalous dimensions of
scalar and fermion matter fields in a chiral multiplet are 
not equal at one-loop level \cite{jones}. 
However, for our case of  the self-energy of the gauge bosons,  
additional auxiliary multiplets required by a manifestly
supersymmetric quantisation will not change the result, 
as discussed  for the holomorphic anomaly to the gauge coupling 
in 4D supersymmetric gauge theory \cite{ah}. 

The paper is organised as follows.  We start with a 
6D ${\cal N}=1$ supersymmetric Abelian gauge theory where the
one-loop correction to the gauge bosons
 is computed. Then we employ the higher dimensional background field method to
 find the one-loop effective action of non-Abelian
gauge theories and  apply this formalism to $T^2/\mathbbm{Z}_2$,
 using orbifold-compatible functional differentiations.
Finally we  discuss  the running of the effective gauge coupling.
Technical details of our calculations are given in the Appendix.

\section{One-loop vacuum polarisation to $U(1)$ gauge bosons
on orbifolds}\label{section2}

We consider the one-loop vacuum polarisation in a 6D 
${\cal N}=1$ supersymmetric Abelian gauge theory coupled to hypermultiplets. 
The two extra dimensions are denoted by the complex coordinate $z=x_5+i x_6$, 
and are  compactified on the orbifold $T^2/\mathbbm{Z}_2$ 
with the  two radii $R_5$ and $R_6$. The torus is modded out 
by the $\mathbbm{Z}_2$ reflection, 
which identifies coordinates of extra dimensions under $z\rightarrow -z$.
Under this $\mathbbm{Z}_2$ action, there appear four fixed points
which transform into themselves.

In a 6D ${\cal N}=1$ supersymmetric gauge theory, a vector multiplet is
composed of gauge bosons $A_M$ and (right-handed) symplectic Majorana gauginos
$\lambda$ while a hypermultiplet is composed of two complex
hyperscalars $\phi_\pm$
with opposite charges and a (left-handed) hyperino $\psi$.
The supersymmetric action 
is given in component fields\footnote{We also included the auxiliary
fields ${\vec D}=(D^1,D^2,D^3)$ for completeness. We have written
gaugino and hyperino in 4D Dirac representations.}  by \cite{lenizu}
\be
S=S_{\rm vector}+S_{\rm hyper}\nonumber
\ee
with
\bea
\!\!S_{\rm vector}&=&\frac{1}{2}\int d^6 X\bigg[\!-\frac{1}{2}F_{MN} F^{MN}
\!+\!{\bar\lambda} i\gamma^M \partial_M\lambda
\!+\!\bar{\lambda}^c i\gamma^M \partial_M\lambda^c 
\!+\!\big\vert D^1+iD^2\big\vert^2
 \!+\!(D^3)^2 \bigg], \,\,\,\,\,\,
\label{vectoraction}\\[6pt]
S_{\rm hyper}&=&\int d^6 X\bigg[\sum_{\pm}|D_M\phi_\pm|^2
+{\bar\psi}i{\bar\gamma}^M D_M\psi+\sqrt{2}g\Big(
{\bar\psi}\lambda\phi^*_- +{\bar\psi}\lambda^c\phi_++{\rm c.c.}\Big) 
\nonumber \\
&&\qquad\qquad\quad -g\Big(
(D^1+iD^2)\phi_+\phi_-+{\rm c.c}\Big)
+gD^3 \Big(\phi^*_+\phi_+-\phi^*_-\phi_-\Big)\bigg], \label{hyperaction}
\eea
where $\lambda^c=C_5{\bar\lambda}^T$ is the five-dimensional charge conjugate
of $\lambda$,
${\mbox {\it D}_M\phi_\pm=(\partial_M\mp ig A_M)\phi_\pm}$, and
$D_M\psi=(\partial_M-ig A_M)\psi$. Details on our conventions 
are given in Appendix~\ref{appendix-a}.

To promote the $\mathbbm{Z}_2$-symmetry of the orbifold to a 
symmetry of our theory, we have to specify the $\mathbbm{Z}_2$ 
parities of the bulk fields. These parities are given by

\be
A_\mu(x,-z)=A_\mu(x,z),\quad
A_{5,6}(x,-z)=-A_{5,6}(x,z), \quad 
\lambda(x,-z)=i\gamma_5\lambda(x,z),  \nonumber
\ee
\be 
\phi_\pm(x,-z)=\pm\eta\, \phi_\pm(x,z), \quad
\psi(x,-z)=i\eta\,\gamma_5\psi(x,z) 
\ee
where $\eta$ can be chosen $+1$ or $-1$. Within this framework, we evaluate the
contributions to the 4D one-loop self-energy of the gauge bosons
induced by bulk fields running in the loop.

\bigskip
\subsection{A bulk fermion contribution}

We consider the one-loop contribution of a 6D left-handed bulk fermion
to the self-energy of the 4D components of the gauge field. 
\FIGURE{
\begin{picture} (300,70)(0,0)
\Photon(90,30)(130,30){2}{4}
\Photon(190,30)(230,30){2}{4}
\Vertex(130,30){2}
\Vertex(190,30){2}
\ArrowArcn(160,30)(30,0,180)
\ArrowArcn(160,30)(30,180,360)
\Text(160,70)[c]{$\psi$}
\Text(80,30)[c]{\small$A_\mu$}
\Text(240,30)[c]{\small$A_\nu$}
\end{picture} \\
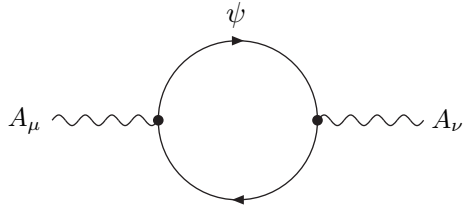
\caption{The Feynman diagram with a bulk fermion $\psi$
contributing to $\Pi_{\mu\nu}$ at one-loop 
order.}\label{fig1}}
The Feynman diagram given in Fig.~\ref{fig1} can be evaluated as
\bea\label{fermion01}
&&\Pi^f_{\mu\nu}(k,\vec k,\vec k')= g^2 \mu^{4-d} \sum_{{\vec p},{\vec p}'}
\int \frac{d^d p}{(2\pi)^d}\,{\rm Tr}\bigg\{
\gamma_\mu
\frac{i}{2}\bigg[\frac{\delta_{{\vec p},{\vec p}'}}{p\hsp+\gamma_5p_5+p_6}
-\eta\frac{\delta_{{\vec p},-{\vec p}'}}{p\hsp+\gamma_5p_5+p_6}i\gamma_5
\bigg]\gamma_\nu 
\nonumber \\
&&\quad\times
\frac{i}{2}
\bigg[\frac{\delta_{{\vec k}'+{\vec p}',{\vec k}
+{\vec p}}}{p\hsp+k\hsp+\gamma_5(k'_5+p'_5)+k'_6+p'_6}
-\eta\frac{\delta_{{\vec k}'+{\vec p}',-{\vec k}
-{\vec p}}}{p\hsp+k\hsp+\gamma_5(k'_5+p'_5)+k'_6+p'_6}
i\gamma_5\bigg]\bigg\}
\\ \nonumber
\eea
where we  used eq.~(\ref{t2z2fermionprop}) for the fermion propagator
in the loop.
Here a sum over discrete momenta $\vec p$
is to be understood as a double sum over integers $n_{1,2}$ such
that for an arbitrary  function~$f$
\bea
\sum_{\vec p} f(\vec p) =\sigma \sum_{n_{1,2}\in\bZ} f(n_1/R_1,
 n_2/R_6),\qquad \sigma\equiv [(2\pi)^2R_5R_6]^{-1}
\eea
where  $\vec p\equiv (p_5,p_6)=(n_1/R_5,n_2/R_6)$. Moreover, we use the
Kronecker delta symbol for discrete momenta, 
whose action and normalisation are 
\bea
\sum_{\vec p}\delta_{{\vec p},{\vec p}'} f({\vec p})=f({\vec
 p}'),\qquad\quad \delta_{\vec p,\vec
 p'}\equiv (2\pi)^2\delta_{p_5,p_5'}\delta_{p_6,p_6'}
=\frac{1}{\sigma} \,\,\delta_{n_1,n_1'} \delta_{n_2,n_2'}
\eea
The integral in (\ref{fermion01}) is continued to $d\equiv 4-\epsilon$
dimensions, with $\epsilon\rightarrow 0$ {\it after} 
performing the double sum; $\mu$ is the finite scale of the DR scheme.
 Note that  {\it both}  the 4D integral {\it and} the double sum 
over the momenta are regularised by the same regulator $\epsilon$. 
That is, $\epsilon$ acts  
essentially as a 6D regulator, as it should be the case. 
  These conventions will be used throughout the paper.
After some standard  calculations, we rewrite  expression (\ref{fermion01}) as
\bea
\Pi^f_{\mu\nu}&=&-\frac{1}{4}g^2\sum_{{\vec p},{\vec p}'}
\int \frac{d^d p}{(2\pi)^d}\frac{\mu^{4-d}}{(p^2-p^2_5)[(p+k)^2
-({\vec p}'+{\vec k}')^2]}
\bigg\{\pi^{(1)}_{\mu\nu}({\vec p}',{\vec k}')\delta_{{\vec k}',{\vec k}}
\nonumber \\[4pt]   
&&
+\pi^{(1)}_{\mu\nu}(-{\vec p}',-{\vec k}')\delta_{{\vec k}',-{\vec k}} 
-\eta\pi^{(2)}_{\mu\nu}({\vec p}',{\vec k}')
\delta_{-2{\vec p}',{\vec k}'-{\vec k}}
-\eta\pi^{(2)}_{\mu\nu}(-{\vec p}',-{\vec k}')
\delta_{-2{\vec p}',{\vec k}'+{\vec k}}\bigg\}\quad\quad
\eea
with 
\bea
\pi^{(1)}_{\mu\nu}({\vec p}',{\vec k}')
&=&4[2p_\mu p_\nu+p_\mu k_\nu+p_\nu k_\mu
+g_{\mu\nu}(-p(p+k)+{\vec p}'\cdot({\vec p}'+{\vec k}'))],
\nonumber 
\\[9pt]
\pi^{(2)}_{\mu\nu}({\vec p}',{\vec k}')
&=&-4i p^\rho k^\sigma \epsilon_{\mu\rho\nu\sigma}.
\eea
Here we note that terms proportional to $\delta_{{\vec k},{\vec k}'}$ or 
 $\delta_{{\vec k},-{\vec k}'}$ conserve 
the external extra momentum $|{\vec k}|$.
 Therefore these terms correspond to bulk terms.
 On the contrary, terms multiplied  
by $\delta_{-2{\vec p}',{\vec k}'-{\vec k}}$ 
 or $\delta_{-2{\vec p}',{\vec k}'+{\vec k}}$ change 
 the external discrete momentum in the compact dimensions, 
and therefore  correspond to brane-localised terms \cite{ggh}.
These momentum non-conserving terms are due to the breaking of translational
invariance along the extra dimensions in the presence of orbifold
fixed points. 
Although the momentum is conserved at each vertex in Feynman diagrams,
extra momenta of ingoing and outgoing
gauge bosons can be different due to 
the momentum non-conserving part $\delta_{{\vec p},-{\vec p}'}$ 
in the propagator of a bulk field running in loops.

After performing the 4D momentum integral, the contribution
involving $\pi^{(2)}_{\mu\nu}$ vanishes. 
Therefore no correction to the localised gauge coupling
 is generated by the bulk fermion. Finally,
after introducing a Feynman parameter and shifting the integration 
momentum as in Appendix~\ref{appendix-c1}, we obtain the correction 
\bea\label{pi-fmunu}
\Pi^f_{\mu\nu}[k,\vec k,\vec k']
&=&-2\, g^2\,\delta_{{\vec k},{\vec k}'}\,\mu^{4-d}
\sum_{{\vec p}'}\int^1_0 dx \int \frac{d^d p}{(2\pi)^d}\frac{1}{(p^2-\Delta)^2}
\nonumber \\
&&\times\bigg[2x(1-x)[(k^2-{\vec k}'^2)g_{\mu\nu}-k_\mu k_\nu] 
+(1-2x){\vec k}'\cdot ({\vec p}'+x{\vec k}')
g_{\mu\nu}\bigg]\qquad
\label{pi-fermion}
\eea
with
$\Delta\equiv -x(1-x)(k^2-{\vec k}'^2)+({\vec p}'+x{\vec k}')^2$.
The first part of this result contains
the familiar tensor structure coming from 6D gauge 
and Lorentz invariance and can be factorised 
out of the momentum integration and the Kaluza-Klein summation.
The second part of (\ref{pi-fermion}) however corresponds to a Lorentz
violating mass term, since 6D Lorentz invariance is broken 
by the compactification. This term leads to radiative 
corrections to the nonzero Kaluza-Klein masses~\cite{schmaltz}.

The current form of the result in eq.~(\ref{pi-fmunu}) is all we need
for our purpose of investigating the one-loop corrections to gauge
couplings in  supersymmetric models. It is nevertheless important to
simplify eq.~(\ref{pi-fmunu}) to identify its
divergences\footnote{The non-zero external momenta $(k, \vec k, \vec k')$
in the Green functions ensure infrared-convergent integrals.}.  
After some algebra we find, in Euclidean 
space\footnote{Denoting by $\Delta_E$  the Euclidean form 
of $\Delta$ we used that: $ \int  d^d p \, {(p^2-\Delta)^{-2}} 
={i\pi^2} \int_0^{\infty} {dt}\, {t^{1-d/2}}\,e^{-\pi \,t\,
    \Delta_E}$. Unless stated  otherwise, our formulae are always written 
using the Minkowskian metric; the
distinction is also obvious by the presence of either $g_{\mu\nu}$ or
$\delta_{\mu\nu}$.}
\begin{align}
&\Pi_{{\mu\nu}}^f[k,\vec k,\vec k']=
-\frac{2 g^2 i \pi^2}{(2\pi)^d}
\,\sigma \,\delta_{\vec k,\vec k'}\,
\Big[[(k^2+{\vec k}'^2) \delta_{\mu\nu}+k_\mu k_\nu]\,\, \Pi_{0}^f
-\delta_{\mu\nu}\,\Pi^{f}_{1}\Big],
\label{pimunuff}\\
&\Pi_{0}^f\equiv \int_0^1\! \! dx\,\, \rho_0(x) 
\,\cJ_0[x(1-x) (k^2+\vec k^{' 2}); x k_5' R_5, x k_6' R_6],\label{pi0ff}
\\
&\Pi^f_{1} \equiv \frac{k_5'}{R_5}
\int_0^1\!\! dx\,\, \rho_1(x) 
\,  \cJ_1 [x(1-x) (k^2+\vec k^{' 2}); x k_5' R_5, x k_6' R_6]\!
+\!\Big(k_5'\!\leftrightarrow\! k_6'; R_5\!\leftrightarrow\! R_6\Big),
\label{pi1ff}
\end{align}
with $\rho_0(x)\equiv 2 x (1-x)$ and $\rho_1(x)\equiv (1-2x)$.
The functions $\cJ_{0,1}[c; c_1,c_2]$ are defined and  studied in detail in
 Appendix~\ref{appendix-d}, eqs.~(\ref{M1v}), (\ref{M1(0)}) to
 (\ref{M1(1)s})  and they can be integrated over $x$,
yielding compact final expressions.  Since these expressions 
 are rather long, we do not present them here.  However, it is 
important for our purpose to notice that $\cJ_{0}$ has a pole, 
while  $\cJ_1$ is actually finite. Using this information, 
the pole structure in $\epsilon$ of the final result  is obtained 
\bea
\Pi_{0}^f=\frac{\pi}{15} (k^2+\vec k^{' 2}) R_5 R_6
\bigg[\frac{-2}{\epsilon}\bigg]+\cO(\epsilon^0),\qquad 
\Pi_1^f=\cO(\epsilon^0)
\eea
with momentum again in  Euclidean space.
The consequence of this 6D  divergence in $\Pi_0^f$ and
thus in $\Pi_{\mu\nu}^f$ is that  a higher derivative 
counterterm is necessary. This is a dimension-six bulk 
counterterm, and its structure would be,
 in a non-susy case, $R_5 R_6  F^{MN} \Box_6 F_{MN}$.
Although each bulk mode brings a pole for the usual gauge
kinetic term, the resummation of infinitely many bulk mode contributions
leads only to a pole for the higher derivative 
term\footnote{As will be discussed in detail in section 5, 
in a regularisation scheme
with a momentum cutoff, note that there is no logarithmically
divergent correction to the $F^{MN}F_{MN}$ operator and this is consistent
with the absence of a $1/\epsilon$ pole to this operator in DR.
In such cutoff regularisation, however, there exists a quadratically
divergent correction to the $F^{MN}F_{MN}$ operator 
(unlike in the 4D gauge theory), discussed in section 5.}.  
A similar result has been obtained in a 6D Abelian gauge theory 
without compactification in \cite{santamaria}. 
We postpone a further discussion on such operators to
Sections~\ref{hypermultiplet1} and \ref{section3} where their role
will be investigated in detail.

\subsection{A bulk scalar contribution}

Now we consider the one-loop contribution of a complex bulk scalar 
with parity $\eta$ to the self-energy of the gauge boson. 
In this case, there are two Feynman diagrams (see Fig.\ref{fig2})
contributing to  the one-loop vacuum polarisation. 

\medskip
\FIGURE{
\begin{picture} (260,80)(0,0)
\Photon(20,35)(50,35){2}{3}
\Photon(110,35)(140,35){2}{3}
\Vertex(50,35){2}
\Vertex(110,35){2}
\DashCArc(80,35)(30,0,360){3}
\Text(80,75)[c]{\small{$\phi_{\pm}$}}
\Text(10,35)[c]{\small{$A_\mu$}}
\Text(150,35)[c]{\small{$A_\nu$}}
\Text(170,35)[c]{$+$}
\Photon(190,5)(230,20){2}{4}
\Photon(230,20)(270,5){2}{4}
\Vertex(230,20){2}
\DashCArc(230,45)(25,0,360){3}
\Text(232,80)[c]{\small{$\phi_{\pm}$}}
\Text(180,10)[c]{\small{$A_\mu$}}
\Text(280,10)[c]{\small{$A_\nu$}}
\end{picture} \\
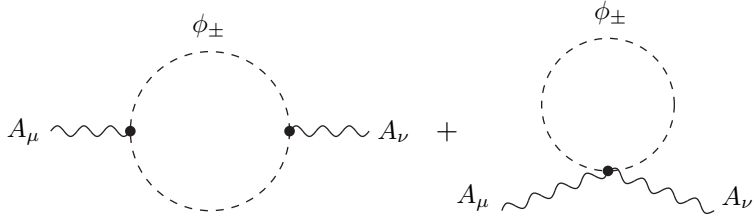
\caption{The Feynman diagrams with the bulk scalar $\phi$ 
contributing to $\Pi_{\mu\nu}$ at one-loop 
order.}\label{fig2}}

\noindent Then the one-loop scalar contribution is 
\be
\Pi^s_{\mu\nu,\pm}[k,\vec k,\vec k']=\Pi^{(1)}_{\mu\nu}[k,\vec
  k,\vec k'] +\Pi^{(2)}_{\mu\nu}[k,\vec k,\vec k']
\ee 
with
\bea
\Pi^{(1)}_{\mu\nu}[k,\vec k,\vec k']&=&(-ig)^2\,\mu^{4-d}
\sum_{{\vec p},{\vec p}'}\int \frac{d^d p}{(2\pi)^d} (2p+k)_\mu (2p+k)_\nu
\frac{i}{2}\bigg[
\frac{\delta_{{\vec p},{\vec p}'}
\pm\eta\delta_{{\vec p},-{\vec p}'}}{p^2-{\vec p}^2}\bigg] 
 \nonumber \\[6pt]
 &&\qquad\quad\times 
\frac{i}{2}
\bigg[\frac{\delta_{{\vec p}'+{\vec k}',{\vec p}+{\vec k}}
\pm\eta\delta_{{\vec p}'+{\vec k}',-{\vec p}-{\vec k}}}
{(p+k)^2-({\vec p}'+{\vec
    k}')^2}\bigg], \\
\Pi^{(2)}_{\mu\nu}[k,\vec k,\vec k']&=&(2ig^2)\, g_{\mu\nu}\,\mu^{4-d}\!\!\!
\sum_{{\vec p},{\vec p}'={\vec p}+{\vec k}-{\vec k}'}
\int \frac{d^d p}{(2\pi)^d} 
\frac{i}{2}\bigg[\frac{\delta_{{\vec p},{\vec p}'}
\pm\eta\delta_{{\vec p},-{\vec p}'}}{p^2-{\vec p}^2}\bigg]\\[-4pt]\nonumber
\eea
where we used  eq.~(\ref{t2z2scalarprop}) for the scalar propagator
 in the loop. After re-arranging the result, we obtain 
 the one-loop vacuum polarisation as
\bea
\Pi^s_{\mu\nu,\pm}[k,\vec k,\vec k']
&=&-\frac{g^2}{2}\mu^{4-d}\sum_{{\vec p}'}\int \frac{d^d p}{(2\pi)^d}
\frac{\delta_{{\vec k},{\vec k}'}\pm\eta
\delta_{-2{\vec p}',{\vec k}'-{\vec k}}}
{(p^2-({\vec p}')^2)[(p+k)^2-({\vec p}'+{\vec k}')^2]} \nonumber \\[6pt]
&&\times
\Big\{-(2p+k)_\mu (2p+k)_\nu 
+2g_{\mu\nu}\Big[(p+k)^2-({\vec p}'+{\vec k}')^2\Big] \Big\}
 \nonumber \\[11pt]
&\equiv&
\Pi^{\rm bulk}_{\mu\nu} [k,\vec k,\vec k']
\pm\eta\Pi^{\rm brane}_{\mu\nu}[k,\vec k,\vec k']
\label{pi-scalar}\\[-3pt]\nonumber
\eea
with the bulk and brane terms easily identified by whether they do
or do not conserve the discrete momenta 
associated with the two compact dimensions.
After using a Feynman parameter and a shift of the integration momentum
we obtain the bulk correction, where a 6D Lorentz violating mass term 
is present again, due to  compactification:
\bea\label{bulk-scalar1}
&&\Pi^{\rm bulk}_{\mu\nu}[k,\vec k,\vec k'] = -\frac{g^2}{2}
\delta_{{\vec k},{\vec k}'}\,\mu^{4-d}
\sum_{{\vec p}'}
\int^1_0 dx \int \frac{d^d p}{(2\pi)^d}\frac{1}{(p^2-\Delta)^2}
\nonumber \\[6pt]
&&\quad\quad\quad \times \Big[
(1-2x)^2[(k^2-{\vec k}'^2)g_{\mu\nu}-k_\mu k_\nu]
+2(2x-1){\vec k}'\cdot({\vec p}'+x{\vec k}')g_{\mu\nu}\Big].
\eea 
As in the fermionic case, the form of the
result in (\ref{bulk-scalar1}) is all we need for our purpose of
investigating one-loop corrections to the gauge couplings in supersymmetric
models. This result can however be evaluated explicitly 
as done in the fermionic case, to identify its
divergences and finite parts\footnote{This is particularly relevant
  in non-supersymmetric models, where similar corrections are
  present.}. One finds, using  an  Euclidean metric 
\bea
&&\Pi_{\mu\nu}^{\rm bulk}[k,\vec k,\vec k']=
-\frac{g^2}{2}\frac{i \pi^2}{(2\pi)^d}
\,\sigma \,\delta_{\vec k,\vec k'}\,
\Big[[(k^2+{\vec k}'^2) \delta_{\mu\nu}+k_\mu k_\nu]\,\, \Pi_{0}^{\rm bulk}
-\delta_{\mu\nu}\,\Pi^{\rm bulk}_{1}\Big]\qquad\qquad\nonumber\\[10pt]
&& \Pi_{0}^{\rm bulk}=\frac{\pi}{30} (k^2+\vec k^{' 2}) R_5 R_6
\bigg[\frac{-2}{\epsilon}\bigg]+\cO(\epsilon^0),\qquad 
\Pi_{1}^{\rm bulk}=\cO(\epsilon^0)
\eea
Here $\Pi^{\rm bulk}_0$ and $\Pi_1^{\rm bulk}$
have an expression identical to that of $\Pi^f_0$ of 
(\ref{pi0ff}) and $\Pi_1^f$ of (\ref{pi1ff}) respectively, 
but with $\rho_0(x)\!=\!(1-2x)^2$, $\rho_1(x)\!=\! 2(2x-1)$.
The  divergence of $\Pi_{\mu\nu}^{\rm bulk}$
requires a higher derivative counterterm,
 of structure identical to that for fermions:
 $R_5 R_6  F^{MN} \Box_6 F_{MN}$. We return to discuss the role
of such operators in Sections~\ref{hypermultiplet1}, \ref{section3}.

For the brane correction
the Kaluza-Klein loop momentum ${\vec p}'$ is fixed by the
difference between  ingoing and outgoing Kaluza-Klein 
momenta ${\vec k}$ and ${\vec  k}'$. After introducing a  
Feynman parameter and shifting the 4D momentum, 
we also find the brane correction as
\begin{align}
\!\!
\Pi^{\rm brane}_{\mu\nu}[k,\vec k,\vec k']\!&=\!
-\frac{g^2}{2}\mu^{4-d}\sum_{{\vec p}'}
\int^1_0 dx \int \frac{d^d p}{(2\pi)^d}\frac{1}{(p^2-\Delta)^2}
\bigg[2(1-3x+2x^2)(k^2-{\vec k}'^2)g_{\mu\nu}  \nonumber \\
&\quad\quad  -(1-2x)^2k_\mu k_\nu 
+\, 4(x-1){\vec k}'\cdot({\vec p}'+x{\vec k}')
g_{\mu\nu}\bigg]\cdot \delta_{-2{\vec p}',{\vec k}'-{\vec k}}\\
&=\!\frac{-i g^2}{2(4\pi)^2}\bigg\{\!
\frac{1}{3}\bigg[\frac{2}{\epsilon}\!+\!\ln 4 \pi \mu^2 e^{-\gamma_E}\bigg]
( g_{\mu\nu} k^2\!-\!k_\mu k_\nu\!-\! 3\vec k.\vec k' g_{\mu\nu})
\!-\!\!\int_0^1\!\!\!\! dx s(x) \ln\Delta\!\bigg\}\qquad \nonumber
\end{align}
with 
\bea
s(x)=2(1\!-\!3x\!+\!2x^2)(k^2\!-\!\vec k^{' 2}) g_{\mu\nu}\!-\!(1\!-\!
2 x)^2 k_\mu k_\nu \!+\!4(x-1) (\vec k/2+(x-1/2)\vec k')^2 g_{\mu\nu}.\qquad
\eea
Therefore, to cancel the one-loop divergence of the brane correction,
brane-localised gauge kinetic terms
containing the derivatives with respect to the extra dimensions are required. 
The remaining integral over $x$ is finite. In conclusion, a
bulk scalar in 6D leads to both bulk higher derivative  
and brane-localised gauge kinetic terms.

\subsection{A hypermultiplet contribution}
\label{hypermultiplet1}

We consider the contribution of a hypermultiplet 
to the vacuum polarisation. A hypermultiplet is composed of one Dirac fermion
and two complex scalars with opposite charges.
Using eqs.~~(\ref{pi-fermion}) and (\ref{pi-scalar}) 
with (\ref{bulk-scalar1}), 
we easily obtain the contribution in a simple form as
\bea
\Pi^{\rm hyper}_{\mu\nu}&=&\Pi^f_{\mu\nu}+\,\Pi^s_{\mu\nu,+}+\,\Pi^s_{\mu\nu,-}
\nonumber \\[6pt]
&=&-g^2 \delta_{{\vec k},{\vec k}'}[(k^2-{\vec k}'^2)g_{\mu\nu}-k_\mu k_\nu]
\mu^{4-d}\sum_{{\vec p}'}\int^1_0 dx \int \frac{d^d p}{(2\pi)^d}
\frac{1}{(p^2-\Delta)^2}.
\label{pi-hyper}
\eea
As indicated, the scalars take opposite $\mathbbm{Z}_2$ parities.
Consequently,  we note that the would-be 
mass corrections to Kaluza-Klein modes of gauge bosons that
we referred to earlier in  the scalar and fermionic  contributions
are cancelled out due to supersymmetry. Also the two would-be brane
contributions of the scalars are cancelled out.
The above result obtained in the component field formalism 
is in agreement with that obtained in a similar calculation
 using instead the superfield approach  \cite{nibbelink}.

The explicit evaluation of $\Pi^{\rm hyper}$ is rather technical
and we  provide the details in Appendix \ref{appendix-d}. Essentially one
 performs the momentum integral in (\ref{pi-hyper}) in the DR scheme, 
then re-writes that result in proper-time representation 
and finally performs the double sum over the
discrete momenta $\vec p\equiv (p_5,p_6)$. Using eqs.~(\ref{M1(0)}),
(\ref{M1(0)s}) for $\cJ_0$, with $a_1\equiv 1/R_5^2$ and $a_2 \equiv 1/R_6^2$,
one finds the contribution of a hypermultiplet in Euclidean
space\footnote{The term $\ln\mu^2$ is made  dimensionless 
by additional logarithmic terms in ${\cal J}^{\rm finite}_0$, 
not shown explicitly.}: 

\begin{eqnarray}\label{pi-hyper0}
\!\Pi^{\rm hyper}(k,\vec k')\!&=&\! \frac{i\,\mu^{4-d}}{(4\pi)^{d/2}}
\sum_{\vec p'} \int_0^1 dx \,\Gamma[2-d/2] \Big[x(1-x) (k^2+ {\vec
    k}^{' 2}) +({\vec p'}+x {\vec k'})^{2}\Big]^{d/2-2}\nonumber\\[7pt]
 &=&
 \frac{i \pi^2\mu^{4-d}}{(2\pi)^d}\sum_{\vec p'}\int_0^1 dx
 \int_0^1 \frac{dt}{t^{d/2-1}} \, e^{-\pi t \big[x(1-x) (k^2+ {\vec
     k}^{' 2}) +({\vec p'}+x {\vec k'})^{2}\big]}\nonumber\\[7pt]
&=&
\frac{i \pi^2 \sigma\,\mu^{4-d}}{(2\pi)^d}
\int_0^1  dx \,  \,
{\cal J}_0 \Big[x(1-x) (k^2+{\vec k}^{\prime 2}); \, x k_5' R_5, \,
 x k_6' R_6\Big]\nonumber\\[9pt]
&=&
\frac{i\sigma}{(4\pi)^2}\bigg\{\!\frac{\pi R_5 R_6}{6} (k^2\!+\!
{\vec k}^{' 2})\bigg[\frac{-2}{\epsilon}-\ln 4\pi^2\mu^2\bigg]\!+\!
\int_0^1\!\! dx \cJ_0^{\rm finite}
\Big]\!\bigg\}\,\,\qquad
\end{eqnarray}
with
\begin{eqnarray}
\cJ_0^{\rm finite}[c; c_1,c_2]\equiv\cJ_0[c;c_1,c_2]
-\pi R_5 R_6 \,c\bigg(-\frac{2}{\epsilon}\bigg).
\end{eqnarray}
The above definition of $\cJ_0^{\rm finite}$ together with
(\ref{M1(0)}), (\ref{M1(0)s}) shows that $\cJ_0^{\rm finite}$
 contains no pole in $\epsilon$. Here
 $c=x(1-x) (k^2+{\vec k}^{\prime 2})$, $c_1=x k_5' R_5$, $c_2= x k_6' R_6$.

Eq.~(\ref{pi-hyper0})  is an important result of this paper.
The presence of the momentum-dependent divergence
 $(k^2\!+\!{\vec k}^{' 2})/\epsilon$ in $\Pi^{\rm hyper}(k,\vec k')$ 
suggests the need for a higher derivative operator as
a counterterm to the one-loop correction.
Note that the counterterm required is actually
a  bulk operator since it is of 6D Lorentz invariant form. Its
form is the supersymmetric version of that already encountered
for bulk scalar and fermion contributions. 
The need for such an operator is ultimately a reflection of
 the fact that the initial theory is non-renormalisable.  
The  divergence found is  due to re-summing the {\it infinitely} 
many bulk mode  contributions in $\cJ_0$, each of them bringing a
 pole $1/\epsilon$,  to obtain instead a $k^2/\epsilon$ pole. 
This means the $k^2/\epsilon$ pole is of  {\it non-perturbative} origin. 
Note that calculations in the past, 
performed for vanishing external momenta, $k^2+\vec k^{' 2}=0$, missed
the presence of such higher derivative operators, since the
coefficient of the pole is then formally\footnote{Strictly speaking
  this should not be the case: even in such limiting cases, 
mathematical consistency would require
  one to introduce an infrared regulator $\lambda_{\rm IR}$ (here replaced 
 by $(k^2+\vec k^{' 2})$) to find a term which ``mixes'' the IR
 ($\lambda_{IR}$)  and UV ($\epsilon$)
 regulators/terms; such unwelcome UV-IR mixing
\cite{Ghilencea:2003xj,Ghilencea:2002ak} would signal a
 non-decoupling of high scale physics from its IR region. This 
 would lead one to conclude that higher derivative counterterms 
are required, if one remembers that the IR regulator 
can be equivalently replaced by non-zero momentum inflow.} 
set to zero.

If one also introduces a non-trivial
complex structure for the underlying torus, $U=R_6/R_5 e^{i\theta}$
(in our case $\theta=\pi/2$),
then the coefficient of the pole in eq.~(\ref{pi-hyper0})
 becomes proportional to $R_5 R_6\sin\theta$. 
For  $\theta=0$, when the two dimensions collapse onto each other, one
obtains the 5D limit \cite{Ghilencea:2005nt} as expected, and no pole is
present anymore in that case. This is consistent with the fact that such
operators are not generated by one-loop gauge corrections in the 5D case
where only a single sum over modes is present.
 However, at two loop order, two sums over the modes are present 
and higher derivative operators will again be generated, even in 5D.
In conclusion  such higher derivative 
operators are usually present in compactifications, being dynamically
generated at the loop level. These operators can also be 
boundary-localised, in the case of localised superpotential
interactions \cite{Ghilencea:2004sq, Ghilencea:2005hm, Ghilencea:2005nt}.

Returning to eq.~(\ref{pi-hyper0}), 
the integral over $x$ contains no poles and can be evaluated
numerically, using our detailed expressions for  $\cJ_{0}$
in Appendix~\ref{appendix-d}.  In specific cases further
 simplifications can occur, for example when $\vec k'=0$. 
The analysis of the higher derivative operator and of $\Pi^{\rm hyper}$
will be further extended to the case of non-Abelian theories, 
where its expression and  properties will  be discussed
in greater detail.

\section{The effective action for non-Abelian gauge theories on orbifolds}
\label{section3}

So far we have considered the case of Abelian gauge theories.
In this section we continue our analysis of one-loop corrections  
and derive the effective action for a non-Abelian gauge theory 
in higher dimensions by developing an approach outlined by Peskin and 
Schroeder \cite{peskin}. 
To this purpose we employ a background field
method applicable to orbifold compactifications.
 First we present the  method and derive  the general form of 
the one-loop effective action, then we apply it to the case of the 
$T^2/\mathbbm{Z}_2$ orbifold.

\subsection{Background field method for gauge theories in
higher dimensions}\label{section3.1}

Let us start with the relevant terms of the 6D supersymmetric action 
with a hypermultiplet in a representation 
of the bulk gauge group

\bea
S=\int d^6X\bigg[ 
\frac{1}{g^2}{\rm Tr}
\bigg(-\frac{1}{2}F_{MN}F^{MN}+2{\bar\lambda} \,i\gamma^M D_M\lambda\bigg) 
+{\bar\psi}\, i{\bar\gamma}^M D_M\psi+\sum_\pm |D_M\phi_\pm|^2\bigg]\quad
\eea
where $F_{MN}=\partial_M A_N-\partial_N A_M-i[A_M,A_N]$, 
$D_M\lambda=\partial_M\lambda-i[A_M,\lambda]$,
 $D_M\psi=(\partial_M-iA_M)\psi$ 
and $D_M\phi_\pm=(\partial_M\mp iA_M)\phi_\pm$.
To introduce the background field  method, we 
split the gauge field into a classical background  and a 
quantum fluctuation:
\be
A^a_M\rightarrow A^a_M+{\cal A}^a_M.
\ee
Then, 
\be
{\bar\psi}i{\bar\gamma}^M D_M\psi\rightarrow 
{\bar\psi}i{\bar\gamma}^M D_M\psi+ {\cal A}^a_M
{\bar\psi}{\bar\gamma}^M t^a\psi,
\ee
where $D_M$ is the covariant derivative with 
respect to the background gauge field.

Likewise, the gauge field strength is  decomposed as
\be
F^a_{MN}\rightarrow F^a_{MN}+D_M {\cal A}^a_N-D_N 
{\cal A}^a_M+f^{abc}{\cal A}^b_M{\cal A}^c_N.
\ee
Considering the higher dimensional generalisation 
of the  Faddeev-Popov procedure for the gauge-fixing, 
the 6D Lagrangian in the Feynman-'t Hooft gauge is given by

\bea
{\cal L}_{FP}&=&
-\frac{1}{4g^2}\Big(F^a_{MN}+D_M {\cal A}^a_N-D_N
{\cal A}^a_M+f^{abc}{\cal A}^b_M{\cal A}^c_N\Big)^2
-\frac{1}{2g^2}(D^M{\cal A}^a_M)^2 \nonumber \\[7pt]
&&
+\frac{1}{g^2}\Big[2{\rm Tr}\Big({\bar\lambda} i\gamma^M D_M\lambda\Big)
+i{\bar\lambda}^a f^{abc}{\cal A}^b_M \gamma^M\lambda^c\Big] 
+{\bar\psi}\Big( i{\bar\gamma}^M D_M+{\cal A}^a_M{\bar\gamma}^M t^a\Big)\psi
\nonumber\\[8pt]
&&
+\sum_\pm \Big(
|D_M\phi_\pm|^2\mp (D^M\phi_\pm)^*i{\cal A}^a_M t^a \phi_\pm
\pm i\phi^*_\pm{\cal A}^{aM} t^a D_M\phi_\pm
+\phi^*_\pm({\cal A}^a_M t^a)^2\phi_\pm\Big) \nonumber \\
&&
+{\bar c}^a \Big(-(D^2)^{ac}-D^M f^{abc}{\cal A}^b_M\Big)c^c,
\\
\nonumber
\eea
where $c^a$ are ghost fields and $D^2=D_M D^M$.

In order to compute the effective action at one-loop order, we shall
ignore terms linear in ${\cal A}^a_M$
and integrate over the  terms which are quadratic in the
gauge fields ${\cal A}^a_M$, gauginos $\lambda$, 
hyperinos $\psi$, hyperscalars $\phi$ and ghost fields $c$.
After integration by parts, the quadratic terms in ${\cal A}^a_M$ 
are simplified to
\be
{\cal L}_{\cal A}=-\frac{1}{2g^2}\Big\{{\cal A}^a_M \Big[-(D^2)^{ac}g^{MN}
-2f^{abc}F^{bMN}\Big]{\cal A}^c_N\Big\}.
\ee
By using the generator of 6D Lorentz transformations on 6-vectors,

\be
\Big({\cal J}^{PQ}\Big)_{MN}=i\Big(\delta^P_M\delta^Q_N-\delta^Q_M\delta^P_N\Big)
\ee
satisfying

\be
{\rm tr}\Big({\cal J}^{PQ}{\cal J}^{MN}\Big)=2\Big(g^{PM}g^{QN}-g^{PN}g^{QM}\Big),
\ee
we can rewrite the above Lagrangian as
\be
{\cal L}_{\cal A}=-\frac{1}{2g^2}\Big\{{\cal A}^a_M\Big[-(D^2)^{ac}g^{MN}
+2\,\Big(\frac{1}{2}F^b_{PQ}{\cal J}^{PQ}\Big)^{MN}(t^b_G)^{ac}\Big]{\cal A}^c_N\Big\}
\ee
with $(t^b_G)^{ac}\equiv if^{abc}$.
Further, the quadratic terms in fermion fields are

\bea
{\cal L}_f=\frac{1}{g^2}{\rm Tr}
\Big(2{\bar\lambda}i\gamma^M D_M\lambda\Big)
+{\bar\psi}i{\bar\gamma}^M D_M\psi.
\eea
Integrating over the fermion fields, we obtain the functional 
determinant of the operator
$(i\gamma^M D_M)$ for the gaugino and $(i{\bar\gamma}^M D_M)$ for the hyperino.
Finally, the quadratic terms in hyperscalars (${\cal L}_s$) and ghost fields
(${\cal L}_g$) are 
\bea
{\cal L}_s&=&\sum_\pm (\phi^a_\pm)^*[-(D^2)^{ac}]\phi^c_\pm ,\\
{\cal L}_g&=&{\bar c}^a[-(D^2)^{ac}]c^c.
\eea
With these findings, after performing the path integral for the
terms quadratic in quantum fluctuations, we obtain the 
effective action for the classical field $A^a_M$ at one-loop order as

\bea
e^{i\Gamma[A]}
&=&{\rm exp}\bigg[i\int d^6X \Big(
-\frac{1}{4g^2}(F^a_{MN})^2+{\cal L}_{\rm c.t.}\Big)\bigg] 
\\[7pt]
&\times&\!\!
 ({\rm det}\Delta_{G,1})^{-\frac{1}{2}}({\rm det}{\cal D}_G)^{+1}
[{\rm det}(-\Delta_{G,0})]^{+1}({\rm det}{\cal D}_r)^{+1} 
[{\rm det}(-\Delta_{r,0})]^{-1}
[{\rm det}(-\Delta_{r^*,0})]^{-1}\nonumber
\eea
with
\bea\label{notation2}
\Delta_{G,1}&=&\frac{1}{g^2}\Big[\Big(-D^2_1g^{MN}+ 2\, 
\big(\frac{1}{2}F^b_{PQ1}{\cal J}^{PQ}\big)^{MN}
t^b_G\Big) \,\delta^{{\cal A}_N}_{12}\Big] ,
\nonumber\\[6pt]
\Delta_{G,0}&=&-D^2_1\, \delta^c_{12}, \qquad
\Delta_{r,0}=-D^2_1\, \delta^{\phi_r}_{12}, \nonumber\\[6pt]
{\cal D}_G&=&\frac{1}{g^2}\Big(
i\gamma^M\partial_{M1}+A^a_{M1} t^a_G \gamma^M\Big)\,
\delta^\lambda_{12}, \nonumber\\[6pt]
{\cal D}_r&=&\Big(
i{\bar\gamma}^M\partial_{M1}+A^a_{M1} t^a_r {\bar\gamma}^M\Big)\,
\delta^\psi_{12},
\eea
where $r$ denotes the corresponding representation and an
extra index ''1''  as in  $f_1$ denotes $f(X_1)$ 
while the $\delta_{12}$'s are defined as functional differentiations
presented below. Finally, as the upper letter on the $\delta_{12}$'s 
imply,  the
above expressions are contributions of the
 gauge bosons, ghosts, hyperscalars, gaugino
and hyperino fields respectively. Further
\bea\label{deltas}
(\delta^{{\cal A}_M}_{12})^a\,_b&\equiv&\frac{\delta {\cal A}^a_M(X_1)}
{\delta {\cal A}^b_M(X_2)},\qquad
(\delta^{\phi_r}_{12})^a\,_b\equiv 
\frac{\delta\phi^a_r(X_1)}{\delta\phi^b_r(X_2)},
\eea
and similar for the remaining fields. Note that as long as there is no
orbifold action present
$\delta^{{\cal A}_M}_{12}=\delta^{\phi_r}_{12}=\delta^\lambda_{12}
=\delta^\psi_{12}=\delta^6(X_1-X_2)$.
With these observations, we have  the full one-loop effective action 

\bea\label{action-1}
\Gamma[A]&=&\int d^6 X \Big(
-\frac{1}{4g^2}(F^a_{MN})^2+{\cal L}_{\rm c.t.}
\Big) 
\nonumber \\
&&+\frac{i}{2}\bigg[\ln\det\Delta_{G,1}-2\,\ln\det{\cal D}_G
-2\,\ln\det(-\Delta_{G,0}) \nonumber \\
&&\qquad-2\,\ln\det{\cal D}_r
+2\,\ln\det (-\Delta_{r,0})+2\,\ln\det (-\Delta_{r^*,0})\bigg].
\eea
This is the general formula 
for the one-loop  effective action in higher dimensions 
with our field content.  
It can be applied to specific
cases, by computing the above determinants, after specifying the
boundary conditions for the fields involved.

\medskip
\subsection{The effective action on the $T^2/\mathbbm{Z}_2$ orbifold }
\label{section-orbifold}

We can now apply the method presented in the
previous section to the case of orbifold compactifications, where 
important changes appear due to the presence of the associated boundary
conditions with respect to the compact dimensions.
On the orbifold $T^2/\mathbbm{Z}_2$, the orbifold 
boundary conditions are given by
\bea\label{bc}
{\cal A}^a_\mu(x,-z)&=&{\cal A}^a_\mu(x,z), \qquad\qquad\,\,
{\cal A}^a_{5,6}(x,-z)=-{\cal A}^a_{5,6}(x,z), \nonumber \\[4pt]
c^a(x,-z)& =&  c^a(x,z), \qquad\qquad \qquad
\lambda(x,-z) =i\gamma^5\,\lambda(x,z),  \\[4pt]
\psi(x,-z)&=&i\gamma^5\,\eta\,\psi(x,z), \qquad\qquad 
\phi_\pm(x,-z)=\pm \eta \phi_\pm(x,z) \nonumber
\eea
where $\eta$ can be chosen either $+1$ or $-1$. 
Taking into account these boundary conditions, 
the functional differentiations defined in
(\ref{deltas}) can be made orbifold-compatible as follows: 
\medskip
\bea\label{delta-again}
\delta^{{\cal A}_\mu}_{12}&=&\frac{1}{2}\Big(\delta^6(X_1-X_2)
+\delta^6(X_1+X_2)\Big)=
\delta^c_{12}\equiv \delta^+_{12}, 
\label{fd1}
\nonumber\\[3pt]
\delta^{{\cal A}_n}_{12}&=&\frac{1}{2}\Big(\delta^6(X_1-X_2)
-\delta^6(X_1+X_2)\Big)
\equiv \delta^-_{12}, \label{fd2}\nonumber
\\[3pt]
\delta^{\phi_\pm}_{12}&=&\frac{1}{2}\Big(\delta^6(X_1-X_2)
\pm \eta\,\delta^6(X_1+X_2)\Big)\\[3pt]
\delta^\lambda_{12}&=&\frac{1}{2}\Big(\delta^6(X_1-X_2)
-i\gamma^5\delta^6(X_1+X_2)\Big)\label{fd3}
\nonumber\\[3pt]
\delta^\psi_{12}&=&\frac{1}{2}\Big(\delta^6(X_1-X_2)
-i\eta \,\gamma^5\delta^6(X_1+X_2)\Big) \nonumber
\eea
where $\delta^6(X_1\pm X_2)\equiv\delta^4(x_1-x_2)\delta^2(z_1\pm z_2)$.
We can now evaluate the determinants in (\ref{action-1}) 
giving the contributions of various  fields to the one-loop effective action.
To second order in the background gauge field we have from 
eq.~(\ref{action-1})

\bea\label{action-2}
\Gamma^{(2)}[A_M]
&=&\frac{1}{2g^2}\sum_{{\vec k}}
\int \frac{d^4 k}{(2\pi)^4}
A^a_M(-k,-{\vec k})A^b_N(k,{\vec k})
(-(k^2-{\vec k}^2)g^{MN}+k^M k^N)
\nonumber\\
&&+\frac{i}{2}\,\Big[
W_{G,1}-2W_{G,0}-2W_{\rm gaugino}+2W_{\rm hypers}
-2W_{\rm hyperino}\Big]
\eea
where each $W$ is the quadratic term of the corresponding log determinant
in (\ref{action-1}).

\bigskip\medskip
\subsubsection{Gauge field contribution $W_{G,1}$}
We start with the contribution of the gauge bosons and 
first introduce the notation:

\bea
{\cal M}\!\equiv \!\left(
\begin{array}{ll} 
-\partial^2_1 g^{\mu\nu}\delta^+_{12}
& \quad 0 \\  \quad 0 &
 -\partial^2_1 g^{mn}\delta^-_{12}
\end{array}
\right),\,\,\,
{\cal N}\!\equiv\! \left(\begin{array}{ll} 
(\Delta_G g^{\mu\nu}+\Delta^{\mu\nu})_1\delta^+_{12} 
& \quad \Delta^{\mu n}_1\delta^-_{12} \\
\quad \Delta^{m\nu}_1\delta^+_{12} & (\Delta_G
g^{mn}+\Delta^{mn})_1\delta^-_{12}
\end{array}
\right)\qquad
\eea
where
\bea\label{DeltaG}
\Delta_G&\equiv & \Delta^{(1)}_G+\Delta^{(2)}_G\nonumber\\[6pt]
\Delta^{(1)}_G&\equiv & i\Big[ 
\partial^M A^a_M t^a_G+A^a_M t^a_G\partial^M
\Big],\qquad
\Delta^{(2)}_G\equiv  A^{aM}t^a_G A^b_M t^b_G,
\\[6pt]
\Delta^{MN}&\equiv & 2 \,\Big(\frac{1}{2}F^b_{PQ}
{\cal J}^{PQ}\Big)^{MN}t^b_G.
\nonumber
\eea 
With this notation and (\ref{notation2}) we obtain 
\bea\label{detDeltaGv1}
\ln\det\Delta_{G,1}&=&
\ln\det\frac{1}{g^2}\Big[ {\cal M} +{\cal N}\Big]
=\ln\det\frac{1}{g^2}{\cal M}
-\sum_{n=1}^\infty\frac{1}{n}\, {\rm tr}\Big[({\cal O}_M\,^N)^n\Big]
 \nonumber \\[7pt]
&=&\ln\det\frac{1}{g^2}{\cal M}
-{\rm tr}({\cal O}_\nu\,^\mu)-{\rm tr}({\cal O}_m\,^n)\nonumber\\[7pt]
&-&
\frac{1}{2}
\bigg[
{\rm tr}({\cal O}_\nu\,^\lambda{\cal O}_\lambda\,^\mu)
+{\rm tr}({\cal O}_m\,^l{\cal O}_l\,^n)
+{\rm tr} ({\cal O}_\nu\,^l{\cal O}_l\,^\nu)
+{\rm tr}({\cal O}_m\,^\lambda{\cal O}_\lambda\,^n)\bigg]+\cdots,
\,\,\qquad\label{detDeltaGv2}
\eea
where we introduced
\begin{small}\bea\label{Odefinition}
\!{\cal O}_M\,^N \! &\!\equiv\!&\!\left(
\begin{array}{ll}\delta^+_{12}
i(-\partial^2_2)^{-1}g_{\nu\lambda} & \quad 0 \\
\quad 0 & \delta^-_{12}i(-\partial^2_2)^{-1} g_{ml}\end{array}\right)
\!\!\left(\begin{array}{ll} 
i(\Delta_G g^{\lambda\mu}\!+\!\Delta^{\lambda\mu})_2\delta^+_{23} 
& \quad i\Delta^{\lambda n}_2\delta^-_{23} \\[10pt]
\quad i\Delta^{l\mu}_2\delta^+_{23} & i(\Delta_G
g^{ln}\!+\!\Delta^{ln})_2\delta^-_{23} 
   \end{array}\right)\qquad
\eea
\end{small}
Therefore, the terms in $\ln\det \Delta_{G,1}$ quadratic in the
 background gauge field are 
\bea\label{Wg}
W_{G,1}[A_M]
= 4\,(T_1^{G+}+T_2^{G+})+2\,(T_1^{G-}+T_2^{G-})+T_3^G+T_4^G+T_5^G+T_6^G.
\eea
Their origin is as follows:  $4( T_1^{G+}+T_2^{G+})$
accounts for part of the term ${\rm tr} ({\cal O}_\nu\,^\lambda
 {\cal O}_\lambda\,^\mu) $
and for the term ${\rm tr} ({\cal O}_\nu\, ^\mu) $, while 
$2( T_1^{G-}+T_2^{G-})$ accounts for 
similar terms but with matrices entries  with extra 
dimensional Lorentz indices. The different factors multiplying them
(4 and 2) arise from the different metric contractions.
Further, $T_3^G$  accounts for
(the remaining part of) ${\rm tr} ({\cal O}_\nu\,^\lambda {\cal O}_\lambda\,^\mu)$
while $T_4^G$ accounts for similar contribution but with all Lorentz indices
extra dimensional. Finally, $T_{5,6}^G$ account for the ``mixed''
indices contributions, the last two terms in the last line of
(\ref{detDeltaGv2}), respectively. All these contributions can be easily
identified by recalling that $\delta^+_{ij}$ ($\delta^-_{ij}$) arise with
contributions from 4D (extra dimensional) Lorentz indices, respectively, 
as seen from the definition of ${\cal O}_M\,^{N}$.
The results of evaluating the terms in (\ref{Wg}) are then 
\bea
T^{G\pm}_1+T^{G\pm}_2
&\equiv& -\frac{1}{2}\, {\rm tr}\Big[\Big(
\delta^\pm_{12}\, i(-\partial^2_2)^{-1}\, 
(i\Delta^{(1)}_{G,2}\, \delta^\pm_{23})\Big)\,\Big(\delta^\pm_{34}
i(-\partial^2_4)^{-1}\, (i\Delta^{(1)}_{G,4}\,
\delta^\pm_{41})\Big)\Big]\nonumber\\[6pt]
&&-
{\rm tr}\Big[\delta^\pm_{12}i(-\partial^2_2)^{-1}
(i\Delta^{(2)}_{G,2}\delta^\pm_{21})\Big]
\nonumber \\[6pt]
&=&
-\frac{1}{2}\,
C_2(G)\sum_{{\vec k},{\vec k}'}
\int \frac{d^4 k}{(2\pi)^4}
A^{aM}(-k,-{\vec k}')\, 
A^{aN}(k,{\vec k}) \,\,\Pi^s_{MN,\pm}.\qquad\qquad\qquad\quad
\label{ta}
\eea
One should consider in (\ref{ta}) either 
the upper or the lower signs  only. Further
$T_3^G$ is generated by  parity-even  gauge fields, as the
presence of $\delta^+_{ij}$ shows and equals
\bea
T_3^G&\equiv&-\frac{1}{2}{\rm tr}\Big[\Big(\delta^+_{12}\,
  i(-\partial^2_2)^{-1}\, 
(i(\Delta_\nu\,^\lambda)_2\, \delta^+_{23})\Big)\,\Big(\delta^+_{34}
i(-\partial^2_4)^{-1} \, (i(\Delta_\lambda\,^\mu)_4\, \delta^+_{41})\Big)\Big] 
\nonumber \\[6pt]
&=&
2\,{\rm tr}\Big[{\cal J}^{\rho\sigma}t^a_G {\cal J}^{\alpha\beta}t^b_G\Big]
\sum_{{\vec k},{\vec k}'}\int \frac{d^4 k}{(2\pi)^4}
\,\, A^a_\mu(-k,-{\vec k}')\, A^b_\nu(k,{\vec k}) \,k_\rho \,g^\mu_\sigma\, k_\alpha
\, g^\nu_\beta \nonumber \\
&&
\times \sum_{{\vec p},{\vec p}'}\int \frac{d^4 p}{(2\pi)^4}
\,\,{\tilde G}_+(p,{\vec p},{\vec p}')\,
\, {\tilde G}_+(p+k,{\vec p}'+{\vec k}',{\vec p}+{\vec k})\nonumber \\
&=& 4C_2(G)\sum_{{\vec k},{\vec k}'}
\int \frac{d^4 k}{(2\pi)^4}
A^a_\mu(-k,-{\vec k}') \, 
A^a_\nu(k,{\vec k})\Big(k^2g^{\mu\nu}-k^\mu k^\nu\Big)
\,\Pi^G_{++}, \qquad\label{tb} 
\eea
$T_4^G$ has similar form, but involves only parity-odd  fields 
(notice the presence of $\delta_{ij}^-$):
\bea
T_4^G&\equiv& -\frac{1}{2}{\rm tr}\Big[\Big(
\delta^-_{12}\, i(-\partial^2_2)^{-1}\, 
(i(\Delta_m\,^l)_2\, \delta^-_{23})\Big)\,\Big( \delta^-_{34}
i(-\partial^2_4)^{-1}\,  (i(\Delta_l\,^n)_4\, \delta^-_{41})\Big)\Big] 
\nonumber \\[6pt]
&=&
2{\rm tr}\Big[{\cal J}^{ij}t^a_G {\cal J}^{kl}t^b_G\Big]
\sum_{{\vec k},{\vec k}'}\int \frac{d^4 k}{(2\pi)^4}
\,\,
A^a_m(-k,-{\vec k}')\, A^b_n(k,{\vec k}) \, k'_i \, g^m_j \, k_k \, g^n_l \nonumber \\
&&
\times \sum_{{\vec p},{\vec p}'}\int \frac{d^4 p}{(2\pi)^4}
\,\,{\tilde G}_-(p,{\vec p},{\vec p}')\,
\, {\tilde G}_-(p+k,{\vec p}'+{\vec k}',{\vec p}+{\vec k}),\qquad\quad
\nonumber \\
&=& 4C_2(G)\sum_{{\vec k},{\vec k}'}
\int \frac{d^4 k}{(2\pi)^4}
A^a_m(-k,-{\vec k}')\, A^a_n(k,{\vec k})
\Big(\!\!-{\vec k}'\cdot {\vec k}g^{mn}-k^m k'^n\Big)\,
\Pi^G_{--}, \label{tc} 
\eea
Finally $T^G_{5}$ and $T_6^G$ have similar structure,  involving 
parity-odd and -even component fields:
\bea
T_5^G&\equiv&-\frac{1}{2}{\rm tr}\Big[
\Big(\delta^+_{12}\, i(-\partial^2_2)^{-1}\,
(i(\Delta_\nu\,^l)_2\, \delta^-_{23})\Big)\,\Big(\delta^-_{34}\,
i(-\partial^2_4)^{-1}\, (i(\Delta_l\,^\mu)_4\,\delta^+_{41})\Big)\Big] 
\nonumber \\[6pt]
&=&
2\, {\rm tr}\Big[
({\cal J}^{\lambda k})_\nu\,^l t^a_G ({\cal J}^{\rho n})_l\,^\mu t^b_G
\Big]%\nonumber\\[6pt]
\Big(k_\lambda A^a_k(-k,-{\vec k}')-k'_kA^a_\lambda(-k,-{\vec k}')\Big) 
\nonumber  \\
&&\Big(k_\rho A^b_n(k,{\vec k})-k_nA^b_\rho(k,{\vec k})\Big) 
\sum_{{\vec p},{\vec p}'}\int \frac{d^4 p}{(2\pi)^4}
\,\,{\tilde G}_-(p,{\vec p},{\vec p}')\,
\, {\tilde G}_+(p+k,{\vec p}'+{\vec k}',{\vec p}+{\vec k}) \nonumber
\\ \nonumber 
&=&-2 C_2(G)\sum_{{\vec k},{\vec k}'}
\int \frac{d^4 k}{(2\pi)^4} 
\Big(
k_\mu A^a_k(-k,-{\vec k}')-k'_k A^a_\mu(-k,-{\vec k}')
\Big) \nonumber \\
&&\quad\qquad\qquad\qquad\quad\times
\Big( k_\rho A^a_n(k,{\vec k})-k_n A^a_\rho(k,{\vec k}) \Big)\,
g^{\rho\mu}g^{kn}
\,\Pi^G_{-+}, \qquad\qquad
\eea
and
\bea
T_6^G
&\equiv&
-\frac{1}{2}{\rm tr}\Big[\Big(
\delta^-_{12}\,  i(-\partial^2_2)^{-1}\, 
(i(\Delta_n\,^\lambda)_2\, \delta^+_{23})\Big)\, \Big(\delta^+_{34}\,
i(-\partial^2_4)^{-1}\, (i(\Delta_\lambda\,^m)_4\, \delta^-_{41})\Big)\Big] 
\nonumber \\[6pt]
&=&
2\,{\rm tr}\Big[({\cal J}^{\mu k})_n\,^\lambda t^a_G 
({\cal J}^{\rho l})_\lambda\,^m t^b_G\Big]%\nonumber\\[6pt]
\sum_{{\vec k},{\vec k}'}\int \frac{d^4 k}{(2\pi)^4} 
\Big(k_\mu A^a_k(-k,-{\vec k}')-k'_kA^a_\mu(-k,-{\vec k}')\Big) \nonumber  \\ 
&&\Big(
k_\rho A^b_l(k,{\vec k})-k_lA^b_\rho(k,{\vec k})\Big) %\nonumber  \\
\sum_{{\vec p},{\vec p}'}\int \frac{d^4 p}{(2\pi)^4}
\,\,{\tilde G}_+(p,{\vec p},{\vec p}')\,
\,{\tilde G}_-(p+k,{\vec p}'+{\vec k}',{\vec p}+{\vec k}) \nonumber \\
&=&-2 C_2(G)\sum_{{\vec k},{\vec k}'}
\int \frac{d^4 k}{(2\pi)^4}
\Big(k_\mu A^a_k(-k,-{\vec k}')-k'_k A^a_\mu(-k,-{\vec k}')\Big)
 \nonumber \\
&&\qquad\qquad\qquad\qquad\quad \qquad\times
\Big(k_\rho A^a_n(k,{\vec k})-k_n A^a_\rho(k,{\vec k}) \Big)
\, g^{\rho\mu}g^{kn}\,\Pi^G_{+-}.\qquad
\label{t6}
\eea
In the equations above we used the notation $C_2(G)$  defined by 
${\rm tr}(t^a_G t^b_G)=C_2(G)\delta^{ab}$. 
In terms of the bulk propagator 
for bosons (See also eq.~(\ref{t2z2scalarprop})),
\bea
{\tilde G}_\pm(p,{\vec p},{\vec p}')&=&\frac{i}{2}
\frac{\delta_{{\vec p},{\vec p}'}\pm \delta_{{\vec p},-{\vec p}'}}
{p^2-{\vec p}^2},
\eea
one has the following expressions for 
 $\Pi^s_{MN,\pm}$ and  $\Pi^G_{\alpha\beta}$ 
used previously
\bea
\Pi^s_{MN,\pm}
&=&\sum_{{\vec p},{\vec p}'}
\int \frac{d^4 p}{(2\pi)^4}\Big[
-(2p'+k')_M(2p+k)_N
{\tilde G}_\pm(p+k,{\vec p}'+{\vec k}',{\vec p}+{\vec k}) \nonumber \\
&&\qquad\qquad\qquad
+2ig_{MN}\delta_{{\vec p}',{\vec p}+{\vec k}-{\vec k}'}\Big]
\cdot{\tilde G}_\pm(p,{\vec p},{\vec p}') \nonumber \\[8pt]
&=&-\frac{1}{2}\sum_{{\vec p}'}\int\frac{d^4 p}{(2\pi)^4}
\frac{-(2p'+k')_M(2p+k)_N+2g_{MN}[(p+k)^2-({\vec p}'+{\vec k}')^2]}
{(p^2-{\vec p}'^2)[(p+k)^2-({\vec p}'+{\vec k}')^2]} \nonumber \\[6pt]
&&\qquad \times
\Big(\delta_{{\vec k},{\vec k}'}
\pm \delta_{-2{\vec p}',{\vec k}'-{\vec k}}\Big), 
\label{pi-sfull} 
\eea
\bea
\Pi^G_{\pm\pm}
&=&\sum_{{\vec p},{\vec p}'}\int
\frac{d^4 p}{(2\pi)^4}\,\, {\tilde G}_\pm(p,{\vec p},{\vec p}')
\,{\tilde G}_\pm(p+k,{\vec p}+{\vec k},{\vec p}'+{\vec k}') 
\nonumber \\[6pt]
&=&-\frac{1}{2}\sum_{{\vec p}'}\int
\frac{d^4 p}{(2\pi)^4}\frac{\delta_{{\vec k},{\vec k}'}
\pm\delta_{-2{\vec p}',{\vec k}'-{\vec k}}}
{(p^2-{\vec p}'^2)[(p+k)^2-({\vec p}'+{\vec k}')^2]} 
\qquad\qquad\qquad\qquad \\[2pt]\nonumber
\label{pi-na}
\eea
and
\bea
\Pi^G_{\pm\mp}
&=&\sum_{{\vec p},{\vec p}'}\int
\frac{d^4 p}{(2\pi)^4}\,\, {\tilde G}_\pm(p,{\vec p},{\vec p}')
\, {\tilde G}_\mp(p+k,{\vec p}+{\vec k},{\vec p}'+{\vec k}')
=\Pi^G_{\pm,\pm}.\qquad\qquad
\eea
To obtain the above results for $T_5^G$ and $T_6^G$ 
we had to change the order of operators in an appropriate way, 
by using
${\cal O}_2\delta^{\pm}_{23}=\delta^{\pm}_{23}{\cal O}_3$
for the $\mathbbm{Z}_2$-even operator $\cal O$
while ${\widetilde{\cal O}}_2\delta^{\pm}_{23}=\delta^{\mp}_{23}
{\widetilde{\cal O}}_3$
for  the $\mathbbm{Z}_2$-odd operator ${\widetilde{\cal O}}$. 
Further, to simplify the Kronecker deltas,
 we have taken into account the $\mathbbm{Z}_2$-parity conditions:
$A^a_\mu(k,{\vec k'})=A^a_\mu(k,-{\vec k'})$ and
$A^a_m(k,{\vec k'})=-A^a_m(k,-{\vec k'})$. This concludes the
evaluation of the gauge fields contribution  $W_{G,1}$ of (\ref{Wg}).

\subsubsection{Ghost field contribution $W_{G,0}$}
Next we evaluate the determinant of the ghost field contribution
 (\ref{notation2}) with (\ref{delta-again})
\bea
\ln\det(-\Delta_{G,0})&=&\ln\det\Big(
(\partial^2-\Delta_G)_1 \delta^+_{12}\Big) \nonumber \\[2pt]
&=&\ln\det(\partial^2_1 \delta^+_{12})
-\sum_{n=1}^\infty\frac{1}{n}{\rm tr }
\Big[\Big(\delta^+_{12}i(-\partial^2_2)^{-1}i(\Delta_G)_2
\delta^+_{23}\Big)^n\Big].
\eea
from which, upon expansion, we isolate the quadratic terms for the
 background field as
\bea\label{Wg0}
W_{G,0}[A_M]&=&T^{G+}_1+T^{G+}_2.
\eea
The sum on the right-hand side  was already computed in (\ref{ta}).

\subsubsection{Hyperscalar contribution $W_\text{hypers}$}

Likewise, the quadratic terms from the determinant for 
hyperscalars are, with (\ref{notation2}), (\ref{delta-again})
\bea
\ln\det(-\Delta_{r,0})&=&\ln\det\Big(
(\partial^2-\Delta_r)_1 \delta^\eta_{12}\Big) \nonumber \\[2pt]
&=&\ln\det(\partial^2_1 \delta^\eta_{12})
-\sum_{n=1}^\infty\frac{1}{n}{\rm tr }
\Big[\Big(\delta^\eta_{12}i(-\partial^2_2)^{-1}
i(\Delta_r)_2\delta^\eta_{23}\Big)^n \Big].\label{rtw}
\eea
with the notation of $\Delta$ as in eq.~(\ref{DeltaG}) with $G \rightarrow r$.
One finds from (\ref{rtw})
\bea\label{Whs}
W_{\rm hypers}[A_M]&=&(T^{r+}_1+T^{r+}_2)+(T^{r-}_1+T^{r-}_2) 
\eea
where $T^{r\pm}_{1,2}=\big[C(r)/C_2(G)\big]\,T^{G\pm}_{1,2}$ and with
$T^{G\pm}_{1}+T^{G\pm}_2$
already evaluated in  eq.~(\ref{ta}). \newline
Here $C(r)$ is defined by
 ${\rm tr}(t^a_r t^b_r)=C(r)\delta^{ab}$.

\subsubsection{Gaugino and hyperino contributions $W_{\text{gaugino}}$ 
and $W_\text{hyperino}$}

Finally, we evaluate
 the determinants for the fermion fields, which 
 are expanded as (using again (\ref{notation2}), (\ref{delta-again}))
\bea
\ln\det{\cal D}_G
&=&\ln\det\Big[\frac{1}{g^2}(i\gamma^M\partial_{M1}
+A^a_{M 1} t^a_G \gamma^M)\delta^\lambda_{12}\Big]
\nonumber \\[6pt]
&=&\ln\det\Big[\frac{1}{g^2}i\gamma^M\partial_{M1}\delta^\lambda_{12}\Big]
-\sum_{n=1}^\infty\frac{1}{n}
{\rm tr}\bigg[\bigg\{\delta^\lambda_{12}\frac{i}{i\gamma^P\partial_{P2}}
(iA^a_{M 2} t^a_G \gamma^M \delta^\lambda_{23})
\bigg\}^n\bigg], 
\eea
\bea
\ln\det{\cal D}_r
&=&\ln\det\Big[(i{\bar\gamma}^M\partial_{M1}
+A^a_{M 1} t^a_r {\bar\gamma}^M)\delta^\psi_{12}\Big] \nonumber \\[6pt]
&=&\ln\det\Big[i{\bar\gamma}^M\partial_{M1}\delta^\psi_{12}\Big]
-\sum_{n=1}^\infty\frac{1}{n}
{\rm tr}\bigg[\bigg\{\delta^\psi_{12}\frac{i}{i{\bar\gamma}^P\partial_{P2}}
(i A^a_{M 2} t^a_r {\bar\gamma}^M\delta^\psi_{23}) \bigg\}^n\bigg].
\eea
with the former (latter) for gaugino  (hyperino) fields, respectively.
From these eqs. the quadratic terms coming from the 
determinants of gaugino and 
hyperino are evaluated to
\bea\label{Wga}
W_{\rm gaugino}[A_M]&=& -\frac{1}{2}{\rm tr}
\bigg[\delta^\lambda_{12}\frac{i}{i\gamma^P\partial_{P2}}
(iA^a_{M 2} t^a_G \gamma^M \delta^\lambda_{23})
\,\,\delta^\lambda_{34}\frac{i}{i\gamma^Q\partial_{Q4}}
(iA^b_{N 4} t^b_G \gamma^N\delta^\lambda_{41})\bigg]
\nonumber \\[5pt]
&=&\frac{1}{2}{\rm tr}(t^a_G t^b_G)\sum_{{\vec k},{\vec k}'}
\int \frac{d^4 k}{(2\pi)^4}
A^{a M} (-k,-{\vec k}')A^{b N} (k,{\vec k})\,
\tilde\Pi_{MN}^f,
\eea
\bea\label{Why}
W_{\rm hyperino}[A_M]&=&-\frac{1}{2}{\rm tr}
\bigg[\delta^\psi_{12}\frac{i}{i{\bar\gamma}^P\partial_{P2}}
(iA^a_{M 2} t^a_r {\bar\gamma}^M \delta^\psi_{23})
\,\,\delta^\psi_{34}\frac{i}{i{\bar\gamma}^Q\partial_{Q4}}
(iA^a_{N 4} t^a_r {\bar\gamma}^N\delta^\psi_{41})\bigg]
\nonumber \\[5pt]
&=&\frac{1}{2}{\rm tr}(t^a_r t^b_r)\sum_{{\vec k},{\vec k}'}
\int \frac{d^4 k}{(2\pi)^4}
A^{a M}(-k,-{\vec k}')A^{b N}(k,{\vec k})\,\Pi_{MN}^f
\eea
Here we introduced the following self-energies
\bea\label{pi-ga}
\tilde\Pi_{MN}^f&\equiv & \sum_{{\vec p},{\vec p}'}\int \frac{d^4 p}{(2\pi)^4}
{\rm Tr}\Big[{\tilde D}_\lambda(p,{\vec p},{\vec p}')\gamma_M
{\tilde D}_\lambda(p+k,{\vec p}'+{\vec k}',{\vec p}+{\vec k})
\gamma_N\Big],\\[6pt]
\Pi_{MN}^f
&\equiv &\sum_{{\vec p},{\vec p}'}\int \frac{d^4 p}{(2\pi)^4}
{\rm Tr}\Big[
{\tilde D}_\psi(p,{\vec p},{\vec p}'){\bar\gamma}_M
{\tilde D}_\psi(p+k,{\vec p}'+{\vec k}',{\vec p}+{\vec k}) 
{\bar\gamma}_N
\Big],
\eea
and used the propagators on $T^2/\mathbbm{Z}_2$ (for details see the
 Appendix, eq.~(\ref{t2z2fermionprop}))
\bea
{\tilde D}_\lambda(p,{\vec p},{\vec p}')&=&\frac{i}{2}
\left(\frac{\delta_{{\vec p},{\vec p}'}}{p\hsp+\gamma_5 p_5-p_6}
-\frac{\delta_{{\vec p},-{\vec p}'}}{p\hsp+\gamma_5
  p_5-p_6}i\gamma_5\right),
\eea
\vspace{-0.4cm}
\bea
{\tilde D}_\psi(p,{\vec p},{\vec p}')&=&\frac{i}{2}
\left(\frac{\delta_{{\vec p},{\vec p}'}}{p\hsp+\gamma_5 p_5+p_6}
-\frac{\eta\delta_{{\vec p},-{\vec p}'}}{p\hsp+\gamma_5
  p_5+p_6}i\gamma_5\right).\\
\nonumber
\eea
This concludes the identification of all component field
 contributions to the effective action. We now have the
necessary technical results eqs.~(\ref{Wg}), (\ref{Wg0}), (\ref{Whs}),
(\ref{Wga}), (\ref{Why}),
to analyse  the one-loop effective action of non-Abelian gauge
theories on $T^2/\mathbbm{Z}_2$.

\subsubsection{The one-loop  effective action on $T^2/\mathbbm{Z}_2$, 
its poles and counterterms}\label{section-one-loop}

In the following we concentrate on the 4D gauge field part of the 
effective action. In this case, we note that $\Pi_{\mu\nu}^f$ 
and $\Pi_{\mu\nu,\pm}^s$  are the same as the ones in 
(\ref{pi-fermion}), (\ref{pi-scalar}), respectively, which
were obtained by using the Feynman diagram approach in the $U(1)$ case.
Therefore, using (\ref{action-2}), 
the 4D gauge field part of the effective action can be written as 
\bea
&&\Gamma^{(2)}[A_\mu]=
\!\frac{1}{2g^2}\sum_{{\vec k}}
\int \!\frac{d^4 k}{(2\pi)^4}
A^a_\mu(-k,-{\vec k})\, A^a_\nu(k,{\vec k})
\Big(-(k^2-{\vec k}^2)g^{\mu\nu}+k^\mu k^\nu\Big)
\qquad\qquad\qquad \nonumber \\[7pt]
&&\qquad\qquad
+\frac{i}{2}\sum_{{\vec k},{\vec k}'}
\int \frac{d^4 k}{(2\pi)^4}
A^{a\mu}(-k,-{\vec k}')\, A^{a\nu}(k,{\vec k})  \\[5pt]
&& \quad\,
 \times  \bigg\{\! C_2(G)\Big[\! -
\!\Pi_{\mu\nu}^{\rm hyper}
\!+ \! 4(k^2g_{\mu\nu}\!-\! k_\mu k_\nu)\Pi^G_{++}
-2{\vec k}\cdot{\vec k}'g_{\mu\nu}(\Pi^G_{+-}\!+\!\Pi^G_{-+})\Big] \!-\! C(r)
\Pi_{\mu\nu}^{\rm hyper} \!\bigg\}\quad \nonumber
\eea
where 
\bea
\Pi_{\mu\nu}^{\rm hyper}\equiv \Pi^s_{\mu\nu,+}
+\Pi^s_{\mu\nu,-}+\Pi^f_{\mu\nu}.
\eea
Then, 
by decomposing this effective action into  bulk  and brane 
parts, we reach the main result of Section \ref{section-orbifold}:
\be
\Gamma^{(2)}[A_\mu]=\Gamma_{\rm bulk}+\Gamma_{\rm brane}
\ee
with 
\bea
\label{gamma1}
\Gamma_{\rm bulk}&=&\frac{1}{2}\sum_{{\vec k},{\vec k}'}
\int \frac{d^4 k}{(2\pi)^4}
A^a_\mu(-k,-{\vec k}') A^a_\nu(k,{\vec k})
\big((k^2-{\vec k}^2)g^{\mu\nu}-k^\mu k^\nu\big) \nonumber \\
&&
\quad \times
\bigg[-\frac{1}{g^2}-i\Big(C_2(G)-C(r)\Big)\, 
\Pi^{\rm hyper}(k,\vec k')\bigg]\delta_{\vec k, \vec k'},
 \\[9pt]\Gamma_{\rm brane}
&=&
\!\frac{1}{2}\!\sum_{{\vec k},{\vec k}'}
\!\int\!\! \frac{d^4 k}{(2\pi)^4}
A^a_\mu(-k,-{\vec k}')A^a_\nu(k,{\vec k})
\big(k^2g^{\mu\nu}\!-\! k^\mu k^\nu\big)
\Big[\!-\!4iC_2(G)\Pi^{\rm local}(k,\vec k,\vec k')\Big]\qquad 
\label{gamma2}
\eea
where 
\bea
\Pi^{\rm hyper}(k,{\vec k}') &\equiv& 
 \mu^{4-d} \sum_{\vec p'}\int \frac{d^d p}{(2\pi)^d}
\frac{1}{(p^2-{\vec p}^{\prime
    2})[(p+k)^2-(\vec p'+\vec k')^2]}, 
\label{pi-hypera} \\[9pt]
\Pi^{\rm local}(k,\vec k,\vec k')&\equiv & 
\frac{\mu^{4-d}}{2} \sum_{{\vec p}'}\int
\frac{d^d p}{(2\pi)^d}\frac{\delta_{-2{\vec p}',{\vec k}'-{\vec k}}}
{(p^2-({\vec p}')^2)[(p+k)^2-({\vec p}'+{\vec k}')^2]}. \label{pi-g2} 
\eea
From the expression of $\Gamma_{\rm bulk}$
 we see that the bulk correction  comes with 
the standard beta function coefficient\footnote{Because the number of modes
is reduced due to orbifolding,
the beta function coefficient is $1/2$ times  that for  a torus
 compactification.} in 6D which is given by $C(r)-C_2(G)$.
Note also that, as in the Abelian  case discussed previously,
 a hypermultiplet does not generate
a boundary-localised gauge coupling.
However, a 6D bulk counterterm  can be present as we already saw in
 the Abelian case (\ref{pi-hyper0}), when  evaluating $\Pi^{\rm hyper}$.
  Unlike the hypermultiplet,
 a vector multiplet does generate  boundary-localised gauge couplings,
 see eqs.~(\ref{gamma2}), (\ref{pi-g2}).
The corresponding (4D) counterterm that we discuss shortly
must then be localised at the fixed points.

The divergent nature of  $\Pi^{\rm hyper}$ of eq.~(\ref{pi-hypera})
was already presented and discussed
to some extent in the Abelian case, Section \ref{section2}, eq.~(\ref{pi-hyper0}).
Since $\Pi^{\rm hyper}$  also appears in
 the bulk correction in the case 
of non-Abelian gauge theories, eq.~(\ref{pi-hypera}), we analyse this in further detail.
From eq.~(\ref{pi-hyper0}), let us recall the following,
\begin{eqnarray}\label{pi-hyper2}
\!\Pi^{\rm hyper}(k, \vec k')
= 
\frac{i  \sigma}{(4\pi)^2} (2\pi \mu)^{\epsilon}
\int_0^1 dx \,\cJ_0\Big[x(1-x)(k^2+\vec k^{' 2}); x k_5' R_5, x k_6' R_6\Big].
\eea
The exact expression of $\cJ_0$ is
needed for  studying the finite effects and the dependence
of the zero-mode gauge coupling
on the momentum $k^2$. This expression would also be needed to 
study dimensional crossover
  effects \cite{O'Connor:zj} of the coupling at $k^2\sim 1/R_{5,6}^2$.
Since $\cJ_0$ is rather complicated, we present $\cJ_0$ below, 
for a somewhat simpler case of $k_5'=k_6'=0$. 
From eqs.~(\ref{M1v}), (\ref{M1(0)}), (\ref{M1(0)s}), (\ref{limit-j0})
and with the following notations
\bea
c\equiv x(1-x) k^2,\quad a_1\equiv \frac{1}{R_5^2},\quad a_2\equiv
\frac{1}{R_6^2},\quad
 s_{\tilde n_1}\equiv 2 \pi \tilde n_1 \sqrt{\frac{c}{a_1}},\quad
\gamma(n_1)\equiv \frac{(c+ a_1 n_1^2\,)^{\frac{1}{2}}}{\sqrt a_2},\qquad
\eea
one has, if $0\leq c/a_1<1$:

\begin{eqnarray}\label{M1(0)text}
&&\! \!\cJ_0[c; 0,0] 
=  \! \frac{\pi c}{\sqrt{a_1 a_2}} \bigg[\frac{-2}{\epsilon}\!+\!
\ln\Big[4\pi \,a_1\, e^{-\gamma_E}\Big]\bigg] 
-\sum_{n_1\in\bZ}\ln\Big\vert 1- e^{-2\pi \,\gamma (n_1)}\Big\vert^2
+ \frac{\pi}{3}\sqrt{\frac{a_1}{a_2}}-2\pi\sqrt{\frac{c}{a_2}}
\nonumber\\[6pt]
&&\qquad\qquad\qquad\qquad\qquad\qquad\qquad
- 2\,\frac{c\, \pi^\frac{1}{2}}{\sqrt{a_1 a_2}}\,
 \sum_{p\geq 1}\,\frac{\Gamma[p\!+\!1/2]}{(p\!+\!1)!}
\bigg[\frac{-c}{a_1}\bigg]^{p} 
\zeta[2p+1]
\end{eqnarray}
with $\gamma_E=0.577216...$. If $c/a_1>1$,  then

\begin{eqnarray}\label{M1(0)stext}
\cJ_0[c; 0,0]
 &=& \!\!
\frac{\pi c}{\sqrt{a_1 a_2}}\bigg[\! \frac{-2}{\epsilon}\!+\!
\ln\Big[
\!\pi \,c \,e^{\gamma_E-1}\Big]\!\bigg] 
\! -\!\sum_{n_1\in\bZ}\!\!\ln\Big\vert
 1\!-\!e^{-2\pi \,\gamma (n_1)}\Big\vert^2\!+\!
4\sqrt{\frac{c}{a_2}}\sum_{\tilde n_1>0}
\frac{K_1(s_{\tilde n_1})}{\tilde n_1}\,.\qquad\,\,\,\,
\\[-3pt]\nonumber
\end{eqnarray}
Here $\zeta[x]$ is
  the Riemann Zeta function; $K_1$ is the modified Bessel function,
  see Appendix~\ref{appendix-f} for definitions.
The pole structure is the same for both expressions of $\cJ_0$.
Regarding  the finite terms,
$\cJ_0$ of eq.~(\ref{M1(0)text}) has power-like terms in $c\sim k^2$ but
these are suppressed by the radii/area of the compactification.
These terms are the counterpart of the term\footnote{This term 
($c\ln c$) will be important for the running of
  the higher derivative operator coupling, see later.} 
$c\ln c$ of eq.~(\ref{M1(0)stext}) in the case $c/a_1\geq 1$. 
Note that in the first square bracket, 
$\cJ_0$ in (\ref{M1(0)stext})  has a power-like dependence on $c\sim k^2$ 
whereas the last two terms in $\cJ_0$ are exponentially
suppressed at large $c/a_1\!\sim\! k^2 R_5^2$ and (given the symmetry
$a_1\!\leftrightarrow \!a_2$) also at large ${\mbox c/a_2\!\sim\! k^2 R_6^2}$. 
The above expressions are  important  when  we discuss
the  running of the effective gauge coupling 
and of the coupling of the higher derivative operator,
after cancelling the divergence in eq.~(\ref{pi-hyper2}).

 Let us  consider some limiting cases.
 If $k^2\!\!\ll\!{\rm min}(1/R^2_5,1/R^2_6)$, 
eqs.~(\ref{pi-hyper2}), (\ref{M1(0)text}) give:
\bea\label{limit}
\Pi^{\rm hyper}(k,0)\!&\approx&\! \frac{i \sigma}{(4\pi)^2} \bigg\{
\frac{\pi}{6} R_5 R_6 k^2 
\bigg[\frac{-2}{\epsilon}
-\ln\Big[\pi e^{\gamma_E}
\mu^2R^2_5\big|\eta(iR_6/R_5)\big|^{-4}\Big]\bigg] \nonumber \\
&&\!-\!\ln\Big[ 4\pi^2 e^{-2}\big\vert \eta(i R_6/R_5)\big\vert^4
  R_6^2\,  k^2\Big]\!\bigg\}\qquad
%\\[1pt]\nonumber
\eea
where we used the Dedekind $\eta$ function, see eq.~(\ref{ddt}). 
This result shows that 
after the addition of  the higher derivative
counterterm which will cancel the pole, the hypermultiplet only brings in a
 {\it logarithmic} dependence with respect to the momentum $k^2$, 
at values of $k^2$ much smaller than $1/R^2_{5,6}$. 
Note that this
 is a {\it low-energy} logarithm, originating from bulk contributions! 
If one evaluated instead  $\Pi^{\rm hyper}(k^2\!=\!0,0)$,
 an IR mass regulator $\mu^2_{IR}$
(replacing $k^2$) would still be required for mathematical consistency.
This would then  lead to a troublesome  UV-IR mixing of type
$\mu_{IR}^2/\epsilon$ in (\ref{limit}), on which the limits
$\mu_{IR}\!\rightarrow\! 0$ and $\epsilon\!\rightarrow\! 0$ do not
 commute. This would simply mean that
 the UV physics does not decouple in the low energy
 limit. This shows, even in the on-shell result for
$\Pi^{\rm hyper}$, that there is a  need for a higher
derivative counterterm, for quantum consistency.
We return to this issue in Section~\ref{other}.

In the case  $k^2\gg {\rm max}(1/R^2_5,1/R^2_6)$, eqs.~(\ref{pi-hyper2}) 
and (\ref{M1(0)stext}) give: 
\bea\label{limit2}
\Pi^{\rm hyper}(k,0)\approx
\frac{i \sigma}{(4\pi)^2} \bigg\{
\frac{\pi}{6} R_5 R_6 k^2 
\bigg[\frac{-2}{\epsilon}-\ln\frac{\mu^2}{k^2}-\ln
\Big(4 \pi e^{8/3-\gamma_E}\Big)\bigg]\bigg\}.\qquad\qquad
\eea

Finally, the brane correction $\Pi^{\rm local}$ of (\ref{pi-g2}) 
also has a divergence. For  any 6D momenta 
\begin{eqnarray}\label{tttt}
\Pi^{\rm local}(k,\vec k, \vec k')\!
\!=\!\frac{i}{32 \pi^2}\,\bigg\{
\frac{2}{\epsilon}\!+\!\ln 4\pi \mu^2 e^{-\gamma_E}\!-\!\!\int_0^1\!\!
dx\! \ln\Big[x(1-x) (k^2\!+\!\vec k^{' 2}) \!+\!
\Big(\frac{\vec k}{2}\!+\!\big(x\!-\!\frac{1}{2}\big)\, 
\vec k'\Big)^2\Big]\!\bigg\}\nonumber\\[-2pt]
\end{eqnarray}
which if $\vec k=\vec k'=0$ simplifies to:
\begin{eqnarray}\label{4drunning}
\Pi^{\rm local}\big(k,0,0\big)
=\frac{i}{32 \pi^2}\bigg\{\frac{2}{\epsilon}+\ln 4\pi e^{2-\gamma_E}
+\ln \frac{\mu^2}{k^2}\bigg\},
\end{eqnarray}
where $\mu$ is the arbitrary (finite) scale introduced by the
regularisation scheme.

The poles in $\Pi^{\rm hyper}$ and $\Pi^{\rm local}$  that we
 identified can be cancelled 
by  introducing the following counterterms in the action:

\bea\label{cct}
\cL_{c.t}\!=\int d^2 z \, d^2 \theta\,
\bigg[\frac{1}{2 h^2} {\rm Tr}\, W^\alpha \Box_6 W_\alpha
\!+\!\frac{1}{2}\sum_{i=1}^4\frac{1}{g^2_{{\rm brane}, i}}
 {\rm Tr} \,W^\alpha\,W_\alpha
 \delta^{(2)}(z\!-\!z_{0}^i)\bigg]
+ {\rm h.c.}
\eea
Here $z_{0}^i(i=1,\cdots,4)$  are the fixed points of the
$T^2/\mathbbm{Z}_2$ orbifold considered.
Further,  $h^2$ is an additional dimensionless bulk coupling 
while $g_{{\rm brane},i}$ is 
a dimensionless brane coupling at the fixed point $z_{0}^i$.
The introduction of such counterterms to cancel the poles
is done up to an overall finite, unknown coefficient.
As a result new parameters (couplings) emerge in the theory.
For small compactification volume (or $k^2 R_{5,6}^2\ll 1$),
 the bulk higher derivative operator is suppressed; however, 
for large radii (or $k^2 R_{5,6}^2\gg 1$)
it is relevant and important for the  overall running of the 
zero-mode gauge coupling.  The  effect of this operator is
largely ignored in the literature, both in effective field theory 
and string theory approaches. The renormalisation and the running
of the coupling $h(k^2)$  will be considered in the next section.

Regarding the coupling $g_{\text{brane}, i}$, after its renormalisation
there will be one additional parameter
for the gauge kinetic term localised at each fixed point.
If one  considers such corrections in
GUT models compactified on orbifolds \cite{SO10},  
brane-localised gauge couplings respecting a  gauge symmetry 
smaller than that in the bulk may be present.
In that case the brane couplings are not universal and 
 can affect the gauge coupling unification in such models~\cite{new}.

\section{``Running'' of the effective  gauge coupling as induced by the
  6D theory}

In this section we consider the one-loop renormalisation and
running of the coefficients  of the higher derivative operator
and of the  gauge kinetic term of the zero-mode gauge field.

To begin with, we consider the running of the bulk coupling
$h$ in (\ref{cct}) for the zero mode of the gauge field.
After subtracting the divergence of the bulk term eq.~(\ref{gamma1})  
with eqs.~(\ref{limit}) and (\ref{limit2}) by a bulk higher derivative 
counterterm,
one has the following momentum dependence of the renormalised $h$:
\bea\label{hcoup}
k^2\ll \frac{1}{R_{5,6}^2}:~~ \frac{4\pi }{h^2(k^2)}&\approx &\frac{4\pi}{h^2_{\rm
 tree}}+\Big[-C_2(G)+C(r)\Big]\frac{1}{96\pi^2}
\ln\Big[\pi e^{\gamma_E} 
\mu^2 R^2_5\big|\eta(iR_6/R_5)\big|^{-4}\Big], \nonumber \\
k^2\gg \frac{1}{R_{5,6}^2}:~~ \frac{4\pi }{h^2(k^2)}&\approx &\frac{4\pi}{h^2_{\rm 
 tree}}+\Big[-C_2(G)+C(r)\Big]\frac{1}{96\pi^2}
\bigg\{\ln\frac{\mu^2}{k^2}+\ln4\pi e^{8/3-\gamma_E}\bigg\}.
\\[-3pt]\nonumber
\eea
After writing each of these equations at two different momentum scales
(for the same renormalisation scale $\mu$) and subtracting them,
we find that above the compactification scales the bulk coupling of the
higher derivative operator runs logarithmically in $k^2$ while below the 
compactification scales it does not run. 
The running of  $h(k^2)$ above the compactification scales is 
a just a bulk effect, little dependent on the details of localised 
singularities associated with the orbifold action\footnote{See also
 the discussion in  \cite{santamaria}.}.  Note that the higher derivative 
counterterm in eq.~(\ref{cct}) ``absorbed'' all linear dependence on $k^2$
in eqs.~(\ref{limit}) and (\ref{limit2}), arising from 
eq.\,(\ref{M1(0)text}), (\ref{M1(0)stext}), and this is relevant for
the discussion below.  For  $k^2 R_{5,6}^2\gg 1$ 
the coupling $h$ is not suppressed,  and this has implications for
the running of  the effective gauge  coupling of the zero-mode gauge
boson above the compactification scales.

Let us now investigate the running of the effective  gauge coupling
$g_{\rm eff}(k^2)$ which is defined as  the coefficient of the
gauge kinetic term of zero-mode  gauge boson.
The tree level value of the effective gauge coupling 
has  contributions from both bulk and branes, including
the bulk higher derivative term. 
It can be read
off from the following gauge kinetic term:
\bea\label{tree}
-\frac{1}{2}
{\rm Tr}\bigg[F_{\mu\nu}\bigg(\frac{1}{g^2_{\rm tree}}
+\frac{1}{\sigma h^2_{\rm tree}}\Box_4\bigg)F^{\mu\nu}\bigg]
\eea
where 
\be
\frac{1}{g^2_{\rm tree}}\equiv\frac{1}{\sigma g^2}
+\sum_{i=1}^4 \frac{1}{g^2_{{\rm brane},i} }, \ \ 
\sigma\equiv \frac{1}{4\pi^2R_5R_6}.
\ee
Here $g^2$ and $g^2_{{\rm brane},i}$ 
are the tree-level gauge couplings
in the bulk and at  the fixed points, respectively.
Note that, although the brane localised couplings $g_{\text{brane}, i}$
are new parameters
introduced in the theory, the coupling $g_{\rm tree}$ 
only depends on their overall 
combination with the bulk gauge coupling $g$. Moreover, due to the new
parameter $h_{\rm tree}$ of the higher derivative 
counterterm, ultimately, there is a momentum dependent contribution to
 the {\it effective} gauge coupling even at tree level.

After taking  into account the radiative corrections (see (\ref{gamma1}),
(\ref{gamma2})) the zero-mode gauge coupling $g_{\rm eff}(k^2)$ is, at 
one-loop\footnote{
Eq.~(\ref{qqq})  can be written in 
a form which separates massive from  massless modes' contributions:
\bea
\frac{1}{g^2_{\rm eff}(k^2)}=\frac{1}{g^2_{\rm tree}}
-\frac{k^2}{\sigma h^2_{\rm tree}}- i
 \Big[-C_2(G)+C(r)\Big]\,\frac{1}{\sigma}\,
\Pi^{\rm hyper}_{m,*}(k,0)
- i \Big[-3 C_2(G)+C(r)\Big] 2 \,\Pi^{\rm local}_*(k,0,0)
\nonumber
\eea
where  $\Pi^{\rm hyper}_{m, *}\equiv \Pi^{\rm hyper}\!\!-
\Pi^{\rm hyper}_{0,0}$, with $\Pi^{\rm hyper}_{0,0}$ the (0,0) mode
  contribution and we used $\Pi_{0,0}^{\rm hyper}/\sigma=2 \Pi^{\rm local}$. 
On this form we see the emergence of 4D ${\cal N}\!=\!2$ and ${\cal N}\!=\!1$
 beta functions of massive and massless sectors.}:
\bea\label{qqq}
\frac{1}{g^2_{\rm eff}(k^2)}=\frac{1}{g^2_{\rm tree}}
-\frac{k^2}{\sigma h^2_{\rm tree}}+ i
 \Big[C_2(G)-C(r)\Big]\,\frac{1}{\sigma}\,
\Pi^{\rm hyper}_*(k,0)
+4 i C_2(G)\Pi^{\rm local}_*(k,0,0).\quad
\eea
The subscript $*$ in the self-energy $\Pi^{\rm local}_*$  means that
only the finite part of $\Pi^{\rm local}$ should be considered,
because its singularity 
(the pole $2/\epsilon$) was cancelled by the tree level coupling 
$g_{\rm tree}$ in eq.~(\ref{cct}).
For the self-energy $\Pi^{\rm hyper}_*$ the subscript $*$
refers to the finite part of $\Pi^{\rm hyper}$  {\it after} the
renormalisation of the coefficient of the
higher derivative counterterm (\ref{hcoup}); therefore 
$\Pi^{\rm  hyper}_*$  does not include the divergence $k^2/\epsilon$
in  $\Pi^{\rm hyper}$ which corresponds to the renormalisation  
of $h_{\rm tree}$ in eq.~(\ref{hcoup}).
With these
considerations, note that $g_{\rm tree}$ and $h_{\rm tree}$ 
in (\ref{qqq}) 
and in the equations to follow denote only the finite part of
tree level couplings.

Let us now address the running of  $g_{\rm eff}(k^2)$ and the relation
connecting it to the tree level coupling $g_{\rm tree}$. To begin
with, consider first the case of $k^2\ll 1/R_{5,6}^2$.
To obtain the running of $g_{\rm eff}(k^2)$ for this region
 one writes  (\ref{qqq}) at two different 
momentum scales  $q^2, k^2\ll 1/R_{5,6}^2$ for the same renormalisation
scale $\mu$ and subtracts them, then uses eqs.~(\ref{limit})
 and (\ref{4drunning}) to find:
\bea\label{rtrt}
\frac{4\pi}{g^2_{\rm eff}(q^2)}\approx\frac{4\pi}{g^2_{\rm
      eff}(k^2)}+\frac{1}{4\pi}
\Big[-3 C_2(G) +C(r)\Big]\ln\frac{k^2}{q^2},\qquad{\rm if}\quad
q^2, k^2 \ll \frac{1}{R_{5,6}^2}.\\[-3pt]\nonumber
\eea
This is an  interesting result: we have  obtained the familiar 
4D logarithmic running of the  effective gauge coupling with the
usual 4D ${\cal N}\!=\! 1$ beta function coefficient given by 
$b_1=-3 C_2(G)+C(r)$. Note that this running was derived from the 
full 6D theory, by taking into account  both bulk and boundary loop effects. 
This is interesting because part of the above  {\it logarithmic} 
running  comes from the bulk\footnote{See  $\Pi^{\rm hyper}$ 
of (\ref{limit}).}, associated with the massless states. More explicitly,
the logarithmic correction  in (\ref{rtrt})
 contains a ``bulk'' part $C(r)\ln k^2$ due to the hypermultiplet, 
while the vector multiplet provides  a ``bulk'' part $-C_2(G)\ln k^2$ 
as well as  a ``brane'' part $-2C_2(G)\ln k^2$, which added together 
give the beta function in (\ref{rtrt}). 
We note that the running of the effective coupling $g_{\rm eff}$
as shown in eq.~(\ref{rtrt}) 
is unaffected by the higher derivative operators 
as long as we are in the region  $k^2\ll 1/R_{5,6}^2$.

The next step in our analysis
is to establish a connection between the tree level coupling $g_{\rm tree}$
and the gauge coupling
at  low momentum  scales well below the compactification scales 
($k^2\ll 1/R_{5,6}^2$),
after integrating out all massive 
Kaluza-Klein modes\footnote{Early studies on this topic can be 
found in \cite{Kaplunovsky:1992vs}, but using instead 
 an {\it on-shell} approach.}.
Using again eq.~(\ref{qqq}) together with
 (\ref{limit}), (\ref{4drunning}), we have
\bea\label{4dr}
\frac{4\pi}{g^2_{\rm eff}(k^2)}\!&\approx&\!
\frac{4\pi}{g^2_{\rm tree}}
\!-\!\frac{b_2}{4\pi}\!\ln\Big[4\pi e^{-\gamma_E}
 \big\vert\eta(i\,u)\big\vert^4\,u\,
(4 \pi^2 R_5 R_6 \,\mu^2)\Big]
\!-\!\kappa\!+\!\frac{b_1}{4\pi} \ln\frac{\xi_1 \mu^2}{k^2}, 
\quad k^2\!\ll\! \frac{1}{R^2_{5,6}},
\nonumber\\[10pt]
&& {\rm with} \ \ 
\kappa\equiv 4\pi^2k^2 R_5 R_6 \bigg[\frac{4\pi}{h^2_{\rm tree}}
+\frac{b_2}{96\pi^2}\ln\Big[\pi e^{\gamma_E}
\mu^2R_5R_6u^{-1}\big|\eta(iu)\big|^{-4}\Big]\bigg]\ll 1.
\\[-3pt]\nonumber
\eea
Here $u\equiv R_6/R_5$ and
 $\xi_1=4\pi e^{2-\gamma_E}$. 
Further $b_1=-3C_2(G)+C(r)$ is the ${\cal N}=1$ beta function
while $b_2=-C_2(G)+ C(r)$ is 1/2 of the $\cN=2$ beta function 
coefficient on the
torus, with 1/2  to account for the fact that the number of modes 
 is reduced on $T^2/\mathbbm{Z}_2$.
As written, eq.~(\ref{4dr}) 
connects $g_{\rm  eff}(k^2\!\ll\!1/R_{5,6}^2)$ to the tree level 
coupling $g_{\rm  tree}$, {\it after} integrating out the massive  
Kaluza-Klein modes. The effect of these modes is accounted for by the
term multiplied by $b_2$ in (\ref{4dr}), as an overall threshold correction. 
It is important to note  from (\ref{4dr}) that the dominant contribution is of
logarithmic dependence on $k^2$ and this is associated with the
massless states only. 
Any power-like dependence  of $g_{\rm eff}(k^2)$
on the momentum scale is suppressed by the compactification 
volume, $\kappa\ll 1$, (i.e. the higher derivative operator 
is also suppressed.) 
This is  the case  after the renormalisation 
of the coupling $h$ of the higher derivative gauge kinetic term, 
eqs.~(\ref{cct}) and (\ref{hcoup}).

Eq.~(\ref{4dr}) can be used to study 
whether the low energy measurements of the couplings, 
e.g. electroweak scale values of the couplings are consistent
with a common value $g_{\rm tree}$, regarded in
this case as the ``unified'' coupling. 
The DR renormalisation  scale $\mu$ is in
 this picture  regarded as the unification scale. 
Eq.~(\ref{4dr}) is the counterpart of that
computed in the (on-shell) string, in various models
\cite{Dixon:1990pc2,Mayr:1993mq,ADB} (see also \cite{Kaplunovsky:1992vs}).
As we shall detail later, our result in (\ref{4dr}) is more in 
agreement with that of the
 4D $Z_N$ orientifold models of type I strings \cite{ADB}, rather than
that of the heterotic string \cite{Dixon:1990pc2,Mayr:1993mq}.

We have so far considered the behaviour of $g_{\rm eff}(k^2)$
at  momentum scales $k^2\!\ll \!1/R_{5,6}^2$ and its relation to the
tree level coupling. At higher momentum scales, the higher derivative 
operator becomes more important and one cannot neglect the presence
of its coupling $h(k^2)$, eq.~(\ref{hcoup}).
The  regime $k^2\sim 1/R_{5,6}^2$ 
is that of dimensional crossover \cite{O'Connor:zj} and is the
most difficult to investigate technically. In this case
eqs.~(\ref{limit}), (\ref{limit2}) provide a rather poor
approximation when used in eq.~(\ref{qqq}) to find $g_{\rm eff}$.
One must use instead the full expressions
of the functions $\cJ_0$,  eqs.~(\ref{M1(0)text}) and (\ref{M1(0)stext}), 
integrated over $x$ as in (\ref{pi-hyper2}). These expressions
converge even in the case  $k^2\sim 1/R_{5,6}^2$ and can be used to find
the running of $g_{\rm eff}$ in this regime.  These expressions are
somewhat complicated and this prevents an intuitive, simple  picture
for this  regime. In this case a full numerical approach based on 
(\ref{M1(0)text}), (\ref{M1(0)stext}) may be more suitable.

Finally, let us consider the case of 
even higher momenta, $k^2\gg 1/R_{5,6}^2$. In this case we find that 
the coupling $h(k^2)$ gives a substantial contribution to the running of
the effective gauge coupling.
From eq.~(\ref{qqq}) together
with eqs.~(\ref{limit2}) and (\ref{4drunning}), 
we obtain the following result: %%% if $k^2\gg 1/R_{5,6}^2$,
\bea\label{ohh}
\!\!\!\!\frac{4\pi}{g^2_{\rm eff}(k^2)}\approx
\frac{4\pi}{g_{\rm tree}^2}-4\pi^2 k^2R_5R_6\bigg[
\frac{4\pi}{h^2_{\rm tree}}+\frac{b_2}{96\pi^2}\ln\frac{\mu^2 \xi_2}{k^2}\bigg] 
-\frac{C_2(G)}{2\pi}\ln\frac{\mu^2 \xi_1}{k^2},
\,\, {\rm if}\,\,
k^2\!\gg \!\frac{1}{R_{5,6}^2}
\\[-5pt]\nonumber
\eea
where $\xi_2=4\pi e^{8/3-\gamma_E}$, $\xi_1=4\pi e^{2-\gamma_E}$ are
subtraction scheme dependent constants for the divergences of the bulk
and brane contributions respectively\footnote{Remember that these are in 
the minimal subtraction scheme, i.e. only the poles in
$\epsilon$ were cancelled by $g_{\rm tree}$ and
$h_{\rm tree}$.}. 
The scale $\mu$ is the familiar
renormalisation scale in the DR scheme, at which a ``boundary''
 value of the coupling is provided.

Eq.~(\ref{ohh}) describes the running of the effective gauge coupling
well above the compactification scales.
The last term in eq.~(\ref{ohh}) is due to massless
states (brane part only), which contribute to the running.
Further, the square bracket accounts for
the contribution coming from the running coefficient 
of the higher derivative term.
Since the square bracket 
involves $k^2 R_5 R_6$ which essentially counts
the number of excited Kaluza-Klein modes, 
we obtain a power-like running with respect to the
{\it momentum} scale, valid above the compactification scales.
Note, however, that the power dependence on $k^2$ is controlled by
the  parameter $h^2_{\rm tree}$ which multiplies it (and is 
also affected by the presence of $\ln\xi_2$  which is a subtraction  
scheme dependent coefficient). 
We therefore  need a  deeper understanding of this
coefficient. 

To this purpose, let us address the origin  of the power-like term 
and explain what ultimately controls it. To do so
we rewrite eq.~(\ref{qqq}) as
\bea
\label{jjj}
\frac{4\pi}{g^2_{\rm eff}(k^2)}=
\frac{4\pi}{g_{\rm tree}^2} - \frac{4\pi}{h^2(k^2)} 
\,(4\pi^2 k^2 R_5 R_6)
+\frac{b_2}{4\pi}\,\delta
-\frac{C_2(G)}{2\pi}
\ln\frac{\mu^2\xi_1}{k^2}. \\[-3pt]\nonumber
\eea
This equation is valid at all values of $k^2$,
large or small relative to $1/R^2_{5,6}$,
provided that other higher dimension operators are negligible.
Here $\delta$ is the integral over $x$ as in (\ref{pi-hyper2}) of
the part in $\cJ_0$ of either (\ref{M1(0)text}) or  (\ref{M1(0)stext}) which
does not contain the first square bracket  in these two equations.
If $k^2\ll 1/R_{5,6}^2$ then $\delta$ gives a log running given 
by the last term in 
(\ref{limit}) while if $k^2\gg 1/R_{5,6}^2$ then  $\delta\approx  0$. 
With these
values  of $\delta$ and with the running of $h(k^2)$ as in (\ref{hcoup})
one recovers the limiting cases of large and small momenta discussed
in (\ref{4dr}) and (\ref{ohh}).

The interpretation of the  result in (\ref{jjj}) is as follows: 
the coefficient of the
power-like term  $k^2\,R_5 R_6$ is ultimately controlled by the
renormalised coupling $h(k^2)$ of the higher derivative term in the
action and by its running. 
In some works the notion "power running" 
refers to  power-like (threshold) corrections in 
the UV  cutoff {\it regulator} as opposed to the power-like 
dependence with respect to the momentum scale that we obtained  here, and these are
not to be confused.  
Our result above clarifies that the power running with respect to the
 momentum scale 
is controlled by the one-loop corrected coupling of the
 higher derivative gauge kinetic term in the action.

In general, in theories with  higher derivative operators additional effects 
are present.  One should essentially start with the full
action {\it including} at the tree level the higher derivative gauge
kinetic term, and quantise the theory in its presence. This is a rather
 difficult problem. Further, in the presence of the higher
 derivative operator, the propagator of the zero-mode gauge boson  
changes into a sum of two terms: one  particle-like propagator 
and one ghost-like propagator, 
respectively\footnote{See Section 4 in \cite{Ghilencea:2005hm} for a similar
  discussion for the case of a massive scalar field.}:
\bea
G(k)=\frac{-ig_{\mu\nu}}{k^2 \Big(\frac{k^2}{h^2}+\frac{1}{g^2}\Big)}
=-ig^2 g_{\mu\nu}\bigg[\frac{1}{k^2}-\frac{1}{k^2+\frac{h^2}{g^2}}\bigg].
\eea 
From the coefficient of each term, one can see 
that both particle and ghost have the same coupling $g$
 to matter fields.
Although the ghost pole is located around the 6D fundamental scale,
the ghost state may give an additional non-vanishing threshold correction to the
gauge coupling. Further, there are many other complications, specific to
higher derivative theories, such as unitarity violation,
non-locality, etc,  see \cite{swh}-\cite{Mannheim:2004qz},
which made the study of these theories less popular. 
Another difficulty that  arises is that
one must also  take into account the effect of brane-localised 
terms on the spectrum of the Kaluza-Klein modes \cite{wagner}, 
not considered in this paper. Therefore, a detailed investigation 
of models with higher derivative operators is far more complicated and 
beyond the purpose of the present work.

To conclude, the higher derivative operator must be included
to ensure the quantum consistency of the model with extra dimensions, 
and therefore plays an important role
in the running of the effective gauge coupling.
After the renormalisation of its coupling $h$ 
there is only a logarithmic dependence on the momentum scale
of the 4D  effective gauge coupling $g_{\rm eff}(k^2\ll 1/R_{5,6}^2)$.
At a higher momentum scale power-like terms in $k^2 R_5 R_6<1$ are
present.
At even higher momentum  scales $k^2\gg 1/R_{5,6}^2$, the higher 
derivative operator is important and its coupling $h(k^2)$ has a logarithmic
running with respect to $k^2$. 
In this case the effective gauge coupling has, after
renormalisation of $h$, a power-like dependence on the momentum scale.
The coefficient of this power-like term in momentum is equal to
 the running coupling of the higher derivative operator. 
These findings provide a clear explanation 
of the power-like running (with respect to the momentum scale) of the gauge
couplings in models with extra dimensions.

\section{Higher derivative operators in other  schemes
and in  string theory}\label{other}

It is interesting to investigate how higher derivative
counterterms emerge in other regularisation schemes and in string theory as
well. This is important because their role in ensuring the quantum
consistency of the models was largely ignored in the
literature. To this purpose, we consider the effects of the massive Kaluza-Klein
modes in a regularisation with a momentum cutoff, i.e. the 
proper-time cutoff regularisation.
Note that a proper-time cutoff is less suitable as a regulator, since
it breaks 4D Lorentz invariance and Ward identities. 
Nevertheless, its use provides a more intuitive picture
and will help our physical  understanding of the important role
of higher derivative operators.

Let us introduce a  cutoff regulator $1/\Lambda^2$ in $\Pi^{\rm hyper}$
of (\ref{pi-hyper2}) and consider this equation for the
 {\it  massive} mode contributions only, denoted $\Pi^{\rm hyper}_{m}$,
i.e. we exclude the $(0,0)$ mode\footnote{The $(0,0)$ mode combines with the
  contribution of $\Pi^{\rm local}$ \!\! to give 4D $\cN\!=\!1$ beta function,
  see  footnote\,[12]}.
One has
\bea \label{res}
\Pi^{\rm hyper}_m(k^2,0)& =& 
\frac{i \pi^2 \sigma}{(2\pi)^4} 
\int_0^1 dx \sum_{n_{1,2}\in\bZ}\hspace{-3mm}\,'
\int_{1/ \Lambda^2}^\infty \frac{dt}{t} \,e^{-\pi \,t\,[k^2
    x(1-x)+n_1^2/R_5^2+n^2_2/R_6^2]}\\
&=&\frac{i \sigma}{(4\pi)^2}\bigg\{\Lambda^2R_5R_6
-\ln\Big[4 \pi e^{-\gamma_E} \,\vert \eta(i \,u)\vert^4 \,u\,
  \big(\Lambda^2 R_5 R_6\big)\Big] \nonumber \\
&&- \frac{\pi}{6} k^2 R_5 R_6 \ln\Big[(4\pi)^{-1} e^{\gamma_E}\Lambda^2R_5R_6 
u^{-1}\big|\eta(iu)\big|^{-4}\Big]\!\bigg\}
\nonumber
\eea
which is valid only if $k^2\!\ll\! {1}/{R_{5,6}^2}\!\ll\! \Lambda^2$.
The prime on the double sum marks the absence of the  $(0,0)$ mode.
The $\ln\Lambda$  term in the square bracket is the counterpart 
of  the $-2/\epsilon$ pole in the DR scheme\footnote{In the DR scheme,
 the massive sector (this excludes the (0,0) mode) gives
for $k^2\ll 1/R_{5,6}^2$  (eq.~(\ref{limit}))
\bea\label{www}
\Pi^{\rm hyper}_m(k^2,0)&=&
\frac{i \pi^2 \sigma\mu^\epsilon}{(2\pi)^{4-\epsilon}}
\int_0^1 dx \sum_{n_{1,2}\in\bZ}\hspace{-3mm}\,'
\int_{0}^\infty \frac{dt}{t^{1-\epsilon/2}} \,e^{-\pi \,t\,[k^2
    x(1-x)+n_1^2/R_5^2+n^2_2/R_6^2]}\nonumber\\
&=&
\frac{i \sigma}{(4\pi)^2}\bigg\{
\!\frac{-2}{\epsilon}
\!-\! \ln\Big[4 \pi e^{-\gamma_E} \,\vert \eta(i \,u)\vert^4 \,u\,
  \big(4\pi^2\mu^2 R_5 R_6\big)\Big] \nonumber \\
&&+\frac{\pi}{6} k^2 R_5 R_6\bigg[\frac{-2}{\epsilon}
-\ln\Big[\pi e^{\gamma_E}\mu^2R_5R_6u^{-1}\big|\eta(iR_6/R_5)\big|^{-4}\Big]\bigg] \!\bigg\}.
\quad 
\eea
}, first term in
(\ref{www}). The $k^2\ln\Lambda$ term corresponds the $k^2/\epsilon$
term in the DR scheme, associated with higher derivative operator.
 These divergences are cancelled  by the bulk kinetic term 
and the higher derivative operator, respectively.
In addition we obtain a quadratic divergence in the {\it regulator}
$\Lambda$ (\ref{res}) which cannot appear in the DR scheme.

To see in more detail the need for a higher derivative operator
in this  regularisation, remember that 
the momentum $k^2$ may be regarded as an  IR
regulator, to ensure the finiteness (at $t\ra \infty$) of
$\Pi^{\rm hyper}$ in (\ref{res}) when  the
massless mode $(n_1,n_2)=(0,0)$ is included. 
One notices  that in the last term of (\ref{res})
the limits $k^2\rightarrow 0$ and $\Lambda^2\rightarrow \infty$
 do not commute \cite{Ghilencea:2002ak}: 
\bea
\Big[k^2\rightarrow 0,\Lambda^2\rightarrow \infty\Big]\not=0.
\eea
We therefore have a rather troublesome UV-IR mixing term
(UV divergent, IR finite) meaning that the two sectors
of the theory are not  decoupled at the quantum level\,!
 As we recall from the comment following (\ref{limit}), a
 similar UV-IR mixing in the DR scheme was cancelled by the 
renormalisation of a higher derivative  counterterm.
In a similar way, the renormalisation of this operator  
cancels the log divergence in the last term of (\ref{res})
so that it enables the decoupling of the IR from the UV regime. 
Finally, the logarithmic and quadratic divergences in the first two terms 
of (\ref{res}) have to be subtracted by the gauge kinetic 
counterterm at a renormalisation point.
However, there remains a  correction $\Lambda^2 R_5 R_6$
with arbitrary coefficient\footnote{One must not forget
 that $\Lambda$ is actually a regulator and $100 \times \Lambda$
 is equally good a choice!}, which may eventually 
 be identified  from a more fundamental theory, e.g. 
from the field theory limit of
the heterotic string  \cite{Ghilencea:2002ak,Ghilencea:2002ff}.

What does  string theory say about these problems or about
 the need for higher derivative operators at the quantum level?
To begin with, it is  interesting to observe that in 4D 
$Z_N$ orientifold models of 
type I strings \cite{ADB}, the one
loop threshold corrections associated with the massive $\cN=2$ sector
are exactly of the type in (\ref{4dr})
{\it after} the tadpole cancellation condition. Note that 
this condition  ``removes'' any
power-like dependence on the  string scale. This similarity of the
results is  interesting, although there does not seem to exist
 a clear field theoretic understanding of  this tadpole cancellation 
condition 
and what that means  for the higher derivative operator that
 we found.  This also raises intriguing issues such as
 whether  the higher derivative counterterm that emerged 
and is relevant at large radii may be related to the
  non-perturbative effects  of D-branes.

Next,  let us consider  the case
of the heterotic string toroidal orbifolds $T^6/Z_N$, $N$ even,
with  ``fixed'' two-torus
 under the orbifold action. This  brings one-loop string
threshold corrections due to the $\cN=2$ massive sector of
 Kaluza-Klein and winding modes \cite{Dixon:1990pc2,Mayr:1993mq}.
In  the limit of large radii (in units $\alpha'$) non-perturbative 
effects (world-sheet instanton effects) are suppressed to give in the
field theory regime:
\bea\label{rtrtrt}
\Pi^{\rm hyper}(k^2=0,0)& \sim & 
-\ln\Big[4\pi e^{-\gamma_E} 
\vert \eta(i u)\vert^4\,u\, T_2 
 \Big]+\frac{\pi}{3}\,T_2
+ \epsilon_{IR} \ln\alpha',
\eea
where $T_2=R_5 R_6/\alpha'$; $u$ is the usual complex structure
(assuming an orthogonal fixed two-torus). 
This result is similar to that in (\ref{res}) for $k^2=0$,
 as discussed in detail in \cite{Ghilencea:2002ak,Ghilencea:2002ff}. 

Although the string provides only an {\it on-shell} result $(k^2=0)$,
the one-loop string nevertheless requires an infrared regulator
denoted  $\epsilon_{IR}$, which plays a role similar to
a small momentum $k^2\rightarrow 0$.
The  last term in (\ref{rtrtrt}) vanishes when 
 the infrared regulator in string is removed
$\epsilon_{IR}\rightarrow 0$, assuming  $\alpha'$ non-zero.
However, $\alpha^{' -1}\sim
M^2_{\text{string}}$ is the string scale, which is 
the counterpart to our UV momentum cutoff  regulator 
$\Lambda^2$ \cite{Ghilencea:2002ff,Ghilencea:2002ak}.
 One immediately observes from the last term in (\ref{rtrtrt})
that the limit of removing the infrared
regulator $\epsilon_{IR}\rightarrow 0$ and the limit of large 
$M_{\text{string}}$ or $\alpha'\rightarrow 0$ which is the effective 
field theory regime, do not commute:
\bea 
\Big[\epsilon_{IR} \rightarrow 0, \alpha'\rightarrow 0\Big]\not=0.
\eea
This is the same problem we encountered in the proper-time cutoff
regularisation scheme, if we regard $\epsilon_{IR}$ as 
$k^2\rightarrow 0$ and $M_{\text{string}}\rightarrow \infty$ as the
counterpart of $\Lambda^2$. Therefore there  is again a UV-IR mixing
 and a non-decoupling of the high scale physics i.e. of massive modes
from the 4D low energy limit
\cite{Ghilencea:2002ak}, also encountered in the DR scheme (see
comment after (\ref{limit})). The reason why such effects are usually not 
discussed in string theory is ultimately related to the underlying 
{\it on-shell} approach,
which ``obscures'' the need for higher derivative counterterms.  
The last term in (\ref{rtrtrt})
is then a  ``remnant'' of such effects, and a reminder of this 
 issue in the heterotic string. This non-decoupling of massive
 modes in the  low-energy (4D) raises questions on the consistency 
 of attempts to match string unification scale
  (in the presence of such thresholds) with MSSM-like unification scenarios.
This underlines   the need for  a study of the higher
derivative operators in  string theory\footnote{For more details on this 
matter see \cite{Ghilencea:2002ak} and Section 3 
in \cite{Ghilencea:2003xj}.}.

\section{Conclusions}

In this paper we performed a general analysis of 
the one-loop corrections to the self-energy of gauge bosons 
in the framework of 6D $\cN=1$ supersymmetric gauge theories on orbifolds.  
We first considered an Abelian gauge theory 
using the Feynman diagram approach 
in the component field formalism.
The analysis was then extended to the case 
of non-Abelian gauge theories on orbifolds.
By employing the background field method in higher dimensions, 
we established the general setup 
for the one-loop effective action for gauge bosons 
and then applied it to the case of the orbifold $T^2/\mathbbm{Z}_2$. 
As a consequence, we have shown that our component field approach  
is consistent with and complementary 
to the superfield calculation \cite{nibbelink,nibbmark}. 
Moreover, the additional benefit of our component field
approach is that
our findings  can be easily used in a non-supersymmetric setup. 

In the case of Abelian theories on $T^2/\mathbbm{Z}_2$ we computed the 
divergent and finite parts of the  one-loop correction to the 
vacuum polarisation tensor.
For the case of a bulk fermion it was shown that only bulk
corrections are present. The bulk
corrections contained a divergence which had to be cancelled 
by the introduction of a
6D  higher derivative counterterm.  
The loop corrections of a bulk scalar to the gauge
boson self-energy were also computed 
to show that there is a bulk (6D) higher derivative
as well as brane localised (4D) gauge kinetic counterterms.
The former is absent in the limit when the two compact dimensions
collapse onto  each other (similar for the bulk fermion),
 in agreement with the result that 
there is no higher derivative counterterm from the gauge interactions 
at one loop in 5D\footnote{Localised superpotential interactions
do bring in one-loop higher derivative counterterms 
in 5D \cite{Ghilencea:2004sq,Ghilencea:2005hm}.}.
Combining the bulk scalar and fermion contributions, 
we showed that a hypermultiplet 
only gives a bulk correction which requires a higher derivative
 counterterm, 
in agreement with other recent studies \cite{nibbmark}.

The above one-loop results were generalised to the case of 
non-Abelian gauge
theories on the $T^2/\mathbbm{Z}_2$ orbifold and many of our results
are expected to apply to other 6D orbifolds as well. 
This generalisation was done by first constructing the effective action 
with a background field  method in higher dimensions, which  was then
applied to 6D orbifolds. To this purpose, we introduced 
functional differentiations compatible with the orbifold actions
 on the fields. 
We found that hypermultiplets provide
only bulk corrections, while vector multiplets bring in both bulk and
boundary-localised corrections. 
The divergence of the bulk correction is cancelled by a 6D higher
derivative counterterm while the divergence of the brane correction
requires 4D boundary-localised gauge kinetic counterterms.
Therefore, after subtraction of divergences, 
there are unknown new parameters (couplings) coming from these
 operators in the
theory.
The bulk correction has a non-perturbative origin since we re-summed
infinitely many individual (divergent) loop contributions of the
 bulk modes.
At the technical level this is related, in part, to a singularity 
(simple pole) 
of the Hurwitz-Riemann Zeta function in the re-summed correction.
We also computed the finite part of the bulk correction which gives 
the momentum
dependence of the self energy of the gauge boson.
After renormalisation of the higher derivative operator,
the finite part of the bulk correction has, at $k^2\ll 1/R^2_{5,6}$,
 a familiar, logarithmic dependence on $k^2$ due to the massless
 states only. 
There are in addition  power-like terms (in $k^2 R_5
 R_6\ll 1$), strongly
 suppressed in this regime, and due to integrated massive modes.
At higher scales the finite  part contains power-like and
exponentially suppressed terms in $k^2 R_5 R_6$.

We then studied the behaviour  of the effective 4D gauge coupling 
$g_{\rm eff}(k^2)$, 
which was defined as the coupling of the zero-mode gauge boson.
After renormalisation of the higher derivative operator coupling,
we discussed in detail the running of the effective gauge coupling
with respect to the momentum scale.  
In the limit of momenta much smaller than the
compactification  scales, the effective coupling 
runs logarithmically  with the 4D ${\cal N}=1$ beta function
and this  low-scale running is induced by  both bulk and brane terms. 

We also analysed in detail the
 threshold corrections to the low energy gauge couplings, 
due to  massive  Kaluza Klein modes with ${\cal N}=2$ beta function
 coefficient.  The relation of the low energy effective coupling 
to the tree level coupling shows that 
there is only a {\it logarithmic} dependence of $g_{\rm eff}(k^2)$ 
on the momentum scale, while power-like terms are strongly suppressed
in the regime $k^2 R_5 R_6\ll 1$. 
This finding has potentially
 interesting consequences for phenomenology, such as the
unification of the gauge couplings. This is the result after the
 renormalisation of the higher derivative coupling, which below
 compactification scale is essentially constant (no running).
  It was observed that this result was in agreement with that of the
4D $Z_N$ orientifolds of the type I string, where no power-like terms are
 present in the one-loop threshold correction to the low-energy coupling.

At higher momentum scales, the higher derivative gauge kinetic term
is more important.  After renormalisation,
its coupling  has a logarithmic  running with respect to the momentum scale.
At $k^2\sim 1/R_{5,6}^2$ we provided technical formulae which allow
the study of the dimensional cross-over regime of the effective gauge
coupling. At larger  momentum scales ($k^2\geq 1/R_{5,6}^2$), 
the initially negligible contribution of the higher derivative term
to the coupling $g_{\rm eff}$ becomes significant  and starts to change 
the running of the effective coupling with respect to momentum scale 
from the logarithmic one to the power-like one. 
This behaviour was studied in detail.
At all momentum scales the coefficient of the power-like term is 
equal to the running coupling of the higher derivative gauge kinetic
term. This is an interesting finding which clarifies the physical
meaning of power-like running (in momentum) in models 
with extra dimensions. 

Finally, the importance of the higher derivative operator
was emphasised by showing the need for them as counterterms
in other regularisation schemes and in (heterotic) string theory.
In particular, it was shown that in these cases there is a UV-IR
mixing (UV divergent, IR finite) at the quantum level, due to ignoring
the quantum role of the higher derivative operator.
In the (on-shell) heterotic string this can be seen from the fact that 
the field theory limit of the one-loop correction from massive states
 does not commute  with the infrared regularisation of the one-loop
 string. 
This underlines the need for the investigation of the role of
higher derivative operators in string theory too.

\bigskip
\noindent
% \begin{acknowledgments}
{\bf Acknowledgements: }
The authors  acknowledge very interesting and useful discussions 
with W.~Buchm\"uller, S.~Y.~Choi, E.~Dudas,
 S.~Groot Nibbelink, M. ~Hillenbach and F.\,Quevedo. 
The work of D. Ghilencea was supported by a post-doctoral research
fellowship from the Particle Physics and Astronomy Research Council 
(PPARC), U.K.. D.G. acknowledges a visiting fellowship from
CERN where this work was completed.
%\end{acknowledgments}

\section{Appendix}
\def\theequation{\thesubsection.\arabic{equation}}
\setcounter{equation}{0}
\def\thesubsection{A}
\appendix
\subsection{Notations and Conventions}\label{appendix-a}

The metric has the signature
$g_{MN}={\rm diag}(+-----)$; $M,N=0,1,2,3,5,6$ are
six-dimensional indices and $\mu,\nu=0,1,2,3$ are four-dimensional ones.
The Clifford algebra in six dimensions is characterised by
\be
\{\Gamma^M,\Gamma^N\}=2g^{MN}, \ \ (\Gamma^M)^T=-C\Gamma^MC^{-1}, \ \
C^T=C, \ \ C^\dagger=C^{-1}.
\ee
An explicit representation for the $8\times 8$ gamma-matrices is
\be
\Gamma^\mu=\left(\begin{array}{ll} 0 & \gamma^\mu \\ \gamma^\mu & 0
\end{array}\right), \ \
\Gamma^5=\left(\begin{array}{ll} 0 & \gamma^5 \\ \gamma^5 & \,\,0
\end{array}\right), \ \
\Gamma^6=\left(\begin{array}{ll} 0 & -\mathbbm{1}_4 \\ \,\mathbbm{1}_4 & \,\,\,0
\end{array}\right)
\ee
where $\gamma^\mu$ and $\gamma^5$
are the four-dimensional gamma matrices, with
\be
\gamma^5=-\gamma^0\gamma^1\gamma^2\gamma^3
=-i\left(\begin{array}{ll} \mathbbm{1}_2 & \,\,\,0 \\ 0 & -\mathbbm{1}_2
\end{array}\right).
\ee
In this basis, the six-dimensional chirality operator is diagonal:
\be
\Gamma^7=\Gamma^0\Gamma^1\Gamma^2\Gamma^3\Gamma^5\Gamma^6
=\left(\begin{array}{ll} -\mathbbm{1}_4 & 0 \\ \,\,\,0 & \mathbbm{1}_4
\end{array}\right).
\ee
The charge conjugation is then
\be
C=\left(\begin{array}{ll} 0 & -C_5 \\ C_5 & \,\,\,\, 0
\end{array}\right)
\ee
where $C_5$ is the five-dimensional charge conjugation.

After imposing the chirality constraint in six dimensions,
the gamma matrices acting
on right-handed or left-handed 6D spinors are reduced to
the following $4\times 4$ matrices, respectively,
\be
\gamma^M\equiv (\gamma^\mu,\gamma^5,-\mathbbm{1}_4) \  \ {\rm and} \ \
{\bar\gamma}^M\equiv (\gamma^\mu,\gamma^5,\mathbbm{1}_4).
\ee
In five dimensions, the gamma matrices $\Gamma^a(a=0,1,2,3,5)$ are given by
\be
\Gamma^\mu=\gamma^\mu, \ \ \ \Gamma^5=\gamma^5
\ee
satisfying the following relations:
\be
(\Gamma^a)^T=-C_5\Gamma^a C^{-1}_5, \ \ C^T_5=-C_5, \ \ C^\dagger_5=C^{-1}_5.
\ee
We note some useful formulae  for the traces, used in the text
\bea
{\rm Tr}[\gamma_\mu\gamma_\nu]&=&4g_{\mu\nu}, \label{str1}\nonumber\\
{\rm Tr}[\gamma_\mu\gamma_\rho\gamma_\nu\gamma_\sigma]
&=&4(g_{\mu\rho}g_{\nu\sigma}-g_{\mu\nu}g_{\rho\sigma}
+g_{\mu\sigma}g_{\rho\nu}),\nonumber \\
{\rm Tr}[\gamma_\mu\gamma_\rho\gamma_5\gamma_\nu\gamma_\sigma]
&=&-4i\epsilon_{\mu\rho\nu\sigma}, \nonumber\\
{\rm Tr}[\gamma_\mu\gamma_\nu\gamma_\sigma]
&=&{\rm Tr}[\gamma_\mu\gamma_\nu\gamma_5]
={\rm Tr}[\gamma_\mu\gamma_\nu\gamma_\nu\gamma_5]=0.\label{str4}
\eea
In the text we also used the following relations on Casimir operators for a
representation $r$ (denoted $G$ ($N$) in the case of the adjoint
(fundamental) representation)
of the group $\cal G$:
\bea {\rm tr}(t^a_G t^b _G)=C_2(G)\delta_{ab}, \,\,\,\,
{\rm tr}(t^a_r t^b_r)=C(r)
\delta^{ab}.
\eea
with $C_2(G)= C(G)=N$, $C(N)=1/2$ and $C_2(N)=(N^2-1)/2N$,
in the case of $SU(N)$.

\def\theequation{\thesubsection.\arabic{equation}}
\setcounter{equation}{0}
\def\thesubsection{B}

\subsection{Propagators of bulk fields on orbifolds}

We present in the following  the propagators on
the $T^2/\mathbbm{Z}_2$ orbifold used in the text. 
On the orbifold $T^2/\mathbbm{Z}_2$, the positions $z\equiv(x_5,x_6)$
in the extra dimensions are identified by $z\rightarrow -z$.
For a bulk fermion,
we impose the boundary conditions as
\bea
P\psi(x,z)&\equiv& i\eta_f\gamma_5\psi(x,-z)=\psi(x,z),  \nonumber \\[7pt]
\psi(x,z)&=&\psi(x,z+2\pi R_5)=\psi(x,z+i2\pi R_6) \label{t2z2fermion}
\eea
with $\eta_f=\pm 1$.
Then, the fermion on the orbifold is written
in terms of a fermion on $T^2$ as
\bea
\psi(x,z)&=&\frac{1}{2}(1+ P)\chi(x,z) \nonumber \\
&=&\frac{1}{2}(\chi(x,z)+i\eta_f\gamma_5\chi(x,-z)).
\eea
By using the fermion propagator on $T^2$ given by

\be
D(x,z;x',z')\equiv\langle\chi(x,z){\overline\chi}(x',z')\rangle
\rightarrow {\tilde D}(p,{\vec p},{\vec p}')\equiv
\frac{i\delta_{{\vec p},{\vec p}'}}{p\hsp+\gamma_5p_5+p_6},
\ee
we find the fermion propagator on the $T^2/\mathbbm{Z}_2$ orbifold as

\bea
&&D_{\eta_f}(x,z;x',z')\equiv\langle\psi(x,z){\overline\psi}(x',z')\rangle
\nonumber \\[10pt]
&&\quad\rightarrow{\tilde D}_{\eta_f}(p,{\vec p},{\vec p}')\equiv\frac{i}{2}
\left(\frac{\delta_{{\vec p},{\vec p}'}}{p\hsp+\gamma_5p_5\pm p_6}
-\eta_f\frac{\delta_{{\vec p},-{\vec p}'}}{p\hsp+\gamma_5p_5\pm p_6}
i\gamma_5\right). \label{t2z2fermionprop}
\eea
Here $\pm$ depends on the $6D$ chirality.
Now we consider a bulk scalar field satisfying the boundary conditions on the
orbifold as
\bea
P\phi(x,z)&\equiv& \eta_s\phi(x,-z)=\phi(x,z), \nonumber \\
\phi(x,z)&=&\phi(x,z+2\pi R_5)=\phi(x,z+i2\pi R_6)\label{s1z2scalar}
\eea
with $\eta_s=\pm 1$.
Similarly to the fermion case, we can write down the scalar on the orbifold
in terms of a scalar on the covering space as
\bea
\phi(x,z)&=&\frac{1}{2}(1+P)\varphi(x,z) \nonumber \\
&=&\frac{1}{2}(\varphi(x,z)+\eta_s\varphi(x,-z)).
\eea
Then,
we obtain the scalar field propagator on the orbifold as
\bea
&&G_{\eta_s}(x,z;x',z')\equiv\langle\phi(x,z){\overline\phi}(x',z')\rangle
\rightarrow 
{\tilde G}_{\eta_s}(p,{\vec p},{\vec p}')\equiv\frac{i}{2}
\frac{\delta_{{\vec p},{\vec p}'}+\eta_s\delta_{{\vec p},-{\vec
      p}'}}{p^2-p^2_5-p^2_6}.
\label{t2z2scalarprop}
\eea

%%%\vspace{1.cm}
\def\theequation{\thesubsection.\arabic{equation}}
\def\thesubsection{C}
\setcounter{equation}{0}

\subsection{Details of the one-loop vacuum polarisation to $U(1)$
gauge bosons}\label{appendix-c}

We discuss in the following the detailed derivation of
the one-loop vacuum polarisation of $U(1)$ gauge bosons 
due to the fermionic and bosonic contributions.

\subsubsection{A bulk fermion contribution}\label{appendix-c1}

After introducing a Feynman parameter and shifting the integration momentum,
we obtain the fermionic correction (\ref{pi-fermion}) as
\bea
\Pi^f_{\mu\nu}&=&-2g^2\delta_{{\vec k},{\vec k}'}\mu^{4-d}
\sum_{{\vec p}'}\int^1_0 dx \int \frac{d^d p}{(2\pi)^d}\frac{1}{(p^2-\Delta)^2}
\Big\{ 2p_\mu p_\nu-2x(1-x)k_\mu k_\nu \nonumber \\[8pt]
&&\qquad
\qquad\quad +g_{\mu\nu}[-p^2+x(1-x)k^2+{\vec p}'\cdot({\vec p}'+{\vec k}')]\Big\}
\eea
with
\bea
\Delta\equiv -x(1-x)(k^2-{\vec k}'^2)+({\vec p}'+x{\vec k}')^2.\\[-3pt]\nonumber
\eea
After re-writing the terms proportional to $g_{\mu\nu}$ as
\bea
-p^2+x(1-x)k^2+{\vec p}'\cdot({\vec p}'+{\vec k}')
&=& -(p^2-\Delta)+2x(1-x)(k^2-{\vec k}'^2)
\nonumber \\[8pt]
&&\quad +(1-2x){\vec k}'\cdot({\vec p}'+x{\vec k}'),
\eea
the correction becomes
\bea
\Pi^f_{\mu\nu}&=&-2g^2\delta_{{\vec k},{\vec k}'}\mu^{4-d}
\sum_{{\vec p}'}\int^1_0 dx \int \frac{d^d p}{(2\pi)^d}\bigg\{
\bigg[\frac{2p_\mu p_\nu}{(p^2-\Delta)^2}-\frac{g_{\mu\nu}}{p^2-\Delta}\bigg]
\nonumber \\[8pt]
&+&\frac{1}{(p^2-\Delta)^2}
\bigg[
2x(1-x)[(k^2-{\vec k}'^2)g_{\mu\nu}-k_\mu k_\nu] 
 +(1-2x){\vec k}'\cdot({\vec p}'+x{\vec k}')g_{\mu\nu}\bigg]\bigg\}.
\eea
By using 
\be
\int \frac{d^d p}{(2\pi)^d}
\bigg[\frac{2p_\mu p_\nu}{(p^2-\Delta)^2}-\frac{g_{\mu\nu}}{p^2-\Delta}\bigg]=0,
\label{cancel}\\[4pt]\nonumber
\ee
we end up with the result
\bea
\Pi^f_{\mu\nu}&=&-2g^2\delta_{{\vec k},{\vec k}'}\mu^{4-d}
\sum_{{\vec p}'}\int^1_0 dx \int \frac{d^d p}{(2\pi)^d}\frac{1}{(p^2-\Delta)^2}
\nonumber \\[6pt]
&&\times\bigg(2x(1-x)[(k^2-{\vec k}'^2)g_{\mu\nu}-k_\mu k_\nu]
+(1-2x){\vec k}'\cdot ({\vec p}'+x{\vec k}')g_{\mu\nu}\bigg).
\eea
used in the text, eq.~(\ref{pi-fmunu}).

%%\vspace{1cm}
\subsubsection{A bulk scalar contribution}\label{appendix-c2}

After using a Feynman parameter and a shift of integration momentum,
the bosonic bulk contribution (\ref{pi-scalar}) is given by
\bea
\Pi^{\rm bulk}_{\mu\nu}
&\equiv&-\frac{1}{2}g^2\delta_{{\vec k},{\vec k}'}\mu^{4-d}
\sum_{{\vec p}'}\int \frac{d^d p}{(2\pi)^d}
\frac{1}{(p^2-\Delta)^2}\Big\{-4p_\mu p_\nu -(1-2x)^2k_\mu k_\nu \nonumber \\[6pt]
&&\quad +2g_{\mu\nu}[p^2+(1-x)^2k^2-({\vec p}'+{\vec k}')^2]\Big\}.
\label{pi-scalar-bulk} \\[3pt]\nonumber
\eea
Rewriting the terms proportional to $g_{\mu\nu}$ as
\bea
p^2+(1-x)^2k^2-({\vec p}'+{\vec k}')^2&=& (p^2-\Delta)
+(1-3x+2x^2)(k^2-{\vec k}'^2) \nonumber \\[6pt]
&&\quad +2(x-1){\vec k}'({\vec p}'+x{\vec k}'),
\eea
the bulk correction becomes
\bea
\Pi^{\rm bulk}_{\mu\nu}&=&-\frac{1}{2}g^2\delta_{{\vec k},{\vec k}'}\mu^{4-d}
\sum_{{\vec p}'}
\int^1_0 dx \int \frac{d^d p}{(2\pi)^d}\bigg\{
-2\bigg[\frac{2p_\mu p_\nu}{(p^2-\Delta)^2}-\frac{g_{\mu\nu}}{p^2-\Delta}\bigg]
\nonumber \\[8pt]
&&\quad +\,
\frac{1}{(p^2-\Delta)^2}\bigg[
2(1-3x+2x^2)(k^2-{\vec k}'^2)g_{\mu\nu}
-(1-2x)^2k_\mu k_\nu \nonumber \\[6pt]
&&\quad  +\, 4(x-1){\vec k}'\cdot({\vec p}'+x{\vec k}')
g_{\mu\nu}\bigg]\bigg\}. \label{pi-scalar-bulka}
\eea
Then, after 4D momentum integration with eq.~(\ref{cancel}),
the first two terms cancel.
Now observe that
\be
\frac{(1-2x)(k^2-{\vec k}'^2)}{(p^2-\Delta)^2}
=-\frac{\partial}{\partial x}\bigg(\frac{1}{p^2-\Delta}\bigg)
+\frac{2{\vec k}'\cdot ({\vec p}'+x{\vec k}')}{(p^2-\Delta)^2}.\\[8pt]\nonumber
\ee
Then from the $x$-integration
\be
\int^1_0 dx \frac{\partial}{\partial x}\bigg(\frac{1}{p^2-\Delta}\bigg)
=\frac{1}{p^2-({\vec p}'+{\vec k}')^2}-\frac{1}{p^2-{\vec p}'^2},\\[8pt]\nonumber
\ee
we note that the surface term for the Feynman parameter
vanishes after the Kaluza-Klein summation with the discrete shift in ${\vec p}'$.
Therefore, we obtain the correction as
\bea
\Pi^{\rm bulk}_{\mu\nu}&=&-\frac{1}{2}g^2\delta_{{\vec k},{\vec k}'}\mu^{4-d}
\sum_{{\vec p}'}
\int^1_0 dx \int \frac{d^d p}{(2\pi)^d}\frac{1}{(p^2-\Delta)^2} \nonumber \\
&&\times \bigg((1-2x)^2[(k^2-{\vec k}'^2)g_{\mu\nu}-k_\mu k_\nu]
+2(2x-1){\vec k}'({\vec p}'+x{\vec k}')g_{\mu\nu}\bigg).
\eea
used in the text, eq.~(\ref{bulk-scalar1}).

%\newpage
\def\theequation{\thesubsection.\arabic{equation}} 
\def\thesubsection{D}
\setcounter{equation}{0}
\subsection{Results and evaluation of series
 $\cJ_{0,1}$  for 6D orbifolds}
\label{appendix-d}

\noindent
We evaluate  (with $c\geq 0$, $a_{1,2}>0$, $0\leq
c_{1,2}< 1$): 
\begin{eqnarray}\label{M1v}
\cJ_{v}[c; c_1,c_2]\!&\equiv & \Gamma[\epsilon/2] \sum_{n_1,n_2\in\bZ}\! (n_1+c_1)^{v}
 \Big[\pi  [c+a_1 (n_1+c_1)^2+a_2
    (n_2+c_2)^2]\Big]^{-\epsilon/2}
\nonumber\\[6pt]
&=&
\!\!\!\sum_{n_1,n_2\in\bZ}\! (n_1+c_1)^{v}
\!\!\int_0^\infty
\frac{dt}{t^{1-\epsilon/2}}\, \,e^{-\pi \,t\, [c+a_1 (n_1+c_1)^2+a_2
    (n_2+c_2)^2]}, \,\,\,\,\, v\!=\! 0, 1...;\quad\,\,\quad\,
\end{eqnarray}
This expression was used in the text for $v=0$ and $v=1$ in  
eqs.~(\ref{pimunuff}), (\ref{pi0ff}), (\ref{pi1ff}), (\ref{pi-hyper}),
(\ref{pi-hyper0}), (\ref{pi-hypera}). 
In these eqs we assumed  $a_i=1/R_{i+4}^2$, $i=1,2$,
 $c_1=x R_5 k_5'$, $c_2=x R_6 k_6'$ and $c=x(1-x)(k^2+\vec k^{' 2})$
in Euclidean metric. Since we can always shift $c_i$ by an integer,
only their fractional part will enter the final result.

The final value of $\cJ_0$ was given in \cite{Ghilencea:2005hm} but in
the text we also  need to  evaluate $\cJ_1$ however. Since the proof
is similar, and to be general, we present the generic steps to evaluate
$\cJ_v$.
The counterpart of $\cJ_v$ with  a factor $(n_2+c_2)^v$ in front of the integral
is obtained from the replacements $c_1\!\leftrightarrow\! c_2$ and
$a_1\!\leftrightarrow\! a_2$. Most important for us is to identify the
poles of $\cJ_v$, (to find the counterterms) but we also 
evaluate the finite part which require us
compute the $\cO(\epsilon)$ term in  the double sum in the first line
in (\ref{M1v}). Notation used:
\begin{eqnarray}\label{notations}
\gamma(n_1)\equiv \frac{\sqrt{z(n_1)}}{\sqrt a_2}-i\, c_2;\qquad \quad
z(n_1)\equiv c + a_1 (n_1+c_1)^2,\qquad u\equiv\sqrt{a_1/a_2}\\[-4pt]\nonumber
\end{eqnarray}
Keeping the sum over $n_1$ fixed, we re-sum (see 
(\ref{p_resumation}))  over $n_2$, so that
\begin{eqnarray}\label{sums}
\!\!\!\!\sum_{n_{1,2}\in\bZ} \!\!
e^{-\pi t \, [ a_2  (n_2\!+\!c_{2})^2\! +a_1 (n_1\!+\!c_{1})^2 ] } &=&
\!\sum_{n_2\!\in\bZ} \!
e^{-\pi\, t \,[ a_2 (n_2\!+\!c_2)^2\!+ a_1 c_1^2]}
+\!\!
\sum_{n_1\!\in\bZ}'\!\sum_{n_2\!\in\bZ} \!\!
e^{-{\pi\, t\, [a_2 (n_2\!+\!c_2)^2\!+\! a_1(n_1\!+\!c_1)^2]}}
\nonumber\\
\nonumber\\
\!&=&\!  \sum_{n_2\in\bZ} 
e^{-{\pi\, t \,[ a_2 (n_2+c_2)^2+ a_1 c_1^2]}}
+ \frac{1}{\sqrt{t\,a_2}} \sum_{n_1\in\bZ}'
e^{-\pi t \, a_1  \, (n_1\!+\!c_1)^2}
\nonumber\\
\nonumber\\
&+&\frac{1}{\sqrt{t\,a_2}}
\sum_{n_1\in\bZ}'\sum_{\tilde n_2\in\bZ}' 
e^{-\frac{\pi {\tilde n_2}^2}{t\,a_2}-\pi t\,
a_1\,(n_1+c_1)^2+ 2 \pi i {\tilde n_2} c_2}\label{d11}
\\
\nonumber
\end{eqnarray}
The first term
has $n_1\!=\!0$, the last two have $n_1\!\not=\!0$. Then
\begin{eqnarray}\label{M1vsum}
\cJ_v=\cK^{(v)}_1+\cK_2^{(v)}+\cK_3^{(v)}
\end{eqnarray}
$\cK^{(v)}_i$, are obtained by integrating term-wise (\ref{d11}) with
 appropriate coefficients and extra $n_1$ dependence, see eqs.~(\ref{K1v}),
 (\ref{rm}), (\ref{def}) below. Their evaluation follows:

\vspace{0.3cm}
\noindent{\bf - Calculation of $\cK_1^{(v)}$:}
\begin{eqnarray}\label{K1v}
\cK_1^{(v)}\equiv 
c_1^v\! \sum_{n_2\in\bZ} \int_0^\infty \frac{dt}{t^{1-\epsilon/2}}
\,e^{-{\pi\, t \,[ a_2 (n_2 + c_2)^2 + a_1 c_1^2]}-\pi
  c t}=
-c_1^v\, \ln\Big\vert 2 \sin (\pi i \gamma(0)) \Big\vert^2\\
\nonumber
\end{eqnarray}
which was computed by first performing a re-summation 
(\ref{p_resumation}) over $n_2$,
and then used the integral representation (\ref{bessel1}) 
of the Bessel function $K_\frac{1}{2}$
its  expression (\ref{K_12}), and (\ref{notations}).

\vspace{0.3cm}
\noindent
{\bf - Calculation of $\cK_2^{(v)}$: }

\vspace{0.2cm}
\noindent
Here we distinguish two cases: if $0<c/a_1<1$ one has:
\begin{eqnarray}
&&
\cK_2^{(v)}\!\equiv\! \frac{1}{\sqrt{a_2}}
\sum_{n_1\in\bZ}' (n_1+c_1)^v \int_0^\infty
\frac{dt}{t^{3/2-\epsilon/2}}\,\,
e^{-\pi t \, a_1  \, (n_1\!+\!c_1)^2 -\pi t\, c}\nonumber\\[4pt]
&&\quad=
\frac{\pi^{\frac{1}{2}-\frac{\epsilon}{2}}}{\sqrt{a_2}} 
\,\Gamma[-1/2+\epsilon/2] \sum_{n_1\in\bZ}' (n_1+c_1)^v 
\Big[c+a_1(n_1+c_1)^2\Big]^{\frac{1}{2}-\frac{\epsilon}{2}}
\nonumber\\[5pt]
&& \!\!=
\frac{(\pi a_1)^{\frac{1}{2}-\frac{\epsilon}{2}}}{\sqrt a_2}
\!\sum_{k\geq 0} 
\bigg[\frac{-c}{a_1}\bigg]^k \frac{\Gamma[k\!-\!1/2\!+\!\epsilon/2]}{k!}
\Big[\zeta[2 k\!-\!q,1\!+\!c_1]\!+\!(-1)^{v}
\zeta[2 k\!-\! q,1\!-\!c_1]\Big]\bigg\vert_{q=v+1-\epsilon}
\label{rm}
\end{eqnarray}
where, in the second line above we used the binomial expansion
\begin{equation}\label{zp}
[a(n+c)^2+q]^{-s}=a^{-s}\sum_{k\geq 0}
\frac{\Gamma[k+s]}{k\,!\, \Gamma[s]} \bigg[\frac{-q}{a}\bigg]^k  
[(n+c)^2]^{-s-k}
\end{equation}
We employed    the Hurwitz
Zeta function,  $\zeta[z,a]=\sum_{n\geq 0} (a+n)^{-z}$,
 $a\not=0,-1,-2,\cdots$  for
$\textrm{Re}(z)\!>\!1$. One has $\zeta[z,1]=\zeta[z]$ where $\zeta[z]$
is the Riemann zeta function.
 Hurwitz zeta-function has one singularity
(simple pole) at $z=1$. Therefore, 
in the last line in (\ref{rm}), under the sum,
a singularity in Zeta functions is present 
for those  $k$  with $2k- v-1=1$. 
When present, this singularity is taken care of  by the  presence
of $\epsilon$ in the argument of Zeta functions. The presence of such
singularity depends on the values of the  parameter $v$.
We therefore distinguish below two situations:

\noindent
(i) $v=-2, 0,2,4,6,8,....$ when such a singularity is present in the term
 with $k=v/2+1$.\newline
\noindent
(ii) when $v$ is different from these values.

In case (ii) the result is already that given by (\ref{rm}) where
one (is allowed to) sets $\epsilon=0$ since the series does not develop
any singularity and converges rapidly under our initial assumption for the
ratio $0\leq c/a_1<1$. For  case (i), when a singularity develops,
we isolate the  corresponding term in the series from the rest, by using
\begin{eqnarray}
\zeta[1+\epsilon,1\pm c_1]& = & \frac{1}{\epsilon}
-\psi(1\pm c_1)+\cO(\epsilon)\nonumber\\[6pt]
\Gamma[v+1/2+\epsilon/2] &=& \Gamma[v+1/2] \
\Big(1+(\epsilon/2)\, \psi(v+1/2)\Big)
+\cO(\epsilon^2)
\nonumber\\[6pt]
x^\epsilon &= & 1+\epsilon \ln x+\cO(\epsilon)
\end{eqnarray}
with $\psi(z)=(d/dz) \ln \Gamma[z]$ the Digamma function.
In the remaining terms in the series we are allowed to take 
$\epsilon\rightarrow0$.
We find that for $v=-2,0,2,4,6,\cdots$

\begin{eqnarray}\label{K2v}
&&\!\!\!\cK^{(v)}_2\!=
\!\sqrt{\pi }u \sum_{k\geq 0}
\frac{\Gamma[k\!-\!1/2]}{k!}
\bigg[\frac{-c}{a_1}\bigg]^k
\Big[\zeta[2 k\!-\! v\!-\!1,1\!+\!c_1]\!+
\!\zeta[2k\!-\!v\!-\!1,1\!-\!c_1]\Big]\bigg\vert_{k\not=v/2+1}
\\[6pt]
&&\!\! -
\sqrt\pi \,u\,
\frac{\Gamma[v/2\!+\! 1/2]}{ (v/2\!+\!1)!} 
\bigg[\frac{-c}{a_1}\bigg]^{v/2+1} 
\bigg[\frac{-2}{\epsilon}+\ln\Big[\pi a_1
e^{-\psi(v/2+1/2)+\psi(c_1)+\psi(-c_1)}\Big]\bigg],\quad 
u\equiv\sqrt{a_1/a_2}
\nonumber\\
\nonumber
\end{eqnarray}
where the series converges quickly  if $\vert c/a_1\vert <1$,  which
justifies our (stronger) initial assumption $0\leq c/a_1 <1$. This
concludes the discussion for case (i).

Replacing now $v=0,1,2$ in the above result, 
one obtains the appropriate expressions
for $\cK^{(0)}$, $\cK^{(1)}$ and  $\cK^{(2)}$, that we need for our
purposes. One has
\begin{eqnarray}
\!\!\!\cK^{(0)}_2\!&=&
\frac{\pi \,c}{\sqrt{a_1 a_2}}\bigg[\frac{-2}{\epsilon}+\ln\Big[4 \pi
    \,a_1\, e^{\gamma_E+\psi(c_1)+\psi(-c_1)}\big]\bigg]
+ 2\pi \,u \Big(\frac{1}{6}+c_1^2\Big)\nonumber\\[8pt]
&+&\!\sqrt \pi \,u \sum_{p\geq 1}\frac{\Gamma[p\!+\!1/2]}{(p+1)!}
\bigg[\frac{-c}{a_1}\bigg]^{p\!+\!1}\!\!\Big(\zeta[2
  p\!+\!1,1\!+\!c_1]\!+\!\zeta[2p\!+\!1,1\!-\!c_1]\Big), \,\,
u\equiv\bigg[\frac{a_1}{a_2}\bigg]^\frac{1}{2}\,\,\label{K20}
\end{eqnarray}
and
\begin{eqnarray}
\cK^{(1)}_2&=& \sqrt\pi \,u \sum_{p\geq 0} 
\frac{\Gamma[p+3/2]}{(p+2)!} \bigg[\frac{-c}{a_1}\bigg]^{p+2}
\Big(\zeta[2p+2,1+c_1]-\zeta[2p+2,1-c_1]\Big)
\nonumber\\[8pt]
&+&2 \pi \,u\,c_1
\Big[\frac{1}{3}\,(1+2c_1^2)+\frac{c}{a_1}\Big],\qquad
u\equiv\sqrt{a_1/a_2}
\label{K21}
\end{eqnarray}
Finally
\begin{eqnarray}
\cK^{(2)}_2&=&
\!\!
\pi u\,\Big[\frac{-1}{30}\!+\! c_1^2 +c_1^4
\Big]\!+\!\frac{\pi \,c}{\sqrt{a_1 a_2}}\Big[\frac{1}{6}\!+\!c_1^2\Big]
\!-\!\frac{\pi\,c^2}{4 a_1 \sqrt{a_1 a_2}}\!
\bigg[\frac{-2}{\epsilon}\!+\!\ln\Big[4\pi a_1\,e^{\gamma_E-2
+\psi(c_1)+\psi(-c_1)}\Big]\bigg]
\nonumber\\[8pt]
&+&
\sqrt\pi \,u\sum_{p>0}\frac{\Gamma[p+3/2]}{(p+2)!}
\bigg[\frac{-c}{a_1}\bigg]^{p+2}\!\!
\Big(\zeta[2p+1,1+c_1]+\zeta[2p+1,1-c_1]\Big),\,\,
\label{K22}
\end{eqnarray}
\noindent
In the remaining case $1 \leq c/a_1$ we examine separately the cases
$v=0,1,2$. One shows:
\begin{eqnarray}\label{K20p}
&&\!\!\!\!\cK^{(0)}_2\equiv\sum_{n_1\in\bZ}' \frac{1}{\sqrt{a_2}}
\int_0^\infty
\frac{dt}{t^{3/2-\epsilon/2}}
\,e^{-\pi t a_1 (n_1+c_1)^2-\pi\, t\, c}\\
&&\!\!=\!
\frac{\pi c}{\sqrt {a_1 a_2}} \Big[\frac{-2}{\epsilon}\!+\!
\ln (\pi \,c\, e^{\gamma_E-1})
\Big]
\!+\! 4 \bigg[\frac{c}{a_2}\bigg]^{\frac{1}{2}}
\!\sum_{\tilde n_1>0}
\frac{\cos (2\pi \tilde n_1 c_1)}{\tilde n_1}
K_1\Big(2\pi \tilde n_1 \sqrt{\frac{c}{a_1}}\Big)
\!+\!
\frac{2\pi}{\sqrt a_2} (c+a_1 c_1^2)^{\frac{1}{2}}\nonumber
\end{eqnarray}
This expression was obtained by firstly  adding and 
subtracting a zero mode, which enabled us to then
re-sum (see (\ref{p_resumation}))
 the series over $n_1\in\bZ$. We then 
used the integral representation of the
modified Bessel functions $K_1$ (\ref{bessel1}). 
The pole present is that of the  initial ``missing''  zero mode.
The presence of the Bessel function $K_1[z]$ which is
exponentially suppressed (\ref{K_1})  
ensures that the result above converges rapidly in this case too.

One also has, for $v=1$   (again $1\leq c/a_1$):
\begin{eqnarray}
\cK_2^{(1)}&\equiv &\frac{1}{\sqrt a_1}
\sum_{n_1\in\bZ}' (n_1+c_1)\int_0^\infty\!
\frac{dt}{t^{3/2-\epsilon/2}}
\,e^{-\pi t a_1 (n_1+c_1)^2 -\pi t \,c}\\[6pt]
&=&
-\frac{1}{2 a_1\pi} \frac{1}{\sqrt a_2}
\frac{\partial}{\partial c_1}
\sum_{n_1\in\bZ}'\int_0^\infty \frac{dt}{t^{5/2-\epsilon/2}}
\,e^{-\pi t a_1 (n_1+c_1)^2-\pi\,t\,c}\nonumber\\[6pt]
&=&
-\frac{1}{2 a_1 \pi}\frac{1}{\sqrt a_2}
\frac{\partial}{\partial c_1}
\bigg\{
-
\frac{\pi^2 c^2}{2\sqrt a_1}\Big[\frac{-2}{\epsilon}
+\ln\big(\pi \, c\,e^{\gamma_E-3/2}\big)\Big]\nonumber\\[10pt]
&+& 4 \,c\,\sqrt{a_1}
\sum_{\tilde n_1>0}
\frac{\cos(2\pi \tilde n_1 c_1)}{\tilde n_1^2}
K_2(s_{\tilde n_1})
-\frac{4\pi^2}{3}(c+a_1 c_1^2)^{\frac{3}{2}}\bigg\}\nonumber
\\[10pt]
&=&
\!\frac{4 c}{\sqrt{a_1 a_2}}
\sum_{\tilde n_1>0}
\frac{\sin (2\pi \tilde n_1 c_1)}{\tilde n_1}
K_2(s_{\tilde n_1})
\!+\!\frac{2\pi c_1}{\sqrt{a_2}} (c+a_1 c_1)^\frac{1}{2},\,\,\,\,
s_{\tilde n_1}\!\equiv\! 2\pi\tilde n_1 \sqrt{c/a_1}\,\,\,\,
\label{K21p}
\end{eqnarray}
where the series converges rapidly, due to exponential suppression
of the Bessel  function $K_2$. To evaluate the integral
over $t$ with denominator $t^{5/2-\epsilon/2}$ one uses 
steps identical  to those for  $\cK_2^{(0)}$ with the only
difference that we encountered an integral representation of
$K_2$ rather than $K_1$.

Finally, for the remaining case $v=2$ ($1\leq c/a_1$):

\begin{eqnarray}
\!\!\cK_2^{(2)}&\!\equiv &\!\frac{1}{\sqrt a_1}
\sum_{n_1\in\bZ}' (n_1+c_1)^2\int_0^\infty
\frac{dt}{t^{3/2-\epsilon/2}}
\,e^{-\pi t a_1 (n_1+c_1)^2 -\pi \,t \,c}\\[6pt]
&=&
-\frac{1}{\pi} \frac{1}{\sqrt a_2}
\frac{\partial}{\partial a_1}
\sum_{n_1\in\bZ}'\int_0^\infty \frac{dt}{t^{5/2-\epsilon/2}}
\,e^{-\pi t a_1 (n_1+c_1)^2-\pi\,t\,c}
=
-\frac{1}{\pi}\frac{1}{\sqrt a_2}
\frac{\partial}{\partial a_1}
\bigg\{
\frac{-4\pi^2}{3}(c+a_1 c_1^2)^{\frac{3}{2}}\nonumber\\[6pt]
&-&
\frac{\pi^2 c^2}{2\sqrt a_1}\Big[\frac{-2}{\epsilon}
+\ln\big(\pi \, c\,e^{\gamma_E-\frac{3}{2}}\big)\Big]
-\frac{4\pi^2}{3}(c+a_1 c_1^2)^{\frac{3}{2}}
+4 c \sqrt{a_1}
\sum_{\tilde n_1>0}
\frac{\cos(2\pi \tilde n_1 c_1)}{\tilde n_1^2}
K_2(s_{\tilde n_1})\bigg\}\nonumber\\[6pt]
&=&
\frac{-\pi\, c^2}{4 a_1 \sqrt{a_1 a_2}}
\Big[\frac{-2}{\epsilon}\! +\! 
\ln\Big(\pi\,c\,e^{\gamma_E-\frac{3}{2}}\Big)\Big]
\!-\! \frac{2 c}{\pi \sqrt{a_1 a_2}}
\sum_{\tilde n_1>0} \!\!
\frac{\cos(2\pi\tilde n_1c_1)}{\tilde n_1^2}\!
\Big[3 K_2 (s_{\tilde n_1})\!
+\!s_{\tilde n_1} K_1(s_{\tilde n_1})\Big]\nonumber\\[6pt]
&+&
\frac{2\pi c_1^2}{\sqrt{a_2}}(c+a_1c_1^2)^\frac{1}{2},\qquad\qquad 
s_{\tilde n_1}\equiv 2\pi \tilde n_1 \sqrt{c/a_1}; \qquad c/a_1\geq 1.
\label{K22p}
\end{eqnarray}
with intermediate steps similar to those for $\cK_2^{(1)}$.

\vspace{0.5cm}
\noindent
{\bf -  Calculation of $\cK_3^{(v)}$:}

\vspace{0.5cm}
\noindent
Finally, we evaluate the remaining:

\begin{eqnarray}\label{def}
\cK_3^{(v)}&\equiv& \frac{1}{\sqrt{a_2}} 
\sum_{n_1\in\bZ}' \sum_{\tilde n_2\in\bZ}'
(n_1+c_1)^v \int_0^\infty
\!\frac{dt}{t^{3/2-\epsilon/2}}
\,\,e^{-\frac{\pi {\tilde n_2}^2}{t\,a_2}-\pi t\,
a_1\,(n_1+c_1)^2+ 2 \pi i {\tilde n_2}\, c_2 -\pi\,t\,c}
\\[6pt]
&=&
 \frac{1}{\sqrt{a_2}} 
\sum_{n_1\in\bZ}' \sum_{\tilde n_2>0}
(n_1+c_1)^v \frac{1}{\tilde n_2}
\, e^{-2\pi \tilde n_2 \,\gamma(n_1)}+c.c.\nonumber\\[6pt]
&=&
-\sum_{n_1\in\bZ}' (n_1+c_1)^v
\ln\Big\vert 1-e^{-2\pi \gamma(n_1)}\Big\vert^2\nonumber\\[6pt]
&=&
-\sum_{n_1\in\bZ} (n_1+c_1)^v
\ln\Big\vert 1-e^{-2\pi \gamma(n_1)}\Big\vert^2
-\frac{2 \pi c_1^v}{\sqrt a_2} (c+a_1 c_1^2)^\frac{1}{2}
+c_1^v\ln\Big\vert 2\sin (\pi i \gamma(0))\Big\vert^2\qquad
\label{K3v}\\[-3pt]
\nonumber
\end{eqnarray}
using the notations in eq.~(\ref{notations}). 
In the last line we re-wrote the result  in a
form which makes explicit the cancellations which occur in the sum
of $\cJ_v=\cK_1^{(v)}+\cK_2^{(v)}+\cK_3^{(v)}$.

The steps in the calculation of $\cK_3^{(v)}$ are
 similar to those so far: we used  
the integral representation of the
 Bessel function $K_{1/2}$ eq.~(\ref{bessel1}), 
then its explicit  expression (\ref{K_12}) and then the
series expansion of the logarithm.
The result for $\cK_3^{(v)}$ is valid for real $v$, not only
 for our cases of interest $v=0,1,2$, regardless of the 
value $c/a_1$ (larger/smaller than 1).

\bigskip
\noindent
We can now add the intermediate eqs to obtain $\cJ_{0,1,2}$
using eq.~(\ref{M1vsum}). 
 $\cJ_0$ quoted below  in (\ref{M1(0)}) and (\ref{M1(0)s}) 
is found from   eqs.~(\ref{K1v}), (\ref{K20}), (\ref{K20p}), (\ref{K3v}).
Further, $\cJ_1$ quoted in (\ref{M1(1)}) and
(\ref{M1(1)s})   is found  using  eqs.~(\ref{K1v}),
(\ref{K21}),(\ref{K21p}), (\ref{K3v}).
Finally $\cJ_2$ quoted in (\ref{M1(2)}) and (\ref{M1(2)s})
is obtained by using  (\ref{K1v}), (\ref{K22}), (\ref{K22p}), (\ref{K3v}).
In conclusion we have the following:

\smallskip
\bigskip
\noindent
{\bf Results:   }
If   $0\!\leq \!c/a_1\!<\! 1$ and with notations (\ref{notations}), 
$\gamma(n_1)\equiv {\sqrt{z(n_1)}}/{\sqrt a_2}-i\, c_2$;
and $z(n_1)\equiv c \!+\! a_1 (n_1\!+\!c_1)^2$,
$u\equiv\sqrt{a_1/a_2}$, 
$s_{\tilde n_1}\! \equiv\! 2\pi \tilde n_1 \sqrt{c/a_1}$,
$\gamma_E=0.577216...$ we obtain  (in the text $a_1=1/R_5^2$,
$a_2=1/R_6^2$

\begin{eqnarray}\label{M1(0)}
&&\! \!\cJ_0[c; c_1,c_2] %%%\equiv \cJ_1 \!
=  \! \frac{\pi c}{\sqrt{a_1 a_2}} \bigg[\frac{-2}{\epsilon}\!+\!
\ln\Big[4\pi \,a_1\,
e^{\gamma_E+\psi({c_1})+\psi(-{c_1})}\Big]\bigg] \!+ 2\pi\, u\,
\bigg[\frac{1}{6}+c_1^2-\big(c/a_1+c_1^2\big)^\frac{1}{2}\bigg]
\nonumber\\[6pt]
&&\!\!\!\!\!\!-\!\! \! \sum_{n_1\in\bZ}
\ln\Big\vert 1\!-\!e^{-2\pi \,\gamma (n_1)}\Big\vert^2
\!\!\! +\! \sqrt\pi\, u\!
\sum_{p\geq 1} \frac{\Gamma[p\!+\!1/2]}{(p\!+\!1)!}
\bigg[\frac{-c}{a_1}\bigg]^{p+1} 
\!\!\!\Big(\zeta[2p\!+\!1,1\!+\!c_1]\!+\!\zeta[2p\!+\!1,1\!-\!c_1]
\!\Big)\qquad\,\,\,
\\[8pt]
\nonumber
\end{eqnarray}
while if  we have   $c/a_1>1$,  then

\begin{eqnarray}\label{M1(0)s}
\,\,\cJ_0[c; c_1,c_2] &=& 
\frac{\pi c}{\sqrt{a_1 a_2}}\bigg[\! \frac{-2}{\epsilon}\!+\!
\ln\Big[
\!\pi \,c \,e^{\gamma_E-1}\Big]\bigg] 
\!\! -\!\!\!\sum_{n_1\in\bZ}\!\!\ln\Big\vert
 1\!-\!e^{-2\pi \,\gamma (n_1)}\Big\vert^2\nonumber
\qquad\qquad\qquad\qquad\\[8pt]
&&\qquad\qquad\qquad+ \,
\frac{4\sqrt c}{\sqrt a_2}\sum_{\tilde n_1>0}
\frac{\cos(2\pi \tilde n_1\,c_1)}{\tilde n_1}\,
K_1(s_{\tilde n_1})\\[-4pt]\nonumber
\end{eqnarray}
The pole structure is the same for both cases; if $c/a_1>1$
and except the first square bracket,
 no power-like terms in $c$ are present (the last one being suppressed
due to $K_1$).

Finally, we quote here a limiting case for the behaviour of the
function $\cJ_0$
\begin{eqnarray}\label{limit-j0}
\cJ_0[c\ll 1; 0,0] 
 &=&  \frac{\pi c}{\sqrt{a_1 a_2}} \bigg[\frac{-2}{\epsilon}
+\ln \Big[4\pi e^{-\gamma_E} 
a_1\big|\eta(i\sqrt{a_1/a_2})\big|^4\Big]\bigg] \nonumber \\
&&-\ln\Big[ 4\pi^2 \, \vert\eta(i \, \sqrt{a_1/a_2}) \vert^4 
\,\,a_2^{-1}\Big]-\ln c
\end{eqnarray}
and this was used in the text in eq.~(\ref{limit}).

\bigskip\bigskip
\noindent
Further, if  $0\leq c/a_1<1$
\begin{eqnarray}\label{M1(1)}
\cJ_1[c,c_1,c_2]&=&
2\pi c_1 \,u\,\bigg[\frac{c\,}{a_1}- 
(c/a_1+c_1^2)^\frac{1}{2} + \frac{1}{3}(1+2 c_1^2)\bigg]-
\!\sum_{n_1\in\bZ}  (n_1+c_1) \,
\ln\Big\vert 1- e^{-2\pi\gamma_{n_1}} \Big\vert^2\quad\nonumber\\[6pt]
&+&
\, \sqrt\pi\, u \sum_{p\geq 0}\frac{\Gamma(p+3/2)}{(p+2)!} 
\bigg[\frac{-c}{a_1}\bigg]^{p+2}\Big(\zeta[2p+2,1+c_1]
- \zeta[2 p+2,1-c_1]\Big)\\[-4pt]\nonumber
\end{eqnarray}
while if  $c/a_1>1$, then

\begin{eqnarray}\label{M1(1)s}
\cJ_1[c,c_1,c_2]
&=&\!- \sum_{n_1\in\bZ}
(n_1\!+\!c_1) \,
\ln\Big\vert 1 - e^{-2\pi\gamma(n_1)}\Big\vert^2
\!+\!
\frac{4\,c\,}{\sqrt{a_1 a_2}}
\!\sum_{\tilde n_1>0} 
\frac{\sin(2\pi \tilde n_1 c_1)}{\tilde n_1}\,
K_2(s_{\tilde n_1})\qquad\qquad\\[3pt]\nonumber
\end{eqnarray}
where $s_{\tilde n_1} \equiv 2\pi \tilde n_1 \sqrt{c/a_1}$.
Note that $\cJ_1$ has no poles in $\epsilon$, unlike the case of $\cJ_{0,2}$.
$K_1$ is exponentially suppressed at large argument.

\bigskip
\noindent
Finally, if  $0\leq c/a_1<1$
\begin{eqnarray}\label{M1(2)}
\cJ_2[c,c_1,c_2]& =&  -
\frac{\pi c^2}{4\, a_1\sqrt{a_1 a_2}} 
\bigg[\frac{-2}{\epsilon} 
+ \ln\Big[ 4\pi \,a_1\,
    e^{\gamma_E+\psi({c_1})+\psi(-{c_1})-2}\Big] \bigg]
\nonumber\\[5pt]
&-& 
\pi\,  u \,\bigg[\frac{1}{30}-\frac{c}{6 a_1}\!-\!c_1^2 
\Big(1- (c/a_1+c_1^2)^{\frac{1}{2}}\Big)^2\bigg]
- \sum_{n_1\in\bZ} (n_1+c_1)^2 \ln\Big\vert 1-e^{-2\pi
  \gamma(n_1)}\Big\vert^2 \nonumber\\[5pt]
&+&\! 
\sqrt\pi \,u\,\sum_{p\geq 1} \frac{\Gamma[p\!+\!3/2]}{(p\!+\!2)!}
\bigg[\frac{-c}{a_1}\bigg]^{p+2}
\!\!\Big(\zeta[2p\!+\!1,1\!+\!{c_1}]\!+\!
\zeta[2p\!+\!1,1\!-\!c_1]\Big).\\[-4pt]\nonumber
         \end{eqnarray}
while if $c/a_1>1$ then:

\begin{eqnarray}
\label{M1(2)s}
\cJ_2[c,c_1,c_2]
&=&
-\frac{\pi c^2}{4 a_1 \sqrt{a_1 a_2}}
\bigg[\frac{-2}{\epsilon}
+\ln\Big[\pi \, c\,\, e^{\gamma_E-3/2}\Big]\bigg]
- \sum_{n_1\in\bZ} (n_1+c_1)^2 \ln \Big\vert 1-e^{2 \pi
  \gamma(n_1)}\Big\vert^2\nonumber\qquad\\[7pt]
&-&
\frac{2 \,c}{\pi \sqrt{a_1 a_2}}
\sum_{\tilde n_1>0} \frac{\cos(2\pi \,\tilde n_1  c_1)}{\tilde n_1^2}
\, \Big[ 3 K_2(s_{\tilde n_1}) + s_{\tilde n_1}
K_1(s_{\tilde n_1})\Big],\\
\nonumber\qquad
\end{eqnarray}
where $s_{\tilde n_1} \equiv 2\pi \tilde n_1 \sqrt{c/a_1}$. 

The series with zeta functions converge under the assumption
$0\leq c/a_1<1$. 
The presence of Bessel functions $K_{1,2}$  (see (\ref{K_12})) 
which are exponentially suppressed with respect to their
 argument (larger than unity) ensures a rapid convergence
of the corresponding series.
Similar expressions exist for 
 $\cI_{v}=\cJ_{v}\vert_{c_1\leftrightarrow c_2;  a_1\leftrightarrow a_2}$;
 and  are obtained from those above
with replacements $a_1\leftrightarrow a_2$, $c_1\leftrightarrow c_2$.

\bigskip
\def\theequation{\thesubsection.\arabic{equation}} 
\def\thesubsection{E} 
\subsection{Definitions of special functions}\label{appendix-f}
\setcounter{equation}{0}

The modified Bessel functions $K_n(z)$ 
used above have the integral representation/definition:
\begin{equation}\label{bessel1}
\int_{0}^{\infty} \! dx\, x^{\nu-1} e^{- b x^p- a
x^{-p}}=\frac{2}{p}\, \bigg[\frac{a}{b}
\bigg]^{\frac{\nu}{2 p}} K_{\frac{\nu}{p}}(2 \sqrt{a \, b}),\quad \textrm{Re}
(b),\, \textrm{Re} (a)>0
\end{equation}
with
\begin{eqnarray}
K_1[x]&=& e^{-x}\sqrt{\frac{\pi}{2
    x}}\left[1+\frac{3}{8 x}-\frac{15}{128 x^2}
+\cO(1/x^3)\right]\label{K_1}\nonumber\\[7pt]
K_2[x]&=& e^{-x}\sqrt{\frac{\pi}{2 x}}\bigg[
1+\frac{15}{8 x}
+\frac{105}{128}\frac{1}{x^2}+\cO(1/x^3)\bigg]\nonumber\\[7pt]
K_{\frac{1}{2}}[x]&=&e^{-x}\sqrt{\frac{\pi}{2x}}\nonumber\\[7pt]
K_{\frac{3}{2}}[x]&=&e^{-x}\sqrt{\frac{\pi}{2x}}\bigg[1+\frac{1}{x}\bigg]
\label{K_12}
\end{eqnarray} 
The definition of the poly-logarithm function used above
\begin{eqnarray}\label{Li}
\Li_\sigma(x)=\sum_{x\geq 1} \frac{x^n}{n^\sigma}
\end{eqnarray}
The one-dimensional Poisson re-summation used in the appendix:
\begin{equation}\label{p_resumation}
\sum_{n\in Z} e^{-\pi A (n+\sigma)^2}=\frac{1}{\sqrt A}
\sum_{\tilde n\in Z} e^{-\pi A^{-1} \tilde n^2+2 i \pi \tilde n
\sigma}
\end{equation}
The Hurwitz Zeta function used in this paper is defined
as  
\bea
\zeta[z,a]=\sum_{n\geq 0} (a+n)^{-z}
\eea
where $a\not=0,-1,-2,\cdots$  for
$\textrm{Re}(z)\!>\!1$. One has 
$\zeta[z,1]=\zeta[z]$ where $\zeta[z]$ is the Riemann zeta function.
 Hurwitz zeta-function has one singularity (simple pole) at $z=1$.

We also used the Dedekind function
\begin{eqnarray}\label{ddt}
\eta(\tau) & \equiv & e^{\pi i \tau/12} \prod_{n\geq 1} (1- e^{2 i
\pi\tau\, n}),
\nonumber\\
\eta(-1/\tau) & = & \sqrt{-i \, \tau}\,\eta(\tau),
\qquad
\eta(\tau+1)=e^{i\pi/12}\eta(\tau).
\end{eqnarray}


\bigskip
\bigskip
\begin{thebibliography}{99}
\def\apj#1#2#3{Astrophys.\ J.\ {\bf #1}, #2 (#3)}
\def\ijmp#1#2#3{Int.\ J.\ Mod.\ Phys.\ {\bf #1}, #2 (#3)}
\def\mpl#1#2#3{Mod.\ Phys.\ Lett.\ {\bf #1}, #2 (#3)}
\def\nat#1#2#3{Nature\ {\bf #1} #2 (#3)}
\def\npb#1#2#3{Nucl.\ Phys.\ {\bf B #1}, #2 (#3)}
\def\plb#1#2#3{Phys.\ Lett.\ {\bf B #1}, #2 (#3)}
\def\prd#1#2#3{Phys.\ Rev.\ {\bf D #1}, #2 (#3)}
\def\prl#1#2#3{Phys.\ Rev.\ Lett.\ {\bf #1}, #2 (#3)}
\def\prt#1#2#3{Phys.\ Rep.\ {\bf #1}, #2 (#3)}
\def\sjnp#1#2#3{Sov.\ J.\ Nucl.\ Phys.\ {\bf #1}, #2 (#3)}
\def\zp#1#2#3{Z.\ Phys.\ {\bf C #1}, #2 (#3)}
\def\jhep#1#2#3{JHEP \ {\bf #1}, #2 (#3)}
\def\epjc#1#2#3{Eur.\ Phys.\ J. \ {\bf C #1}, #2 (#3)}


\bibitem{add}
  N.~Arkani-Hamed, S.~Dimopoulos and G.~R.~Dvali,
  {\it ``The hierarchy problem and new dimensions at a millimeter,''}
  Phys.\ Lett.\ B {\bf 429} (1998) 263
  [arXiv:hep-ph/9803315];\newline
I.~Antoniadis, N.~Arkani-Hamed, S.~Dimopoulos and G.~R.~Dvali,
 {\it  ``New dimensions at a millimeter to a Fermi and superstrings at a TeV,''}
  Phys.\ Lett.\ B {\bf 436} (1998) 257
  [arXiv:hep-ph/9804398]. See also \newline
  J.~D.~Lykken,
  {\it ``Weak Scale Superstrings,''}
  Phys.\ Rev.\ D {\bf 54} (1996) 3693
  [hep-th/9603133].

\bibitem{randall}
L.~Randall and R.~Sundrum,
  {\it ``An alternative to compactification,''}
  Phys.\ Rev.\ Lett.\  {\bf 83} (1999) 4690
  [arXiv:hep-th/9906064].

\bibitem{ggh} H.~Georgi, A.~K.~Grant and G.~Hailu,
  {\it ``Brane couplings from bulk loops,''}
  Phys.\ Lett.\ B {\bf 506} (2001) 207
  [arXiv:hep-ph/0012379].

\bibitem{schmaltz}
H.~C.~Cheng, K.~T.~Matchev and M.~Schmaltz,
  {\it ``Radiative corrections to Kaluza-Klein masses,''}
  Phys.\ Rev.\ D {\bf 66} (2002) 036005
  [arXiv:hep-ph/0204342].


%\cite{Ghilencea:2004sq}
\bibitem{Ghilencea:2004sq}
D.~M.~Ghilencea, 
{\it  ``Higher derivative operators as loop counterterms in one-dimensional
 field theory orbifolds''}
JHEP {\bf 0503} (2005) 009 [arXiv:hep-ph/0409214]
%%CITATION = HEP-PH 0409214; %%

%\cite{Ghilencea:2005hm}
\bibitem{Ghilencea:2005hm}
  D.~M.~Ghilencea and H.~M.~Lee,
  {\it ``Higher derivative operators from transmission
  of supersymmetry breaking  on
  S(1)/Z(2),''}
  JHEP {\bf 0509} (2005) 024  [arXiv:hep-ph/0505187].
  %%CITATION = HEP-PH 0505187;%%

%\cite{Ghilencea:2005nt}
\bibitem{Ghilencea:2005nt}
  D.~M.~Ghilencea and H.~M.~Lee,
  {\it ``Higher derivative operators from Scherk-Schwarz supersymmetry breaking on
  T**2/Z(2),''}
  JHEP {\bf 0512} (2005) 039  [arXiv:hep-ph/0508221]; 

\bibitem{santamaria}
%\cite{Oliver:2003cy}
%\bibitem{Oliver:2003cy}
  J.~F.~Oliver, J.~Papavassiliou and A.~Santamaria,
  {\it ``Can power corrections be reliably computed in models with extra
  dimensions?,''}
  Phys.\ Rev.\ D {\bf 67} (2003) 125004
  [arXiv:hep-ph/0302083].
  %%CITATION = HEP-PH 0302083;%%


\bibitem{nibbelink} S.~Groot Nibbelink and M.~Hillenbach,
  {\it ``Renormalization of supersymmetric gauge theories on orbifolds: Brane gauge
  couplings and higher derivative operators,''}
  Phys.\ Lett.\ B {\bf 616} (2005) 125
  [arXiv:hep-th/0503153].

\bibitem{nibbmark}
 S.~Groot Nibbelink and M.~Hillenbach,
  {\it ``Quantum Corrections to Non-Abelian SUSY Theories on Orbifolds,''}
  arXiv:hep-th/0602155.

%\cite{Ghilencea:2003xj}
\bibitem{Ghilencea:2003xj}
  D.~M.~Ghilencea,
  {\it ``Compact dimensions and their radiative mixing,''}
  Phys.\ Rev.\ D {\bf 70} (2004) 045018
  [arXiv:hep-ph/0311264].
  %%CITATION = HEP-PH 0311264;%%

%\cite{Alvarez:2006we}
\bibitem{Alvarez:2006we}
  E.~Alvarez and A.~F.~Faedo,
  {\it ``Renormalized Kaluza-Klein theories,''}
  arXiv:hep-th/0602150.
  %%CITATION = HEP-TH 0602150;%%



%\cite{Taylor:1988vt}
\bibitem{Taylor:1988vt}
 I.~Antoniadis,
  {\it ``A Possible New Dimension At A Few Tev,''}
  Phys.\ Lett.\ B {\bf 246} (1990) 377.\newline
T.~R.~Taylor and G.~Veneziano,
{\it ``Strings And D = 4,''}
Phys.\ Lett.\ B {\bf 212} (1988) 147.\newline
%%CITATION = PHLTA,B212,147;%%
%\cite{Lanzagorta:1995gp}
%\bibitem{Lanzagorta:1995gp}
M.~Lanzagorta and G.~G.~Ross,
{\it ``Infrared fixed points revisited,''}
Phys.\ Lett.\ B {\bf 349} (1995) 319
[arXiv:hep-ph/9501394].\newline
%%CITATION = HEP-PH 9501394;%%
%\bibitem{Dienes:1998vg}
K.R. Dienes, E. Dudas, T. Gherghetta,
{\it ``Grand unification at intermediate mass scales through extra dimensions,''}
Nucl.Phys.B {\bf 537} (1999) 47
[hep-ph/9806292];\newline
%\cite{Contino:2001si}
%\bibitem{Contino:2001si}
  R.~Contino, L.~Pilo, R.~Rattazzi and E.~Trincherini,
  {\it ``Running and matching from 5 to 4 dimensions,''}
  Nucl.\ Phys.\ B {\bf 622} (2002) 227
  [arXiv:hep-ph/0108102];\newline
  %%CITATION = HEP-PH 0108102;%%
%\cite{Hall:2001xb}
%\bibitem{Hall:2001xb}
  L.~J.~Hall and Y.~Nomura,
  {\it ``Gauge coupling unification from unified theories 
in higher dimensions,''}
  Phys.\ Rev.\ D {\bf 65} (2002) 125012
  [arXiv:hep-ph/0111068];\newline
  %%CITATION = HEP-PH 0111068;%%
%\cite{Kim:2001at}
%\bibitem{Kim:2001at}
  H.~D.~Kim, J.~E.~Kim and H.~M.~Lee,
  {\it ``Top-bottom mass hierarchy, s - mu puzzle and gauge coupling unification
  with split multiplets,''}
  Eur.\ Phys.\ J.\ C {\bf 24} (2002) 159 [hep-ph/0112094];\newline
  %%CITATION = HEP-PH 0112094;%%
%%\bibitem{Ghilencea:200dg}
D.~M.~Ghilencea, G.~G.~Ross, 
{\it ``On gauge unification in type I/I'  models''}
Phys.~Lett.~B {\bf 480} (2000) 355. [arXiv:hep-ph/0001143]; 
%%CITATION = HEP-PH 0001143;%%
{\it ``Unification and extra space-time dimensions''}, 
Phys.~Lett. B {\bf 442} (1998) 165, [arXiv:hep-ph/9809217]\newline
%% CITATION = HEP-PH 9809217;%%
%\cite{Choi:2002wx}
%\bibitem{Choi:2002wx}
  K.~w.~Choi, H.~D.~Kim and I.~W.~Kim,
  {\it ``Gauge coupling renormalization in orbifold field theories,''}
  JHEP {\bf 0211} (2002) 033
  [arXiv:hep-ph/0202257];
  %%CITATION = HEP-PH 0202257;%%
%\cite{Choi:2002zi}
%\bibitem{Choi:2002zi}
  %K.~w.~Choi, H.~D.~Kim and I.~W.~Kim,
  {\it ``Radius dependent gauge unification in AdS(5),''}
  JHEP {\bf 0303} (2003) 034 [arXiv:hep-ph/0207013];\newline
  %%CITATION = HEP-PH 0207013;%%
%\cite{Choi:2002ps}
%\bibitem{Choi:2002ps}
  K.~w.~Choi and I.~W.~Kim,
  {\it ``One loop gauge couplings in AdS(5),''}
  Phys.\ Rev.\ D {\bf 67} (2003) 045005
  [arXiv:hep-th/0208071];\newline
  %%CITATION = HEP-TH 0208071;%%
%\cite{Hebecker:2002vm}
%\bibitem{Hebecker:2002vm}
  A.~Hebecker and A.~Westphal,
  {\it ``Power-like threshold corrections to gauge unification in extra
  dimensions,''}
  Annals Phys.\  {\bf 305} (2003) 119
  [arXiv:hep-ph/0212175];
  %%CITATION = HEP-PH 0212175;%%
%\cite{Hebecker:2004xx}
%\bibitem{Hebecker:2004xx}
  %A.~Hebecker and A.~Westphal,
  {\it ``Gauge unification in extra dimensions: Power corrections vs.
  higher-dimension operators,''}
  Nucl.\ Phys.\ B {\bf 701} (2004) 273
  [arXiv:hep-th/0407014];
  %%CITATION = HEP-TH 0407014;%%
%\bibitem{Hebecker:2004ce}
  A.~Hebecker and M.~Trapletti,
  {\it ``Gauge unification in highly anisotropic string compactifications,''}
  Nucl.\ Phys.\ B {\bf 713} (2005) 173
  [arXiv:hep-th/0411131].


%\cite{Ghilencea:2002ak} 
\bibitem{Ghilencea:2002ak}
  D.~M.~Ghilencea,
  {\it ``On gauge couplings, 'large' extra-dimensions
and the limit $\alpha'\to 0$
  of the string,''}
  Nucl.\ Phys.\ B {\bf 653} (2003) 27
  [arXiv:hep-ph/0212119].
  %%CITATION = HEP-PH 0212119;%%


%\cite{O'Connor:zj}
\bibitem{O'Connor:zj} 
D.~O'Connor, C.~R.~Stephens,
{\it ``Renormalization Group Theory Of Crossovers,''}
Phys.\ Rept.\  {\bf 363} (2002) 425.
%%CITATION = PRPLC,363,425;

\bibitem{ADB}
I.Antoniadis, C.Bachas, E. Dudas,  
{\it ``Gauge couplings in four-dimensional type I string orbifolds,''}
Nucl. Phys. B {\bf 560} (1999) 93; [arXiv:hep-th/9906039].

 \bibitem{Dixon:1990pc2} 
 L.~J.~Dixon, V.~Kaplunovsky and J.~Louis,
 {\it ``Moduli Dependence Of String Loop Corrections To Gauge 
 Coupling Constants,''}
Nucl.\,Phys.\,B {\bf 355} (1991) 649.
 \, See also ref.\cite{Mayr:1993mq}.

 \bibitem{Mayr:1993mq}
   P.~Mayr and S.~Stieberger,
  {\it ``Threshold corrections to gauge couplings in orbifold compactifications,''}
  Nucl.\ Phys.\ B {\bf 407} (1993) 725
  [arXiv:hep-th/9303017].

\bibitem{wz}
%\cite{Wess:1974jb}
%\bibitem{Wess:1974jb}
  J.~Wess and B.~Zumino,
  {\it ``Supergauge invariant extension Of quantum electrodynamics,''}
  Nucl.\ Phys.\ B {\bf 78} (1974) 1;
  %%CITATION = NUPHA,B78,1;%%
J.~Wess and J.~Bagger,
{\it Supersymmetry and supergravity} (1992), 
Princeton Series in Physics.

\bibitem{jones}
J.~R.~T.~Jones, {\it Supersymmetric gauge theories},
in Proceedings, 
TASI lectures in elementary particle physics 1984, Ann Arbor, p291.

\bibitem{ah}
%\cite{Arkani-Hamed:1997mj}
%\bibitem{Arkani-Hamed:1997mj}
  N.~Arkani-Hamed and H.~Murayama,
  {\it ``Holomorphy, rescaling anomalies and exact beta functions in  supersymmetric
  gauge theories,''}
  JHEP {\bf 0006} (2000) 030
  [arXiv:hep-th/9707133].
  %%CITATION = HEP-TH 9707133;%%


\bibitem{lenizu}  
H.~M.~Lee, H.~P.~Nilles and M.~Zucker,
  {\it ``Spontaneous localization of bulk fields: The six-dimensional case,''}
  Nucl.\ Phys.\ B {\bf 680} (2004) 177
  [arXiv:hep-th/0309195].

\bibitem{gr}
I.S. Gradshteyn, I.M. Ryzhik, {\it ``Tables of Integral, Series and
  Products''}, Academic Press Inc., New York/London, 1980.

\bibitem{peskin}
M.~E.~Peskin and D.~V.~Schroeder, {\it An Introduction to Quantum Field
Theory}, Addison Wesley Publishing Company (1995) 535p.



\bibitem{SO10} 
%\cite{Kawamura:2000ev}
%\bibitem{Kawamura:2000ev}
  Y.~Kawamura,
  {\it ``Triplet-doublet splitting, proton stability and extra dimension,''}
  Prog.\ Theor.\ Phys.\  {\bf 105} (2001) 999
  [arXiv:hep-ph/0012125];\newline
  %%CITATION = HEP-PH 0012125;%%
%\cite{Altarelli:2001qj}
%\bibitem{Altarelli:2001qj}
  G.~Altarelli and F.~Feruglio,
  {\it ``SU(5) grand unification in extra dimensions and proton decay,''}
  Phys.\ Lett.\ B {\bf 511} (2001) 257
  [arXiv:hep-ph/0102301];\newline
  %%CITATION = HEP-PH 0102301;%%
%\cite{Kobakhidze:2001yk}
%\bibitem{Kobakhidze:2001yk}
  A.~B.~Kobakhidze,
  {\it ``Proton stability in TeV-scale GUTs,''}
  Phys.\ Lett.\ B {\bf 514} (2001) 131
  [arXiv:hep-ph/0102323];
  %%CITATION = HEP-PH 0102323;%%
%\cite{Hall:2001pg}
%\bibitem{Hall:2001pg}
  L.~J.~Hall and Y.~Nomura,
  {\it ``Gauge unification in higher dimensions,''}
  Phys.\ Rev.\ D {\bf 64} (2001) 055003
  [arXiv:hep-ph/0103125].
  %%CITATION = HEP-PH 0103125;%%
%\bibitem{Hebecker:2001wq}
%  A.~Hebecker and J.~March-Russell,
  {\it ``A minimal S(1)/(Z(2) x Z'(2)) orbifold GUT,''}
  Nucl.\ Phys.\ B {\bf 613} (2001) 3
  [arXiv:hep-ph/0106166];\newline
 L.~J.~Hall, Y.~Nomura, T.~Okui and D.~R.~Smith,
  {\it``SO(10) unified theories in six dimensions,''}
  Phys.\ Rev.\ D {\bf 65} (2002) 035008
  [arXiv:hep-ph/0108071].\newline
% \bibitem{Hebecker:2001jb}
  A.~Hebecker and J.~March-Russell,
  {\it ``The structure of GUT breaking by orbifolding,''}
  Nucl.\ Phys.\ B {\bf 625} (2002) 128
  [arXiv:hep-ph/0107039];
  {\it ``Proton decay signatures of orbifold GUTs''}
  Phys.\ Lett.\ B {\bf 539} (2002) 119
  [arXiv:hep-ph/0204037];\newline
T.~Asaka, W.~Buchmuller and L.~Covi,
  {\it ``Gauge unification in six dimensions,''}
  Phys.\ Lett.\ B {\bf 523} (2001) 199
  [arXiv:hep-ph/0108021];
  {\it ``Bulk and brane anomalies in six dimensions,''}
  Nucl.\ Phys.\ B {\bf 648} (2003) 231
  [arXiv:hep-ph/0209144];
  {\it ``Quarks and leptons between branes and bulk,''}
  Phys.\ Lett.\ B {\bf 563} (2003) 209
  [arXiv:hep-ph/0304142];\newline
W.~Buchmuller, L.~Covi, D.~Emmanuel-Costa and S.~Wiesenfeldt,
  {\it ``Flavour structure and proton decay in 6D orbifold GUTs,''}
  JHEP {\bf 0409} (2004) 004
  [arXiv:hep-ph/0407070]; 
W.~Buchmuller, J.~Kersten and K.~Schmidt-Hoberg
  {\it "Squarks and sleptons between branes and bulk,''}
  JHEP {\bf 02} (2006) 069 
  [arXiv:hep-ph/0512152]. 




\bibitem{new}
D.~M.~Ghilencea, H.~M.~Lee, K.~Schmidt-Hoberg, work in progress.

\bibitem{swh}
S. W. Hawking, {\it ``Quantum Field theory and Quantum Statistics:
Essays in Honour of the 60 th Birthday of E.S. Fradkin''},
eds. A.~Batalin et al, Bristol 1987;\newline
S.~W.~Hawking and T.~Hertog,
{\it ``Living with ghosts,''}
Phys.\ Rev.\ D {\bf 65} (2002) 103515 [hep-th/0107088]. 

 \bibitem{Bergshoeff:1986jm}
  E.~Bergshoeff, M.~Rakowski and E.~Sezgin,
  {\it ``Higher Derivative Super Yang-Mills Theories,''}
  Phys.\ Lett.\ B {\bf 185} (1987) 371;

\bibitem{Simon:1990ic}
J.~Z.~Simon,
{\it ``Higher Derivative Lagrangians, Nonlocality, Problems And Solutions,''}
Phys.\ Rev.\ D {\bf 41} (1990) 3720; 

\bibitem{Smilga:2005gb}
A.~V.~Smilga,
{\it ``Ghost-free higher-derivative theory,''} hep-th/0503213; 
{\it ``Benign vs. malicious ghosts in higher-derivative theories,''}
Nucl.\ Phys.\ B {\bf 706} (2005) 598
[hep-th/0407231]; 

\bibitem{Pais}
  A.~Pais and G.~E.~Uhlenbeck,
{\it ``On Field Theories With Nonlocalized Action,''}
  Phys.\ Rev.\  {\bf 79} (1950) 145; 

\bibitem{deUrries:1998bi}
F.~J.~de Urries and J.~Julve,
{\it ``Ostrogradski formalism for higher-derivative scalar field theories,''}
J.\ Phys.\ A {\bf 31} (1998) 6949
[hep-th/9802115]; 

\bibitem{Mannheim:2004qz}
P.~D.~Mannheim and A.~Davidson,
{\it ``Dirac quantization of the Pais-Uhlenbeck fourth order oscillator,''}
hep-th/0408104; 
{\it ``Fourth order theories without ghosts,''}
hep-th/0001115.


\bibitem{wagner}
%\cite{Carena:2002me}
%\bibitem{Carena:2002me}
  M.~Carena, T.~M.~P.~Tait and C.~E.~M.~Wagner,
  {\it ``Branes and orbifolds are opaque,''}
  Acta Phys.\ Polon.\ B {\bf 33} (2002) 2355
  [arXiv:hep-ph/0207056];
  %%CITATION = HEP-PH 0207056;%%
%\cite{Carena:2002dz}
%\bibitem{Carena:2002dz}
  M.~Carena, E.~Ponton, T.~M.~P.~Tait and C.~E.~M.~Wagner,
  {\it ``Opaque branes in warped backgrounds,''}
  Phys.\ Rev.\ D {\bf 67} (2003) 096006
  [arXiv:hep-ph/0212307].
  %%CITATION = HEP-PH 0212307;%%
%\cite{delAguila:2003bh}
%\bibitem{delAguila:2003bh}
  F.~del Aguila, M.~Perez-Victoria and J.~Santiago,
{\it  ``Bulk fields with general brane kinetic terms,''}
  JHEP {\bf 0302} (2003) 051
  [arXiv:hep-th/0302023];
  %%CITATION = HEP-TH 0302023;%%
%\cite{Chaichian:2002uy}
%\bibitem{Chaichian:2002uy}
  M.~Chaichian and A.~Kobakhidze,
  {\it ``Kaluza-Klein decomposition and gauge coupling unification in orbifold
  GUTs,''}
  arXiv:hep-ph/0208129.
  %%CITATION = HEP-PH 0208129;%%


\bibitem{Kaplunovsky:1992vs}
%\cite{Kaplunovsky:1987rp}
%\bibitem{Kaplunovsky:1987rp}
V.~S.~Kaplunovsky,
{\it ``One Loop Threshold Effects In String Unification,''}
Nucl.\,Phys.\,B {\bf 307} (1988) 145;
[Erratum-ibid.\,B {\bf 382} (1988) 436] \newline [arXiv:hep-th/9205068];
%%CITATION = HEP-TH 9205068;%%
(For a completely revised version see hep-th/9205070). 

%\cite{Ghilencea:2002ff}
\bibitem{Ghilencea:2002ff}
D.~Ghilencea, S.~Groot-Nibbelink,
{\it ``String threshold corrections from field theory,''}
Nucl.\ Phys.\ B {\bf 641} (2002) 35
[arXiv:hep-th/0204094].
%%CITATION = HEP-TH 0204094;%%

\end{thebibliography}
\end{document}